\documentclass[epj]{svjour}
\usepackage{graphics}
\usepackage{graphicx}
\usepackage{xcolor}
\usepackage{sidecap}
\usepackage{microtype}
\usepackage{url}
\usepackage{lineno}

\begin{document}
%
\title{The BGOOD experimental setup at ELSA}
\author{%
S.~Alef\inst{1}%
\and P.~Bauer\inst{1}%
\and D.~Bayadilov\inst{2,3}%
\and R.~Beck\inst{2}%
\and M.~Becker\inst{2,}\thanks{No longer employed in academia}%
\and A.~Bella\inst{1,\mathrm{a}}%
\and J.~Bieling\inst{2, \mathrm{a}}
\and S.~B\"ose\inst{2, \mathrm{a}}%
\and A.~Braghieri\inst{4}%
\and \newline K.-Th.~Brinkmann\inst{5}%
\and P.L.~Cole\inst{6}%
\and R.~Di Salvo\inst{7}%
\and D.~Elsner\inst{1}%
\and A.~Fantini\inst{7,8}%
\and O.~Freyermuth\inst{1}%
\and F.~Frommberger\inst{1}%
\and G.~Gervino\inst{9,10}%
\and F.~Ghio\inst{11,12}%
\and S.~Goertz\inst{1}%
\and A.~Gridnev\inst{3}%
\and E.~Gutz\inst{5}%
\and D.~Hammann\inst{1, \mathrm{a}}%
\and J.~Hannappel\inst{1,}\thanks{Currently, DESY Research Centre, Hamburg, Germany} 
\and W.~Hillert\inst{1,\mathrm{b}}%
\and O.~Jahn\inst{1,\mathrm{a}}%
\and R.~Jahn\inst{2, }\thanks{Retired}%
\and J.R.~Johnstone\inst{1,\mathrm{a}}
\and R.~Joosten\inst{2}%
\and T.C.~Jude\inst{1}
\and H.~Kalinowsky\inst{2,\mathrm{c}}
\and V.~Kleber\inst{1,}\thanks{Currently, Forschungszentrum J\"ulich, Germany}
\and F.~Klein\inst{1}
\and K.~Kohl\inst{1}
\and K.~Koop\inst{2,\mathrm{a}}%
\and N.~Kozlenko\inst{3}%
\and B.~Krusche\inst{13}%
\and A.~Lapik\inst{14}%
\and P.~Levi Sandri\inst{15}
\and V.~Lisin\inst{14}%
\and I.~Lopatin\inst{3}%
\and G.~Mandaglio\inst{16,17}%
\and M.~Manganaro\inst{16,17,}\thanks{Currently, University of Rijeka, Croatia}
\and F.~Messi\inst{1,}\thanks{Currently, Lund University \& ESS, Sweden}
\and R.~Messi\inst{7,8}%
\and D.~Moricciani\inst{7}%
\and A.~Mushkarenkov\inst{14}%
\and V.~Nedorezov\inst{14}%
\and D.~Novinskiy\inst{3}%
\and P.~Pedroni\inst{4}%
\and A.~Polonskiy\inst{14}
\and B.-E.~Reitz\inst{1}%
\and M.~Romaniuk\inst{7,18}%
\and T.~Rostomyan\inst{13}%
\and G.~Scheluchin\inst{1}%
\and H.~Schmieden\inst{1}%
\and A.~Stugelev\inst{3}%
\and V.~Sumachev\inst{3}%
\and V.~Tarakanov\inst{3}%
\and V.~Vegna\inst{1}%
\and D.~Walther\inst{2,\mathrm{c}}%
\and H.-G.~Zaunick\inst{2,5}%
\and T.~Zimmermann\inst{1,\mathrm{a}}%
}
\institute{%
Rheinische Friedrich-Wilhelms-Universit\"at Bonn, Physikalisches Institut, Nu\ss allee 12, 53115 Bonn, Germany%
\and Rheinische Friedrich-Wilhelms-Universit\"at Bonn, Helmholtz-Institut f\"ur Strahlen- und Kernphysik, Nu\ss allee 14-16, 53115 Bonn, Germany%
\and Petersburg Nuclear Physics Institute, Gatchina, Leningrad District, 188300, Russia%
\and INFN sezione di Pavia, Via Agostino Bassi, 6 - 27100 Pavia, Italy%
\and Justus-Liebig-Universit\"at Gie\ss en, II. Physikalisches Institut, Heinrich-Buff-Ring 16, 35392 Gie\ss en, Germany%
\and Lamar University, Department of Physics, Beaumont, Texas, 77710, USA
\and INFN Roma ``Tor Vergata", Via della Ricerca Scientifica 1, 00133, Rome, Italy%
\and Universit\`a di Roma ``Tor Vergata'', Dipartimento di Fisica, Via della Ricerca Scientifica 1, 00133, Rome, Italy%
\and INFN sezione di Torino, Via P.Giuria 1, 10125, Torino, Italy%
\and Universit\`a di Torino, Dipartimento di Fisica,  via P. Giuria 1, 10125, Torino, Italy%
\and INFN sezione di Roma La Sapienza, P.le Aldo Moro 2, 00185, Rome, Italy %
\and Istituto Superiore di Sanit\`a, Viale Regina Elena 299, 00161, Rome, Italy %
\and Institut f\"ur Physik, Klingelbergstrasse 82, CH-4056 Basel, Switzerland%
\and Russian Academy of Sciences Institute for Nuclear Research, Prospekt 60-letiya Oktyabrya 7a, 117312, Moscow, Russia%
\and INFN - Laboratori Nazionali di Frascati, Via E. Fermi 54, 00044, Frascati, Italy%
\and INFN sezione Catania, 95129, Catania, Italy%
\and Universit\`a degli Studi di Messina, Dipartimento MIFT,  Via F. S. D'Alcontres 31, 98166, Messina, Italy%
\and Institute for Nuclear Research of NASU, 03028, Kyiv, Ukraine
}

%
\mail{\newline T.C.~Jude, \url{jude@physik.uni-bonn.de}\newline
P.~Levi Sandri, \url{paolo.levisandri@inf.infn.it}}
\date{Received: date / Revised version: date}
%
\abstract{
The BGOOD experiment at the ELSA facility in Bonn has been commissioned within the framework of an international collaboration.
The experiment pursues a systematic investigation of non-strange and strange meson photoproduction, in particular $t$-
channel processes at low momentum transfer. The setup uniquely combines a central almost $4\pi$ acceptance BGO crystal calorimeter with a large aperture forward magnetic spectrometer providing excellent detection
of both neutral and charged particles, complementary to other setups such as Crystal Barrel, Crystal Ball, LEPS and CLAS.
\PACS{
      {13.60.Le} {Photoproduction of mesons}
      {25.20.-x} {Photonuclear reactions}
     } 
} 
\maketitle
\setcounter{tocdepth}{10}

\vfill\null 

\tableofcontents

\section{Introduction}
\label{sec:introduction}
Photoproduction of mesons as a probe of the excitation structure of the nucleon has been exploited since the 1960s.
The use of GeV-range energy tagged photon beams at high
duty cycle electron accelerators in combination with large
acceptance detectors, like CLAS and GlueX at Jefferson Laboratory \cite{Mecking:2003zu}, A2 at MAMI \cite{Schumann2010}, Crystal Barrel at ELSA \cite{Aker:1992ny,Gabler:1994ay,Novotny:1991ht}, GrAAL at ESRF~\cite{Bartalini:2005wx}, and LEPS at SPring8~\cite{Nakano:2001xp}, has in recent years put the technique on par with pion scattering to
unravel the complex nucleon excitation patterns. 
The experiments have significantly added to our understanding of baryon excitations. Nevertheless, crucial features of the spectrum still remain unresolved. These are often associated with close-by photoreaction thresholds, for example the $S_{11}(1535)$ resonance close to the $\eta N$ 
and $\bar{K}\Lambda$ thresholds, or the $\Lambda(1405)$ at the $\bar{K}$N threshold. Moreover, unexpected and hitherto unexplained structures are observed in photoproduction: examples are a narrow peak in $\eta$ photoproduction off the neutron at the almost degenerate $\bar{K}\Sigma$ and $\omega$N thresholds~\cite{kuznetsov07,jaegle08,jaegle11,miyahara07}, or a cusp-like fall off of the forward (and total) cross section in $K^0_s\Sigma^+$ photoproduction at the $K^*\Sigma$ threshold~\cite{ewald12}.


The BGOOD detector is especially designed to investigate meson photoproduction at thresholds and at
low momentum transfer, {\it t}, to the residual hadronic system. It consists of two main parts: a forward large aperture magnetic spectrometer, with an {\it Open Dipole} magnet, and a central detector with a {\it BGO} crystal calorimeter, both eponymous for the whole experiment. BGOOD is situated at the ELSA electron accelerator facility at the {\it Rheinische Friedrich-Wilhelms-Universit\"at}, Bonn, Germany. Using the ELSA electron beam, an energy tagged bremsstrahlung photon beam is produced, impinging upon either a cryogenic liquid hydrogen or deuterium target, or alternatively, a solid state target such as carbon, at the centre of the \textit{BGO Rugby Ball}.

An amorphous bremsstrahlung radiator yields unpolarised photon beams, with linearly polarised photon beams generated by
coherent bremsstrahlung using a diamond crystal radiator. Instantaneous tagged photon intensities of 25\,MHz are routinely achieved in the energy range (10$\div$90)\% of the incident electron beam energy. 


This paper is organised as follows:  sect.~\ref{sec:ELSA}  describes the ELSA accelerator, and sects.~\ref{sec:Detector} and \ref{sec:Tagger}, describe the setup and electronics for the  BGOOD detector components and the photon tagging system.  The trigger system and data acquisition are detailed in sects.~\ref{sec:TriggerDAQ} and \ref{sec:DAQ}.  The performance of the beam and detector components are discussed in sects.~\ref{sec:Beam_perf} and \ref{sec:Det_perf}.  Finally, ``bench mark" physics measurements are presented in sect.~\ref{sec:Examples_results}.

\section{The ELSA Accelerator}
\label{sec:ELSA}

\begin{figure*}[hbt]
	\centering
	{\includegraphics[width=\textwidth,trim={0cm 0 0 0cm},clip=true]{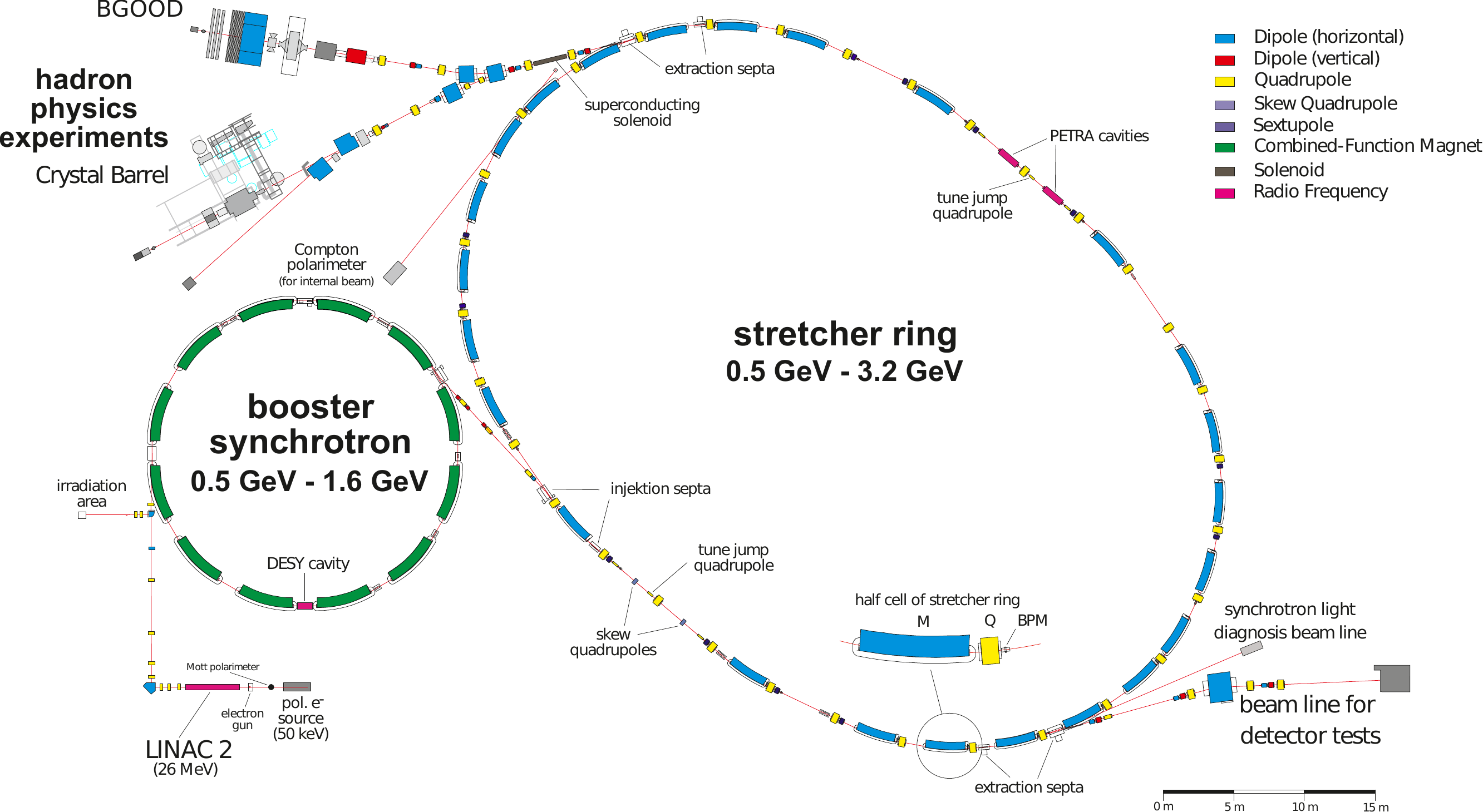}}
	\caption{Overview of the ELSA accelerator facility.  Figure adapted from ref.~\cite{Hillert:20017}.}
	\label{fig:ELSA}       
\end{figure*}

The ELectron Stretcher Accelerator (ELSA)~\cite{Hillert:2006yb,Hillert:20017}, shown
in fig.~\ref{fig:ELSA}, consists of three stages. Electrons are released 
either from a source of spin polarised electrons, or 
by a thermionic electron source. These electrons are injected 
into the LINAC and accelerated up to an energy of 26\,MeV.
The electron beam is then transferred to the second stage, the 
booster synchrotron (combined function machine) of 69.9\,m circumference.
The booster synchrotron runs
with a fixed cycle time of 20\,ms, corresponding to a 50\,Hz repetition rate, accelerating electron beams typically up to 1.2\,GeV (a maximum of 1.6\,GeV is possible). 
When the maximum energy is reached, the beam
is extracted within one revolution and transferred 
to the third stage, the stretcher ring.

The stretcher is an ovally shaped ring of 164.4\,m circumference comprised of 16 FODO\footnote{Focussing quadrupole, Open drift space, Defocussing quadrupole, Open drift space.}  cells.
Shown in fig.~\ref{fig:ELSA} as the blue and yellow components,
these are magnet lattices consisting of
horizontally focusing quadrupoles,
drift space (which can contain dipole magnets) and vertically focusing, horizontally defocussing quadrupoles.
 
Sextupole magnets are installed 
for chromaticity correction and for driving
the resonance during the extraction phase (described below).
Several injections from the booster synchrotron are accumulated in the
stretcher ring in such a way that a homogeneous filling of the ring
is obtained. When the stored current reaches a given upper threshold (typically
20 to 25\,mA for hadron physics experiments), the injection process is stopped and
the energy of the stored electron beam is accelerated up
to the desired energy (maximum 3.2\,GeV). The beam is then extracted slowly
by means of resonance extraction at a third integer betatron resonance (4
2/3) to either an area for detector tests (bottom right of fig.~\ref{fig:ELSA}), or 
one of the two hadron physics experiments, Crystal Barrel~\cite{Aker:1992ny,Gabler:1994ay,Novotny:1991ht} or BGOOD (top
left of fig.~\ref{fig:ELSA}).

\begin{figure*}[hbt]
	\centering
	{\includegraphics[width=1\textwidth,trim={0cm 0 0 0cm},clip=true]{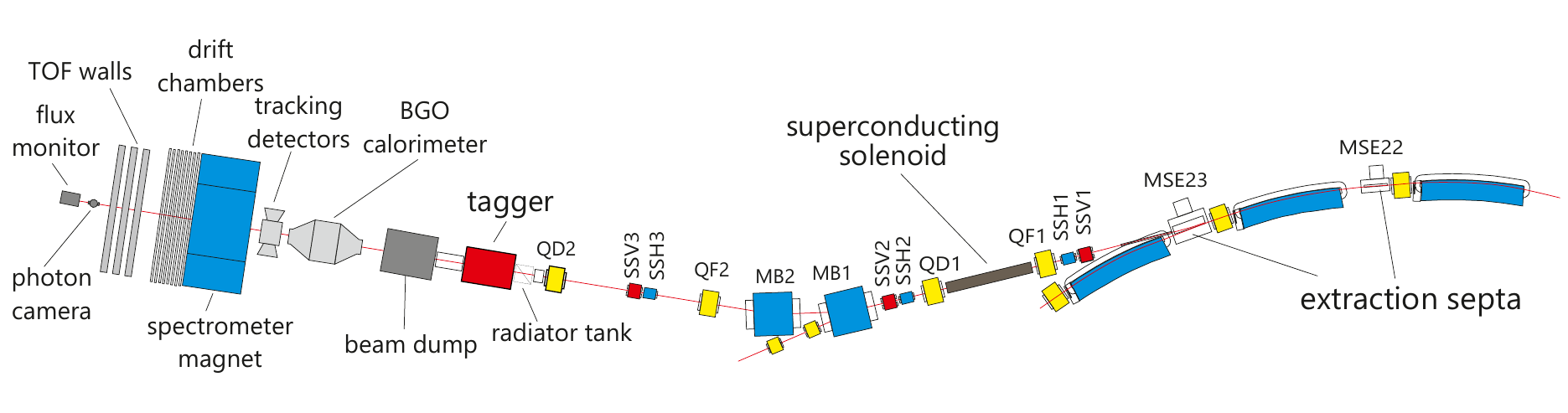}}
	\caption{The BGOOD beam line, showing the extraction of the electron beam from ELSA to the experimental area.}
	\label{fig:BGOODbeamline}       
\end{figure*}

The electron beam extraction from ELSA to the BGO\-OD experimental area is shown in fig.~\ref{fig:BGOODbeamline}.  This is achieved using two extraction septa, MSE22 and MSE23.
 Two dipole magnets, MB1 and MB2 are installed in the experimental beam line.  MB1 is used as a switch to deflect the beam to one of the hadron physics experiments. Two horizontally focussing quadrupole magnets, QF1/2 and two vertically focussing quadrupole magnets, QD1/2 are used for beam focussing.  Three horizontal steerer magnets, SSH1/2/3 and 3 vertical steering magnets, SSV1/2/3 are used for adjustment of the beam position.

 Typical extracted currents for BGOOD are 0.5 to 1.0\,nA.
The stretcher ring has a flexible timing, allowing for, in principle an
unlimited length of the extraction phase.
For the BGOOD experiment the extraction phase is 4 to 15\,s, with 1.5 to 2\,s between spills, 
giving a macroscopic duty factor
between 70\,\% and 90\,\%. 
Due to the accelerating 500\,MHz radio frequency,
the extracted beam has a micro-structure of electron 
bunches separated by 2\,ns.
The extracted electron beam at the tagger position in fig.~\ref{fig:BGOODbeamline} has a Gaussian distribution with sigmas of typically 1.2 and 0.3\,mm in the horizontal and vertical directions respectively.  Measurements of these distributions are described in sect.~\ref{subsubsec:Radiators}.

\begin{figure}[htbp]
	\begin{center}
		\includegraphics[trim=7cm 15cm 8cm 7cm,clip=true,width=0.25\textwidth]{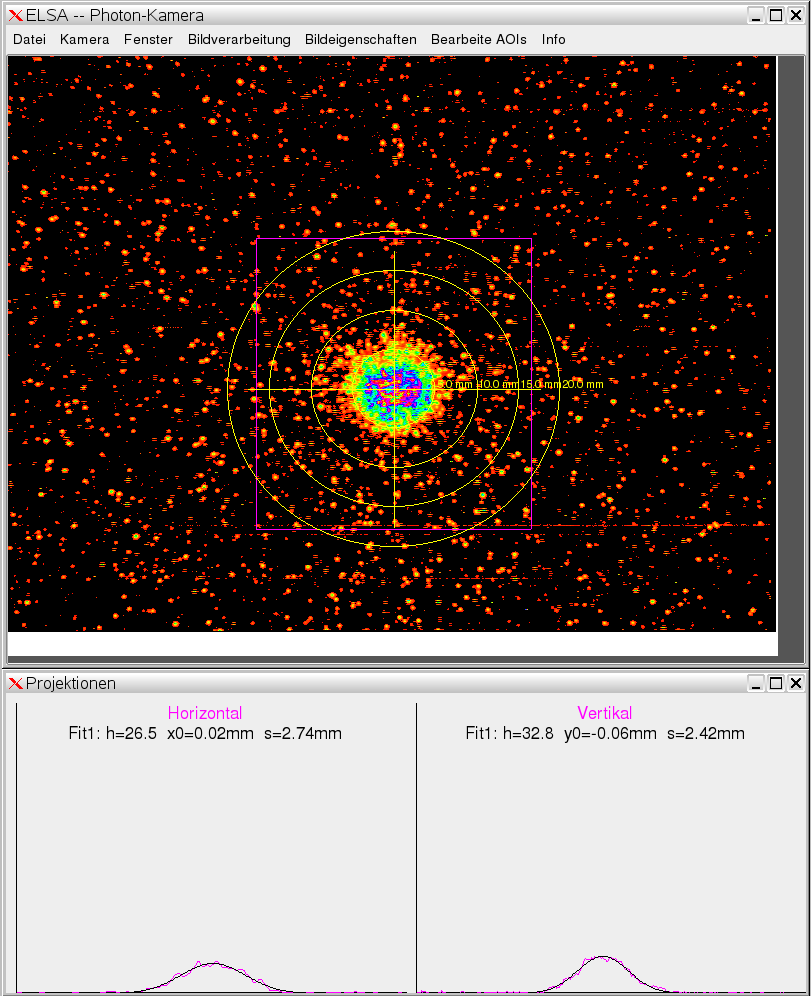}\includegraphics[trim=7cm 15cm 8cm 7.15cm, clip=true,width=0.25\textwidth]{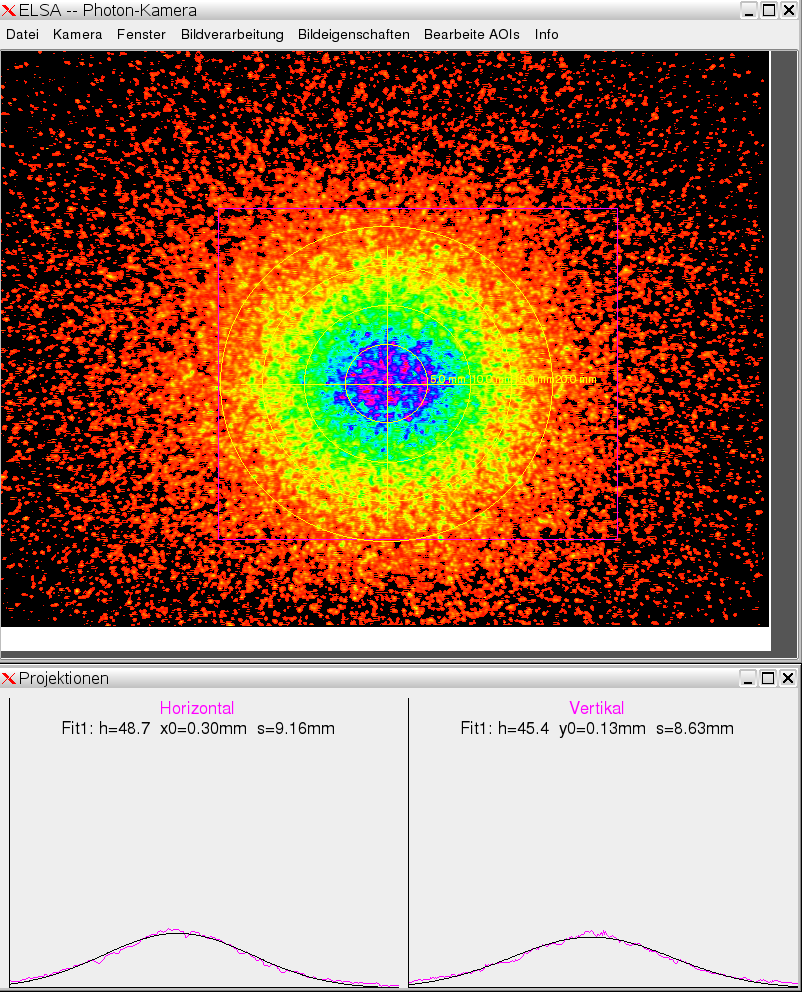}
		\caption{Distribution of the photon beam at the photon camera for collimated and uncollimated beam (left and right respectively).  The yellow circles indicate 5\,mm intervals.}
		\label{fig:beamspot}
	\end{center}
\end{figure}

Images from the photon camera depicted in fig.~\ref{fig:BGOODbeamline} downstream from the ToF Walls, are shown in fig.~\ref{fig:beamspot} for both collimated and uncollimated beams.  Approximate Gaussian sigma distributions of 2.5 and 9.0\,mm respectively are determined.  This measured beam spot is significantly larger than measured at the target, due to the divergence of the beam as it passes approximately 7\,m from the goniometer to the photon camera.

\section{The BGOOD detector}
\label{sec:Detector}

\begin{figure*} [htbp]
  \centering
\vspace*{0cm}
\resizebox{0.9\textwidth}{!}{%
 \includegraphics{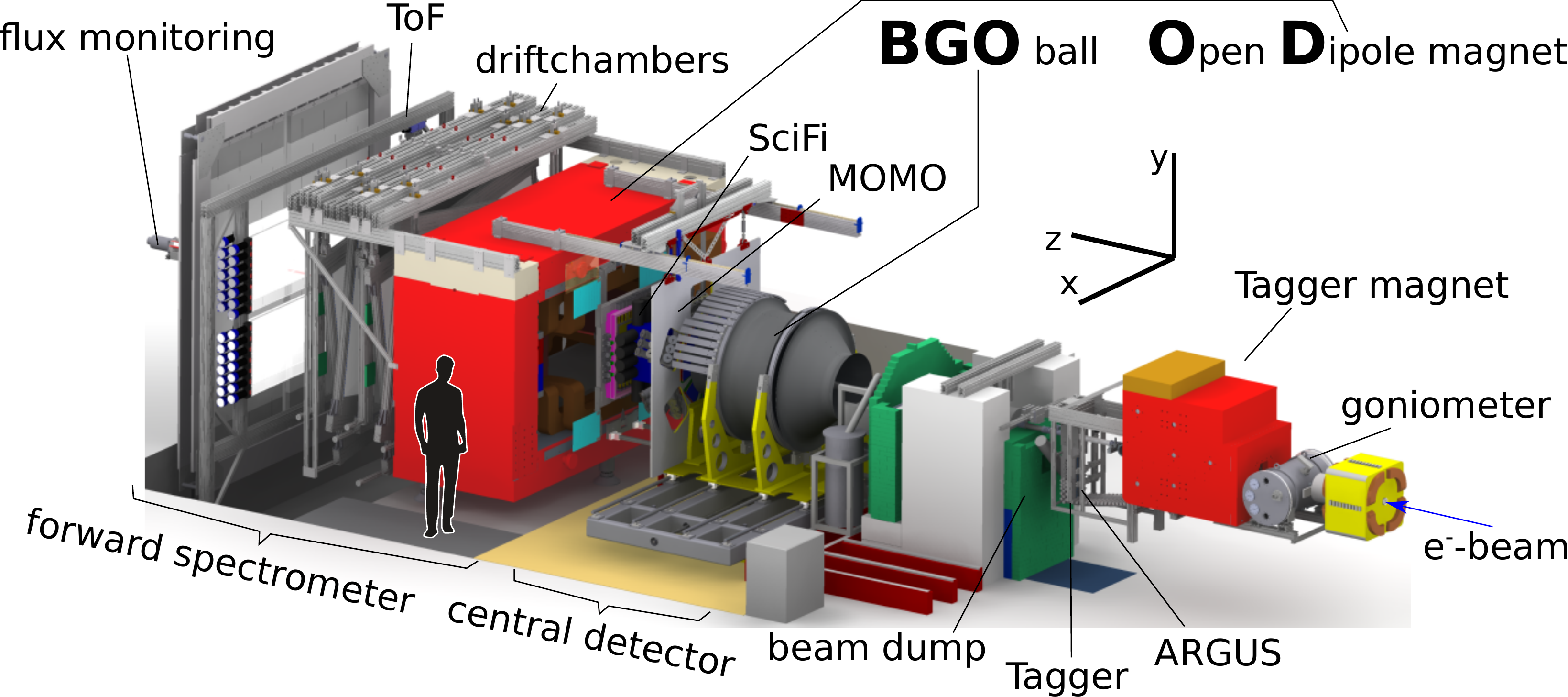}
}
   \caption{Overview of the BGOOD setup.}
   \label{fig:BGOODsetup}
\end{figure*}

The BGOOD detector shown in fig.~\ref{fig:BGOODsetup} is ideal to investigate low
momentum transfer kinematics, corresponding to very forward going mesons. The residual (excited) hadronic systems
then decay almost at rest and, consequently,
into full $4\pi$ solid angle. To cater for this situation, BGOOD consists
of a central detector enclosing the target in the polar angle
range $(25\div 155)^\circ$.  A large
aperture magnetic spectrometer covers the more forwards angular 
range $(1\div 12)^\circ$. The setup is complementary to the other experiments mentioned in the introduction.

The main component of the central detector is a highly
segmented {\it BGO} crystal calorimeter ({\it BGO Rugby Ball}). This is complemented
by a segmented plastic scintillator barrel for charged particle 
identification, and by two coaxial cylindrical multiwire proportional chambers (MWPCs)
for charged particle tracking and reaction vertex reconstruction.

The intermediate range between the open $25^{\circ}$ forward cone
of the BGO Rugby Ball and the rectangularly shaped gap
of the spectrometer magnet is covered by the {\it SciRi} detector which is comprised of a series of segmented scintillator rings.
A Multi-gap Resistive Plate Chamber (MRPC) developed as a Time-of-Flight detector for the ALICE experiment~\cite{Cortese:2002kf} is currently under construction to complement the SciRi detector, with an anticipated time resolution of 40~ps and $2^\circ$ angular resolution.

The forward spectrometer consists of a large aperture
dipole magnet, sandwiched between tracking detectors.
Tracking upstream from the magnet is performed with two sets of scintillating fibre
detectors, {\it MOMO} and {\it SciFi}. Eight double
layers of drift chambers serve for tracking downstream from the
magnet. The spectrometer is completed by three layers of
scintillator time-of-flight (ToF) walls. Measuring the velocity in addition to the momentum enables unambiguous charged
particle identification over a wide momentum range.

This section first describes the in-house electronics which are common to most BGOOD detector components.  The central detector region, forward spectrometer and target system are subsequently described in sects. \ref{subsec:CentralDetector1}, \ref{subsec:ForwardSpectrometer} and \ref{subsec:Target}.


\subsection{In-house readout electronics}
\label{subsec:IHRE}

The BGOOD experiment uses both commercial and in-house readout electronics.
The main advantage of custom electronics is that any preprocessing or preselection to be done in real time can be implemented in the hardware itself.
Furthermore, the integration of the trigger logic into the same hardware module as the TDC lowers the number of required components. Most of the in-house electronics are realised using the Elektroniklaboratorien Bonn (ELB) Field Programmable Gate Arrays (FPGAs) VME-boards~\cite{ELB}.  
The FPGAs are integrated circuits that feature a set of logic elements such as flip-flops, loadable Look up Tables (LUTs) and inverters which can be flexibly interconnected by applying a user-defined hardware description.  This is usually complemented by additional elements such as phase-locked loops or delay-locked loops for clock generation and control.  The programmability of this integrated circuit allows the implementation of logic circuits directly without manufacturing an expensive, dedicated ASIC if the offered resources are sufficient. The input and output of the ELB FPGA-board can be configured using daughter boards (mezzanines), to specialise the board for specific tasks. One FPGA board can be
equipped with up to three mezzanines.  Fitting the board with three LVDS-input mezzanines and the jTDC firmware, described in sect. \ref{subsubsec:jTDC}, creates a 100 channel TDC. Additionally, the LVDS mezzanines can be exchanged for discriminator mezzanines (described in sec.~\ref{subsubsec:jDisc}) to create a 48 channel board, which discriminates the signal and acquires the time with the TDC implemented on the FPGA. Such a  board is referred to as a jDisc.

\subsubsection{FPGA-based TDCs: The jTDC}
\label{subsubsec:jTDC}


The jTDC firmware is upon an FPGA based high resolution TDC, implemented on a Xilinx\textsuperscript{\textregistered}Spartan\textsuperscript{\textregistered} 6.
Each input channel is directed to one of the carry chains on the FPGA. The signal is then sampled in time using a 400\,MHz clock by using the delay in the carry chain between the individual flip-flops.  This is known as the \textit{tapped delay line technique}. A time resolution of better than 40\,ps RMS can be achieved. The general features of the jTDC firmware are the following:

\begin{itemize}
\item Up to 100 TDC channels per board with 40\,ps average bin size.
\item Scalers (32\,bit at 200\,MHz) for every input channel.
\item Acquisition dead time of 5\,ns.
\item A minimum 3\,ns length of input signals.

\item A maximum number of 155 hits per TDC channel are stored in DATA-FIFO.

\item The TDC time window is 775\,ns.

\item Two individual configurable trigger outputs available (logical OR of all input signals or subsets).
\end{itemize}

The jTDC therefore provides good time resolution with a high rate stability. Since the firmware was
developed in-house it is readily available for extension with customised trigger logic
and scalers as necessary for each detector, removing any complications of combining separate
electronic modules with different properties for these tasks.
Further details can be found in refs.~\cite{bieling12,github}.

\subsubsection{ELB Discriminator Mezzanine}
\label{subsubsec:jDisc}


The full input electronics is contained on a
daughter board (the mezzanine) and the signal directly enters the FPGA
which can handle all roles simultaneously, provided the necessary firmware is developed: TDC,
time over threshold extraction, rate monitoring and trigger logic. 
This discriminator provides the following features and key specifications (more details in ref.~\cite{ELB}):
\begin{itemize}
\item 16 inputs per mezzanine, offered with various connectors.
\item Time over Threshold (ToT) information.
\item Analogue input bandwidth larger than 550\,MHz.
\item 14-bit DAC for threshold setting between -4\,V and 4\,V.
\item 8-bit DAC for hysteresis setting between 0\,mV and 80\,mV.
\item 16-bit ADC for readback of thresholds and hysteresis.
\item A pass-through time of 500\,ps.
\item A pass-through time jitter of 30\,ps.
\end{itemize}

The key features for our application are the finely adjustable
thresholds and the ToT information. The ToT is extremely useful not only
to provide a crude signal integral information for gain matching, but also to improve the achieved
time resolution via time walk corrections.

\subsection{Central detector}
\label{subsec:CentralDetector1}


\begin{figure} [h]
  \centering
  \resizebox{0.5\textwidth}{!}{ \includegraphics[trim={3cm 10cm 4.5cm 1cm},clip]{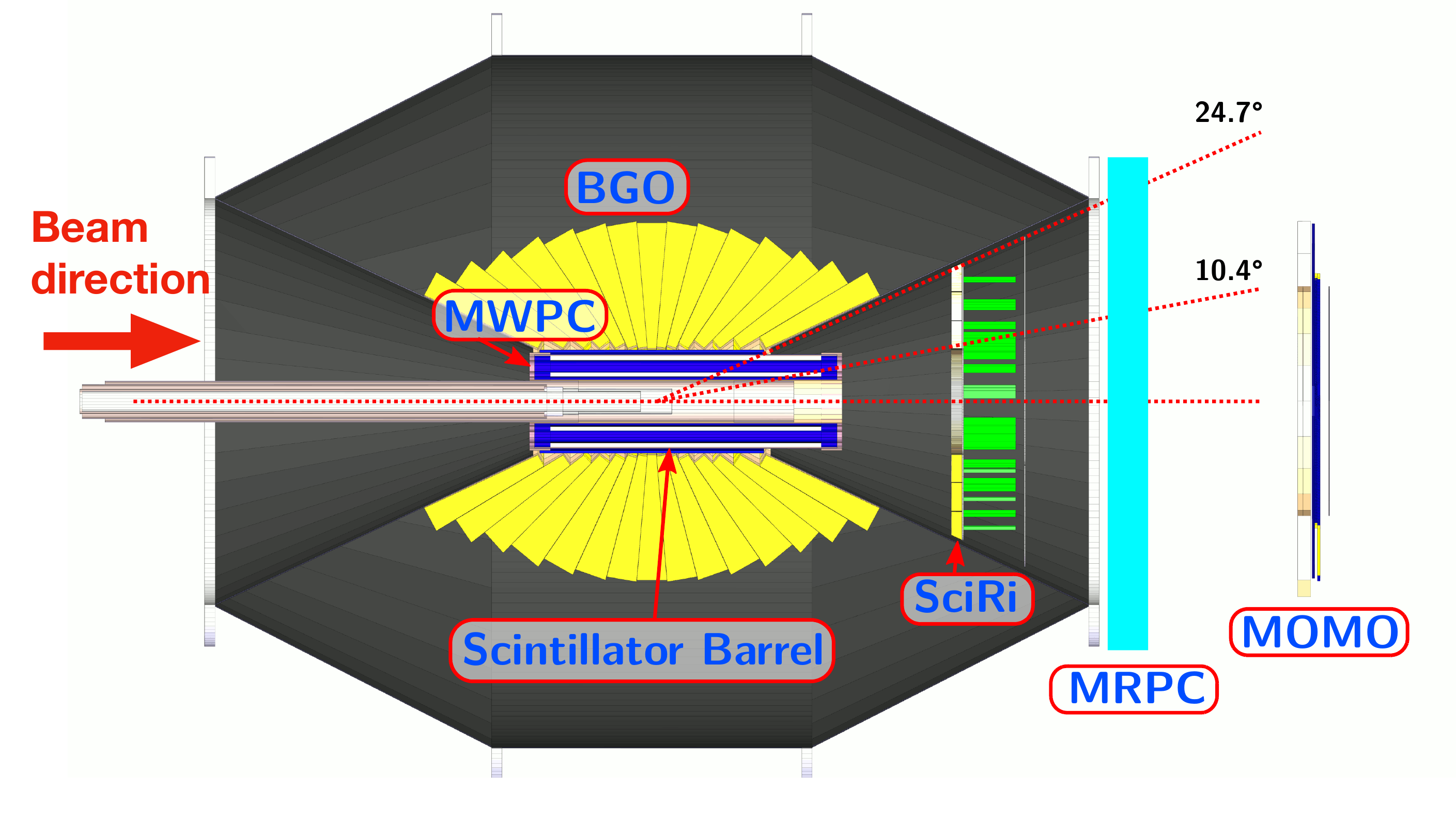}
 
}
   \caption{Slice view of the central detector consisting of the segmented BGO Rugby Ball, cylindrical inner scintillator barrel and MWPC, and
the SciRi scintillator ring detector in forward direction. Also depicted are MOMO and SciFi, the front tracking fibre detectors
of the forward spectrometer and the MRPC detector under construction.  Figure adapted from ref.~\cite{Freyermuth}.}
   \label{fig:central}
\end{figure}

\subsubsection{BGO {Rugby Ball} calorimeter}
\label{subsubsec:BGO}

The main component of the central detector is the BGO calorimeter\footnote{BGO is short for the crystal material, 
	Bismuth Germanate $\mathrm{Bi_4(GeO_4)_3}$.},
previously used at the GrAAL experiment at ESRF~\cite{Bartalini:2005wx}.
The geometry is illustrated in the slice view of fig.~\ref{fig:central}.
The BGO Rugby Ball is segmented into 480 crystals of  24\,cm depth, corresponding to over $21$ radiation lengths.
They are arranged in 15 sectors (called crowns) of 32 crystals each.
Each crown covers $\Delta\theta = (6\div10)^{\circ}$ in polar angle and the full azimuthal ring,
corresponding to $\Delta\phi = 11.25^\circ$ per crystal. 
The total polar angular range covered is $\theta = (25\div155)^{\circ}$.

The geometric arrangement requires the crystals to be shaped as pyramidal sectors 
with trapezoidal basis, of which eight different dimensions are used.
Each crystal is wrapped in a 30\,$\mu$m aluminised Mylar foil,
and coupled to a photomultiplier tube (linearity selected Hamamatsu R580 or R329-02).
The individual detector elements sit in a basket structure made of carbon fibre.
To separate the crystals mechanically and optically, 
each basket is divided into 20 individual cells with $0.38$ and $0.54\,$mm
thick internal and external walls, respectively.
The carbon fibre structure is held by a steel frame and is mechanically separated into two halves.
A precision rail system opens the BGO Rugby Ball laterally and provides  access to the 
scintillator barrel, the MWPCs, the SciRi and the cryogenic target system inside.

The photomultiplier tubes are shielded with $\mu$-metal,  which is sufficient for low magnetic fields.  An iron cover in the forward hemisphere provides additional shielding from the significant fringe field of the Open Dipole magnet, which is typically 3\,mT at the most downstream end of the BGO Rugby Ball.  

\begin{figure} [h]
  \centering
\resizebox{0.45\textwidth}{!}{%
  \includegraphics{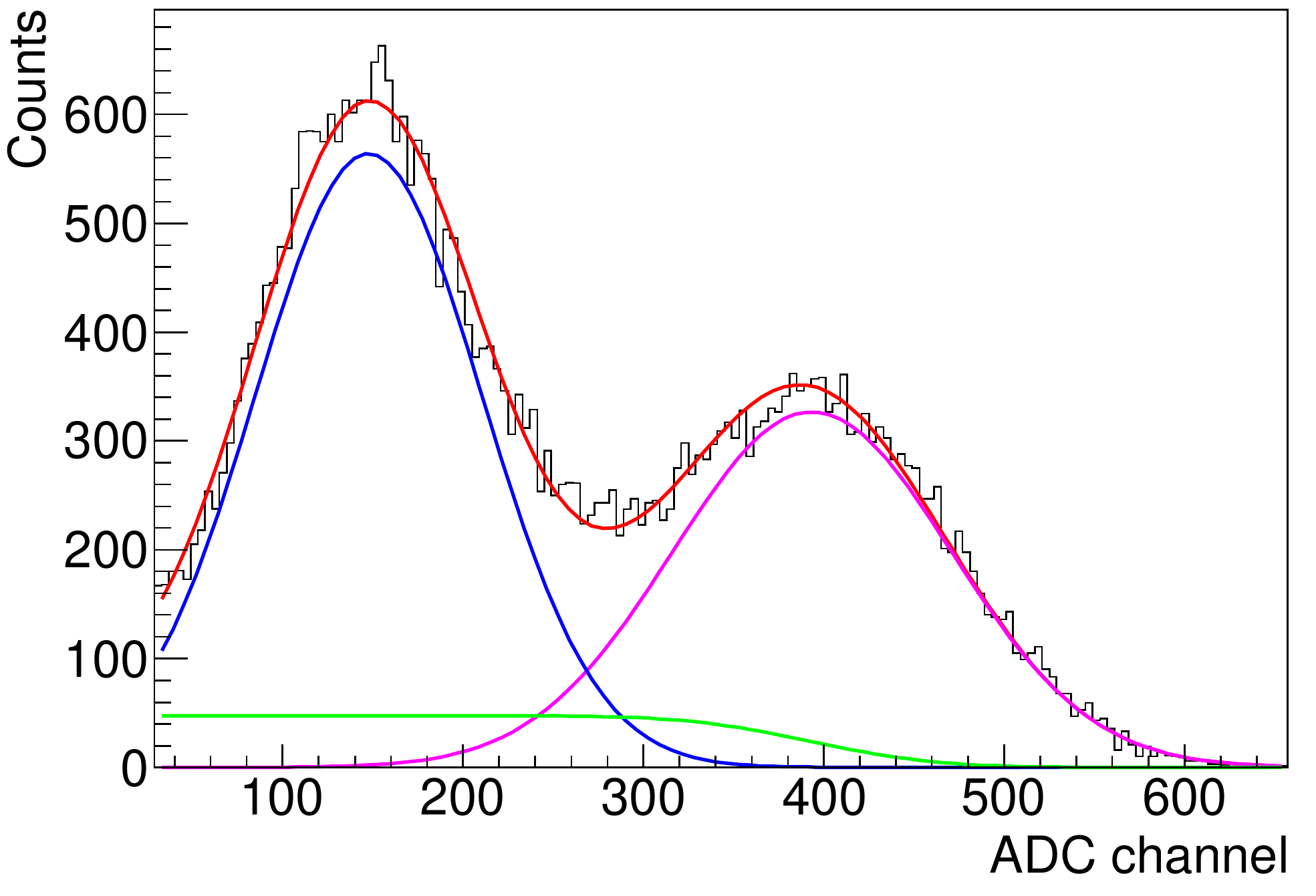}
}
   \caption{$^{22}$Na energy spectrum in one BGO crystal.  The fitted function in red is the sum of two Gaussian functions and the error function of the Gaussian function at higher channels (shown in blue, magenta and green respectively).  $^{22}$Na decays via $\beta^+$ emission to  $^{22}$Ne$^*$.  The lower energy peak at approximately 140\,channels corresponds to a 511\,keV photon from the annihilation of the $\beta^+$ with an atomic electron.  The higher energy peak at approximately 400\,channels is the 1.275\,MeV $\gamma$ decay excitation of $^{22}$Ne$^*$ to the ground state.}
   \label{fig:Na-22}
\end{figure}

The photomultiplier signals are sent to 15 modules (one per crown) where
each signal is split into two. One is sent to an analogue adder that sums the 32 signals
of a crown; the other signal, delayed and attenuated, is the input
for the W-IE-NE-R AVM16 sampling ADCs.
The employed programmable attenuators ensure the $\simeq 100$\,MeV signals during data taking
are in the same electronic range as the $1.27$\,MeV signals from 
$^{22}$Na radioactive sources permanently placed inside the BGO Rugby Ball
and used for calibration.
An additional sum mixer adds up the crown sums to provide a total calorimeter 
sum signal for trigger purposes.
The ADC sampling rate is 160\,MHz. 
A time resolution of 2\,ns is achieved, while the energy resolution for 1\,GeV photons, measured in a test beam experiment with the old electronics was 3\,\%
(FWHM)~\cite{LeviSandri:1996tk}. 

The BGO Rugby Ball response for protons~\cite{Zucchiatti:1992isx}, neutrons~\cite{Bartalini:2006ci}, pions and deuterons~\cite{Anghinolfi:1995qz} has also been measured.  Although not explicitly measured, it is anticipated that the energy resolutions achieved at the BGOOD setup are equal or better with the new electronics.  The time measured with the BGO Rugby Ball also provides the time information for particles passing through the MWPCs and Scintillator barrel.  

Three steps are made for an accurate energy calibration,
optimised for an electromagnetic shower from incident photons.
The radioactive sources enable an initial energy calibration of the 
BGO Rugby Ball (see fig.~\ref{fig:Na-22}). 
After this, the calibration proceeds by fitting to the $\pi^0$
mass from the abundantly observed $\pi^0 \rightarrow 2\gamma$ decay.
A small run-by-run correction is applied to move the measured invariant mass
peak to the correct position.  This compensates for short term gain fluctuations, predominantly caused by temperature changes.
Finally, an iterative correction is made for each crystal.  Events are selected where at least 50\% of the electromagnetic shower from the decay photon was deposited in a given crystal.  A correction is applied per crystal to ensure the measured $\pi^0$ invariant mass spectrum is at the correct position.  As each correction affects the spectra for the other crystals, the procedure requires several iterations.



More details on the BGO Rugby Ball and the initial energy calibration can be found in ref.
\cite{Ghio:1997yw}. 

\subsubsection{Scintillator barrel}
\label{subsubsec:Barrel}

The cylindrical scintillator barrel is intended to distinguish between neutral and charged 
particles and to provide $\Delta E/\Delta x$ energy loss measurements of the latter.
It is placed between the BGO Rugby Ball and the MWPCs and consists of 32 
plastic scintillator bars, 5\,mm thick, made from BC448 (former Bicron, now Saint-Gobain).
The active length is 43\,cm, and the scintillator barrel mean radius is 9.75\,cm.

Each bar is upstream-side connected to a Hamamatsu H3164-10 photomultiplier tube.
The signals are electronically processed with jDisc.
The detection efficiency for charged particles is $\simeq 98\%$.
In contrast, both photon and neutron detection efficiencies are $< 1$\,\% 
over a wide energy range.

\subsubsection{Cylindrical MWPCs}
\label{subsubsec:MWPC}

\begin{figure}[htbp]
\begin{center}
\vspace{-0cm}
\resizebox{0.3\textwidth}{!}{\includegraphics{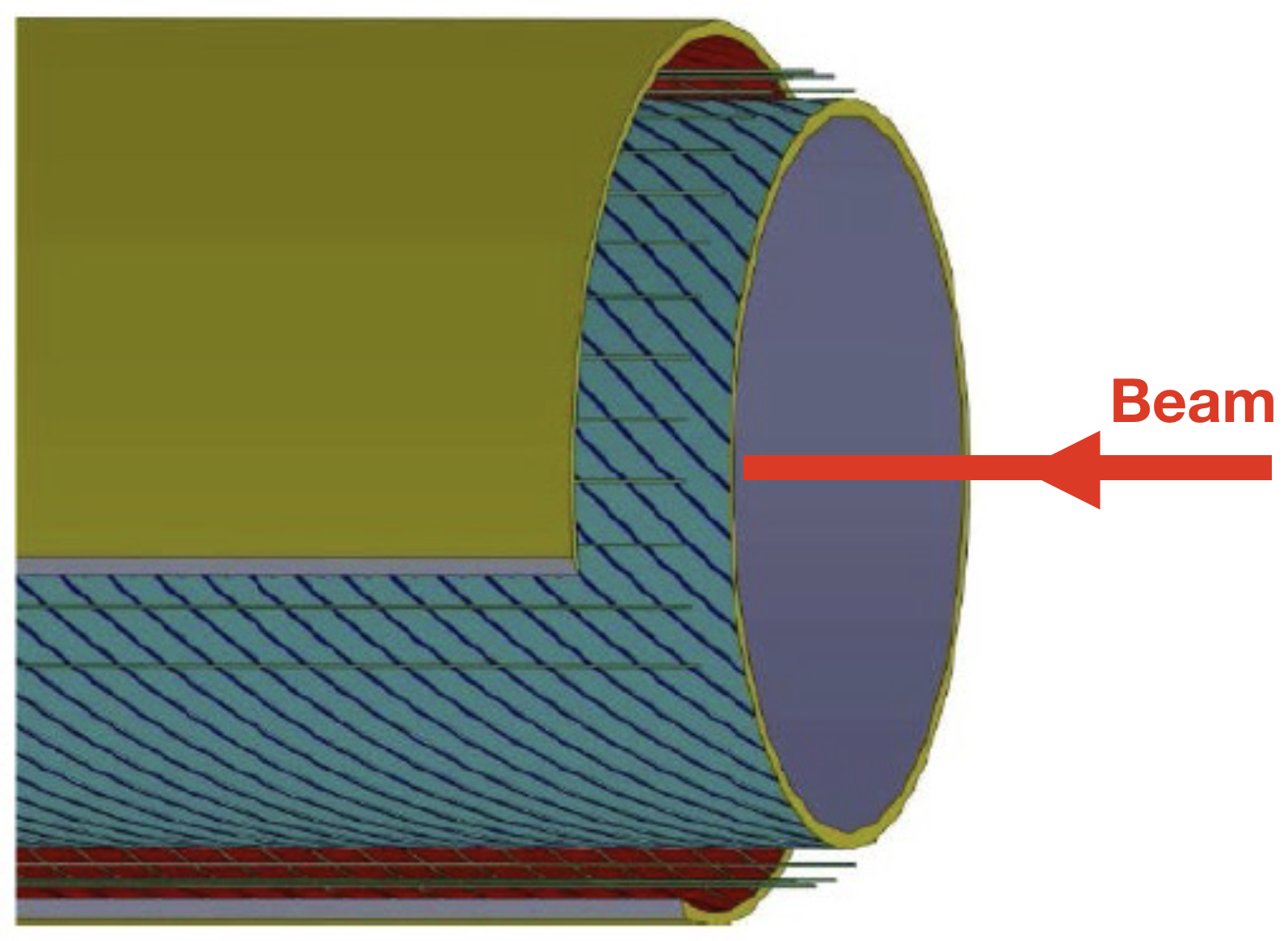}}\vspace{-0.1cm}
  \caption{Schematic view of one cylindric MWPC. Visible are the horizontally stretched
signal wires (parallel to the beam axis), and the helically wound cathode strips
on the inner cathode layer.}
 \label{fig:MWPC}
\end{center}
\end{figure}

Two MWPCs are used for charged particle tracking inside the BGO Rugby Ball.
Each of them consists of three separate cylindrical layers of the same general design.
One of them is schematically depicted in fig.~\ref{fig:MWPC}.

\begin{table}[htbp]
\centering\begin{tabular}{ l l l }
\hline
Item                          &  Inner   		& Outer    \\ 
\hline
Length /mm               &  520             &  520            \\
Inner radius  /mm     &  $48.2$        &  $79.5$       \\
Outer radius  /mm     &  $56.2$        &  $87.5$       \\
Anode wires              &  160             &  256            \\
Inner strips                &    48            &     80            \\
Outer strips                &    56            &     88            \\
\hline
\end{tabular}
\caption{Geometrical characteristics of the MWPCs.}
\label{tab:MWPC-dimensions}
\end{table}



Each chamber has inner and
outer cylindrical walls made from 2\,mm thick Rohacell
covered in 50\,$\mu$m Kapton film. The interior surfaces are
laminated with copper strips (0.1\,$\mu$m thick, 4.5\,mm
wide and separated by 0.5\,mm) which form the cathode. 
For the anode, 20\,$\mu$m diameter gold-plated tungsten wires are 
stretched along the longitudinal axis with a tension of 60\,g. 
There is one anode wire per 2\,mm spacing 
and the anode-to-cathode gap is 4\,mm.
The inner and outer strips are wound in opposite directions
at $\pm 45^\circ$ with respect to the wires.  These intersect two or three times over the length of the chamber.
 For a single charged particle, the impact point can be unambiguously reconstructed from the intersection of fired cathode strips and anode wires.
Combined information from the two chambers then yields the desired 
particle track.

Geometrical dimensions and number of wires and strips of both MWPC chambers are given
in table\,\ref{tab:MWPC-dimensions}.
These result in a polar angular coverage of $\theta = (8\div163)^{\circ}$.
The  plastic flanges providing mechanical stability partially shadow
the forward polar angular region from $8^\circ$ to $18^\circ$ with a maximum 
material budget of about 0.07 radiation lengths. This corresponds   
to a maximum multiple scattering rms width of approximately $0.8^\circ$ and $0.2^\circ$ for 
protons with momenta 0.5 and 1.0 \,GeV/c respectively.
When requiring signals from all MWPC detector components, resolutions of 
$\delta z \simeq 500$\,$\mu$m along the beam axis,
$\delta \phi \simeq 1.5^{\circ}$ in azimuthal and 
$\delta \theta \simeq 1^{\circ}$ in polar angle can be achieved.

The MWPCs are operated at 2500\,V with a gas mixture of
$69.5$\,\% argon,
$30.0$\,\% ethane, and
$0.5$\,\%  halocarbon 14 (CF$_4$).

Before the gas passes through the chambers, 
it is bubbled through alcohol in a tank within a small refrigerator. In this way, 
an adjustable amount (between 3\,\% to 10\,\%) 
of alcohol vapour is added to the mixture to prevent current instabilities.

Wire signals are pre-amplified by in-house circuitry based on the FEC2
chip~\cite{abbrescia00} developed for the CMS experiment, while for the strips, the
AD8013 chip from Analog Devices is used.
Digital wire signals from the preamplifiers are then further processed
using jTDC modules for timing information.
Cathode strip signals are evaluated using the same type of sampling ADC 
as for the BGO Rugby Ball.

\subsubsection{Scintillator ring}
\label{subsubsec:SciRi}

\begin{figure}[htbp]
\begin{center}
\vspace{-0cm}
\resizebox{0.1\textwidth}{!}{\includegraphics{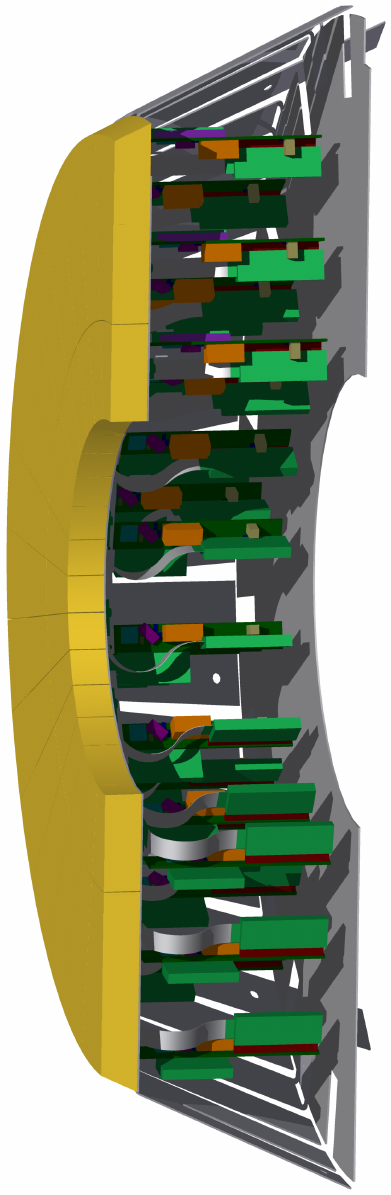}}\vspace{-0.1cm}
  \caption{CAD drawing of one half of the SciRi detector. The plastic scintillator material is shown
in yellow, additionally, the aluminium holding and shielding structure containing the readout electronics
can be seen. Each half of the detector is fixed to the corresponding
half of the BGO Rugby Ball.}
 \label{fig:SciRi}
\end{center}
\end{figure}

The Scintillating Ring detector ({\it SciRi}, shown in fig.~\ref{fig:SciRi}) covers the small
acceptance gap between the open forward cone of the BGO Rugby Ball and the rectangular magnet gap of the forward spectrometer. 
It also partly overlaps with the scintillator barrel and MWPCs.

SciRi is a segmented plastic scintillator detector used for direction reconstruction of charged particles. The
individual 20\,mm thick elements are arranged 
in three rings. Within the total polar angular range of $\theta = (10\div25)^{\circ}$,
each ring covers $\Delta \theta = 5^{\circ}$. One ring consists of 32 segments, yielding an
azimuthal coverage of $\Delta \phi = 11.5^{\circ}$, which is the
same as for the BGO Rugby Ball. The whole detector is mounted
inside the forward opening cone of the BGO Rugby Ball.

The scintillator elements 
are individually read out by Hamamatsu S11048(X3) avalanche photo diodes
(APDs).  Using jDisc modules for signal processing as for the scintillator barrel,
time resolutions of approximately 3\,ns are achieved. Pulse heights are
derived by time-over-threshold measurements.
Further details are found in ref.~\cite{georgmasters}.

%
%

\subsection{Forward spectrometer}
\label{subsec:ForwardSpectrometer}

The forward spectrometer is arranged around the large
open dipole magnet. Charged particles are tracked in front
of the magnet by means of two scintillating fibre detectors,
{\it MOMO} and {\it SciFi}. Behind the magnet, particle trajectories 
are determined through eight double layers of drift
chambers. The photon beam passes through small holes at
the centre of the fibre detectors, and $5\times 5$\,cm$^2$
insensitivity
spots in the large area drift chambers. Thus, very forward
angles down to approximately $1.5^{\circ}$ can be reached. The setup 
enables momentum reconstruction with $\delta p/p \simeq 3\%$ over a range of approximately 400 to 1100\,MeV/c with a polar angle resolution better than $1^\circ$. Particle identification is accomplished through
velocity measurements with the ToF Walls. 
The components of the forward 
spectrometer are described in the following.

\subsubsection{Open Dipole magnet}
\label{subsubsec:Magnet}

The central part of the forward spectrometer is the dipole
magnet\footnote{The magnet is on loan from DESY.} which provides the magnetic field to measure the
momentum of charged particles (and determine the sign
of charge).

The magnet has outer dimensions of 280\,cm (height) $\times$ 390\,cm (width) $\times$ 150\,cm (length),
with a total weight of 94\,tons. Particles pass through the central gap
with dimensions 54\,cm (height) $\times$ 84\,cm (width) $\times$ 150\,cm (length).
From the target, this allows an angle of acceptance up to $12.1^{\circ}$ and $8.2^{\circ}$ in horizontal and vertical 
direction, respectively.
At the maximum current of 1340\,A the integrated field along the beam axis is approximately 0.71\,Tm.

The vertical gap between the poles was extended for use at BGOOD, and three dimensional field maps at three different currents were measured at GSI Darmstadt.  These were performed in 1\,cm steps through the magnetic field, extending through the fringe field (130\,cm downstream and 180\,cm upstream of the centre of the magnet) and over the full X-Y plane normal to the beam direction.  Upon installation of the Open Dipole magnet at BGOOD, an additional three dimensional field map was measured.  This was performed to determine the effect that nearby detectors, photomultiplier tubes and mounting had upon the fringe magnetic field.  A simulation of the field using the CST Studio Suite\textsuperscript{\textregistered} was concurrently made, with agreements between the field maps at a level of approximately 1\%.  For more details see refs.~\cite{torstenthesis,philippthesis}.

\subsubsection{Front tracking: {MOMO} and {SciFi}}
\label{subsubsec:MOMO-SciFi}

\begin{figure}[htbp]
\begin{center}
\vspace{-0cm}
\resizebox{0.3\textwidth}{!}{\includegraphics{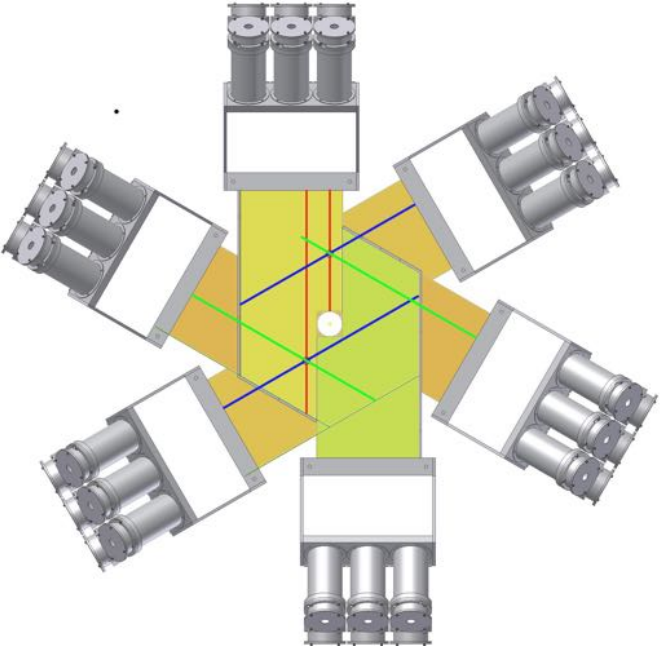}}\vspace{-0.1cm}
  \caption{Schematics of the MOMO detector,  viewed along
the beam axis.}
 \label{fig:MOMO}
\end{center}
\end{figure}

MOMO is a 
scintillating fibre detector that was originally
built for the MOMO\footnote{Monitor of Mesonic Observables facility.}  experiment at COSY~\cite{Bellemann99,Bellemann07}.
 It is composed of 672 fibres, 
which are each 2.5\,mm in diameter. The fibres are
arranged in three layers, with each layer consisting of two
``closed packed" planes of fibres, so that the fibres of one plane lie in the gap between the fibres of the other plane. The layers are arranged so that there
is a $60^{\circ}$ angle between fibres of adjacent layers, as shown
in fig.~\ref{fig:MOMO}. The fibres are coupled directly and read out by 16 channel Hamamatsu R4760 
phototubes. Cylinders of 1\,mm Permenorm
surround each tube to shield the fringe magnetic field
from the open dipole magnet. The photomultiplier signals
are connected to leading edge discriminators and to commercial TDCs
(CAEN V1190A-2eSST)  with about 80\,ps (RMS)
time resolution.

MOMO covers a diameter of 44\,cm, with a central
4.5\,cm hole to allow the photon beam to pass. Fibres
within each layer overlap slightly so that charged particles
pass either one or two fibres per layer. Charged particle
positions are then determined by coincidence hits between
either two or three of the different layers. A timing coincidence 
of 20\,ns is required between the fibres used to
reconstruct particle positions.

A comparison between real and simulated data showed
that MOMO has a particle position reconstruction efficiency 
of $\simeq$ 80\,\% due to the fibre arrangement.
 MOMO was optimised for the larger particle
penetration angles at the MOMO experiment~\cite{Bellemann07} and the small
angles at BGOOD induce a larger fraction of fibre edge
hits.  MOMO is a vital component however for monitoring efficiencies of other detectors, 
and the reduced efficiency is not usually a problem for track reconstruction.  
If MOMO is not used,
 the centre of the target and the SciFi detector  (the second tracking station) can be used to define the trajectory prior to the magnetic field.  For the reconstruction of most hadronic reactions, this results in only a small loss in angular resolution and the rejection of background.
The default track finding algorithm (see sect. \ref{subsubsec:Tracking-MomentumReconstruction}) however uses both detectors.

\begin{figure}[htbp]
\begin{center}
\vspace{-0cm}
\resizebox{0.3\textwidth}{!}{\includegraphics{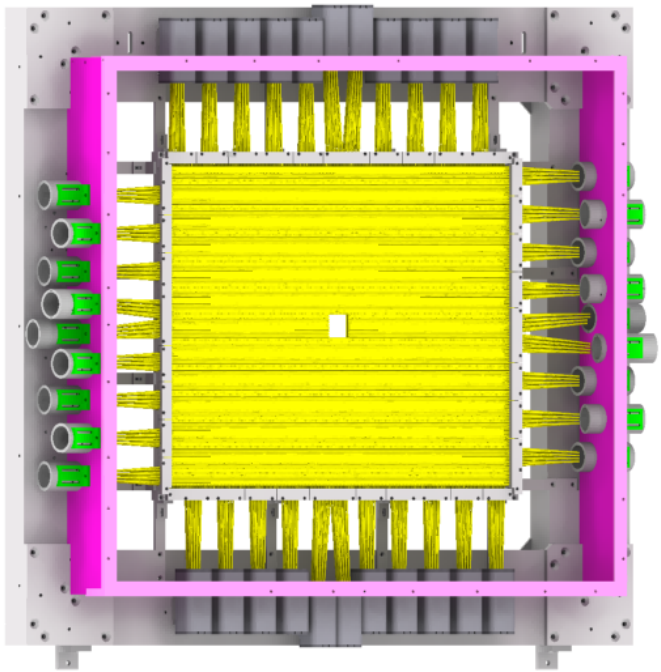}}\vspace{-0.1cm}
  \caption{Schematics of the SciFi detector with its two crossed
planes of in total 640 scintillating fibres.}
 \label{fig:SciFi}
\end{center}
\end{figure}

SciFi, a scintillating fibre detector consisting of 640 BC440
 fibres of 3\,mm diameter, is directly mounted to the upstream
surface of the spectrometer magnet. A schematic of it is
shown in fig.~\ref{fig:SciFi}. The fibres are arranged to form two
320 fibre layers, at an angle of $90^{\circ}$ to each other, covering
an active region of 66\,cm in width and 51\,cm in height.

A 4$\times$4\,cm$^2$ hole in the centre allows the photon beam
to pass. Groups of 16 fibres are arranged into modules,
with each module coupled directly and read out by a 16 channel Hama-
matsu H6568 photomultiplier. The photomultipliers are
magnetically shielded against the dipole's fringe field using
$\mu$-metal, and soft iron tubes. The latter
have quadratic cross section for the bottom and top rows
of photomultipliers, but are circularly shaped at the left
and right sides. This requires a relative displacement of adjacent photomultiplier tubes, as can be seen in fig.~\ref{fig:SciFi}.

The fibres in each layer are separated by 1\,mm and
thus overlap with fibres in the other layer slightly. 
A charged particle therefore passes
through one or two fibres per layer. The time resolution of
2\,ns at FWHM per fibre allows a tight time cut of 2.7\,ns
between adjacent fibres. The combination of both layers
determines the particle position. Comparison between real
and simulated data demonstrated that SciFi has a particle position reconstruction efficiency of 97.5\,\%.

\subsubsection{Rear tracking: Drift chambers}
\label{subsubsec:DCs}

Charged particle tracking
is performed behind the open dipole magnet by eight double layer drift chambers
which cover a sensitive area of $246 \times 123$~cm$^2$.
For unambiguous track reconstruction, the wires are oriented in four
different directions in the plane perpendicular to the beam
axis, labelled X (vertical), Y (horizontal), U ($+9^{\circ}$ tilt
against vertical), and V ($-9^{\circ}$ against vertical).
X and Y chambers have equal outer dimensions and almost equal 
sensitive area, but different wire orientation.
U and V chambers have the wires stretched along
the shorter edge direction. The whole chambers then are installed tilted 
to achieve the desired wire orientation in the
experiment. To preserve the total active area, U and V
chambers have slightly larger dimensions than X and Y.
The main features of the drift chambers are summarised
in table \ref{tab:DC-features}.

\begin{table}[htbp]
\hspace*{-0.2cm}
\centering\begin{tabular}{l l l l}
\hline
 Feature 	&  X 			& Y				& U  V  \\ 

\hline
Orientation                 &vertical 		&horizontal 	& $\pm9^{\circ}$\\
Width /mm               	&  	2867 		&3053			& 3139\\		    
Height /mm     		&  1965 		&1779 			&2335 \\			    
Sens. area /mm     			&  2456/1396	&2483/1232 	&2592/1765\\ 	       
Anode wires              	&  288 			&144 			&304\\		            
Pot. wires                	&    868 		&436 			&916\\		           
Gas volume /l         &   284.3 		&289.9			& 398.2\\	            
\hline
\end{tabular}
\caption{Main features of the rear tracking drift chambers.}
\label{tab:DC-features}

\end{table}

\begin{figure}[htbp]
\begin{center}
\vspace{-0cm}
\resizebox{0.4\textwidth}{!}{\includegraphics{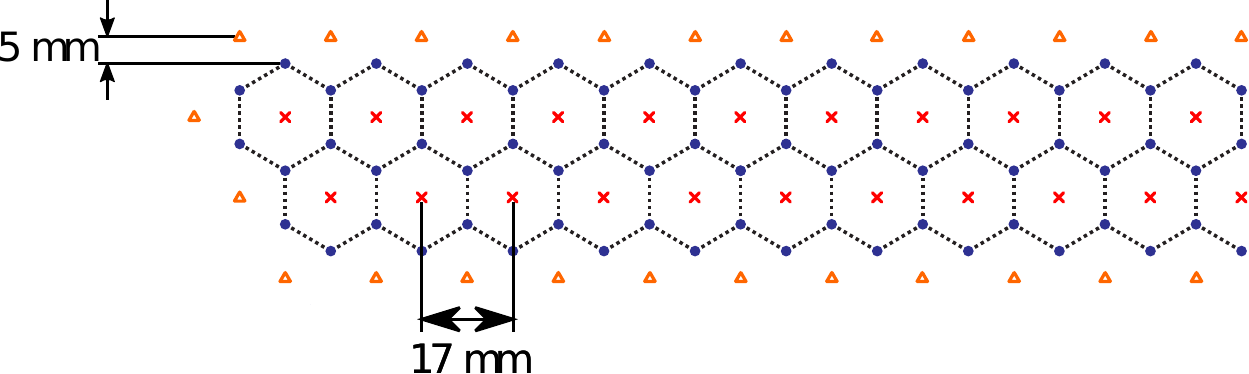}}\vspace{-0.1cm}
  \caption{Schematic of the double layer drift cell geometry. Signal
(anode) wires are shown as red crosses, potential wires as blue circles and field-forming wires at the edges as orange triangles.  Figure from ref.~\cite{Freyermuth}.}
 \label{fig:d_cells}
\end{center}
\end{figure}

The drift cells have a hexagonal structure with a width of 17\,mm, 
arranged in two layers at 15\,mm
distance to resolve the left-right ambiguity of particle intersections. 
The wires are gold plated tungsten. The anode
(signal) wires are 25\,$\mu$m in diameter and kept on a ground potential.
Drift cells are defined by the hexagonally arranged cathode wires of 25\,$\mu$m 
diameter, typically set at high voltages around U\,$ = -2800$\,V. The efficiency plateau is reached
at U\,$\simeq -2700$\,V. To assure an equal field distribution in
each cell and shield against external field distortions, additional field-forming wires 
surround the double-layer of
drift cells, shown in fig.~\ref{fig:d_cells}. These are 200\,$\mu$m diameter gold
plated beryllium bronze. All wires are soldered to PCB
boards and additionally glued with epoxy glue to increase
the mechanical stability.

Contrary to the other tracking detectors, the drift chambers do not have a central hole. 
The photon beam directly passes through. To avoid signal overflow and even damage
of the chambers through ionisation by the direct photon
beam, central spots of $\simeq 5\times5$\,cm$^2$ have been made insensitive. 
This is achieved by galvanising additional gold within
this area to the signal wires (six in each layer), thus increasing the total 
wire diameter to approximately 100\,$\mu$m.
This sufficiently reduces the gas amplification within the spot to prevent signals from the beam.

The gas is a mixture of argon (70\,\%) and CO$_2$ (30\,\%),
which is easy to operate, as it is non-toxic and not inflammable. Over the whole efficiency plateau the drift velocity
is practically constant, $v_D \simeq$ 7\,cm/$\mu$s. This is a rather
high value compared to other gas mixtures using organic
quenchers, however it provides no limitation to the
required position resolution $\delta x$, where $\delta x < 300$\,$\mu$m is achieved.
This is more than sufficient for the spectrometer's design momentum resolution of 3\,\%. 
In addition, polymerisation
of organic gas components on the signal wires is avoided,
which would reduce the gas amplification and lead to signal deterioration.

\begin{figure}[htbp]
\begin{center}
\vspace{-0cm}
	\includegraphics[width=\columnwidth,trim={0cm 0 0 11cm},clip=true]{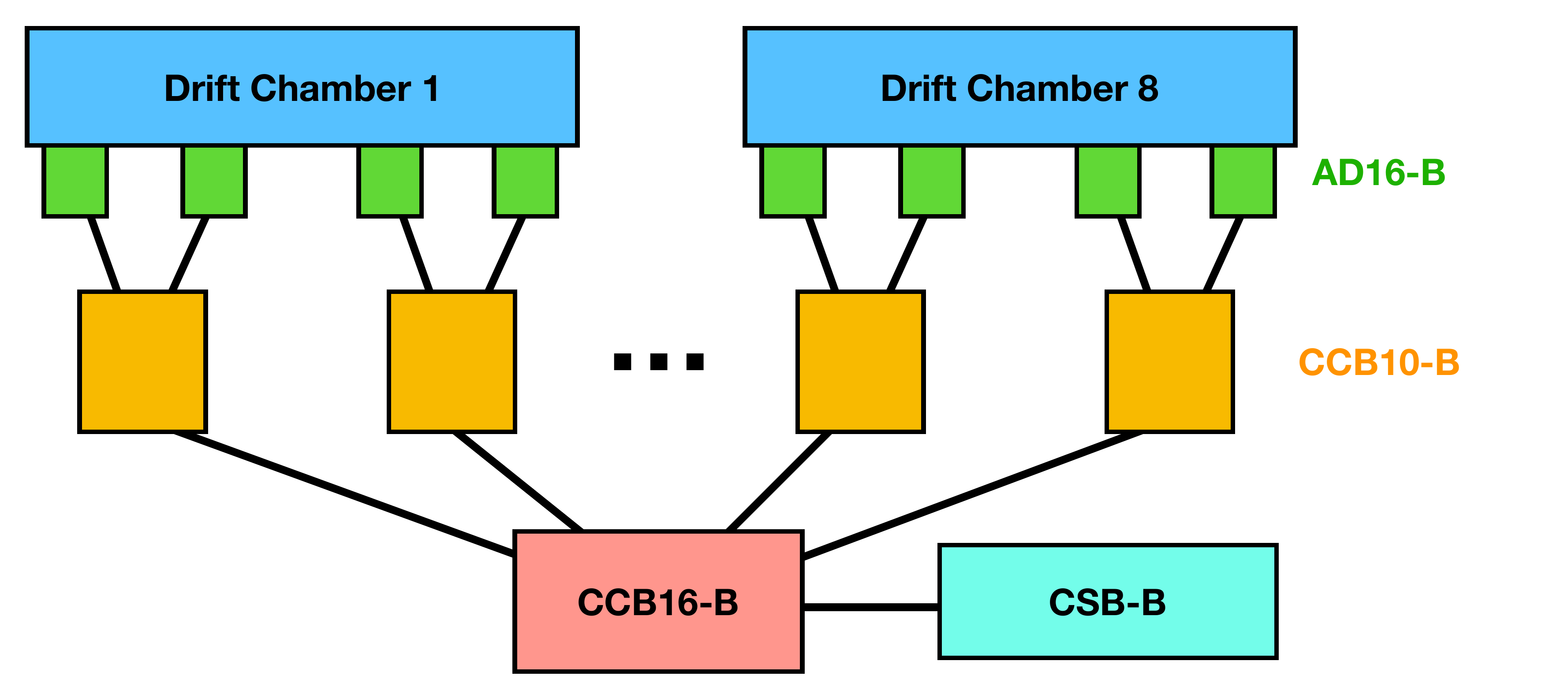}\vspace{-0.1cm}
  \caption{Schematic of the drift chamber readout system. It consists of front end amplifier/discriminator cards AD16-B, two levels of concentrators boards CCB10-B or CCB16-B, and system buffer PCI boards CSB-B.}
 \label{fig:d_readout}
\end{center}
\end{figure}

The anode signals are read out using the CROS-3B1\footnote{Coordinate Read Out System, 3rd generation, version B.}
system developed and built at the PNPI, Gatchina.  The main
components are frontend amplifier/discriminators, concentrators, and a system  buffer. 
A schematic of the readout is depicted in fig.~\ref{fig:d_readout}.

The frontends labelled AD16-B are directly attached
to the wire PCBs. Each board amplifies and discriminates signals of 16 anode wires. 
An on-board Xilinx\textsuperscript{\textregistered}   FPGA
produces an LVDS output of the discriminated signals. This
provides remote threshold adjustment and programmable
output delay.  The integrated TDC functionality for the individual wires for drift time measurements has a resolution of 2.5\,ns.  All TDC information derives from the same main clock,  such that identification of the clock cycle in which the trigger signal was seen is sufficient.
Signals are serially sent to one of the
first (low) level concentrator boards. Each CCB10-B low
level concentrator collects signals of 10 AD16-B cards and
transmits a single serial signal to a top level concentrator CCB-16B. The top level concentrator collects up to
16 CCB10-B outputs.  14 outputs are required for the BGOOD drift chambers. The top level concentrator then transmits a single signal to the CSB-B system buffer via fibre link. This is a PCI card with interface to the trigger electronics that collects all readout data. The CSB-B also provides convenient initialisation
and configuration of the frontend and concentrator cards.


\subsubsection{Time-of-flight spectrometer}
\label{subsubsec:ToF}

The final part of the forward spectrometer is the time-of-flight spectrometer,
5.6\,m downstream from the target.  The spectrometer measures
particle $\beta$ and thus, particle
identification is achieved by combining this with the
momentum determined from the deflection in the magnetic field.

The spectrometer consists of 3 walls of plastic scintillators, each segmented into individual horizontally oriented
bars. We used scintillator bars from the earlier GrAAL
and SAPHIR experiments.  The basic features of the ToF bars
are listed in table \ref{tab:TOF-features}.

\begin{table}[htbp]
\centering\begin{tabular}{l l l l}
\hline
 Wall 	&  1 			& 2				& 3  		\\ 

\hline
Origin                 &GrAAL 		&GrAAL	& SAPHIR	\\
Number of bars			&14		&14		&8			\\
Length /cm		&300		&300		&340		\\
Height /cm     		&11.5		&11.5		&21.5		\\
Depth /cm 		&3			&3			&5			\\
Time res. /ps		&350		&350		&700		\\
Distance /cm			&558		&571		&596.5		\\

\hline
\end{tabular}
\caption{Features of ToF spectrometer walls.  The time resolution is a combination of the light output from the bars, the photomultiplier tubes and corresponding electronics.}
\label{tab:TOF-features}

\end{table}

The ToF spectrometer covers a total area of $3\times3$\,m$^2$.
Each wall has a 10 to 22\,cm horizontal gap in the centre, both for the photon beam to pass through
and to avoid signals from $e^+e^-$ pairs
abundantly produced in the target and air, and swept over the
central dispersive plane by the open dipole magnet.

The readout of the scintillator bars of the ToF Walls is performed with photomultiplier tubes. Bars from GrAAL experiment
are equipped with Philips XP2282/B tubes. The readout electronics are composed of FPGA based jTDC and jDisc.

The crucial property of the ToF detector is the time resolution which is strongly affected by time walk when using leading edge discriminators. This effect can be taken into account and minimized with the Time over Threshold (ToT) technique~\cite{Freyermuth}, achieving a time resolution of \nolinebreak{$\sigma=$\,0.34\,ns} when combining information from multiple walls.

\subsection{Cryogenic target system}
\label{subsec:Target}

The BGOOD experiment uses either solid state
targets, for example, carbon, or cryogenic liquid targets, positioned at the centre of the BGO Rugby Ball.
Mostly cryogenic liquid hydrogen $\mathrm{(LH_2)}$ or deuterium $\mathrm{(LD_2)}$ targets are used.

The cryogenic target cell consists of an aluminium hollow cylinder of 4 cm diameter,  0.5\,mm wall thicknesses and, normally, 6 cm length.  Beam entrance and exit windows
are made of  0.1\,mm thin Mylar foil. For $\mathrm{LH_2}$ this yields a 
target area density of 0.425\,g\,cm$^{-2}$, corresponding to $2.53 \times 10^{-7}$\,$\mu$b$^{-1}$.

The target cell is surrounded by an outer vacuum shield, an aluminium pipe with a wall thickness of $0.66\pm 0.1$\,mm. The target system is connected to the beam pipe via a fast closing valve. The target is mounted on a rail system, enabling the target to be moved in and out of the beam line. This operation requires breaking the last part of the beam line vacuum.

The target cell is 
interchangeable and can be replaced by 
cells of different length (an 11\,cm length cell is also available). 
The hydrogen or deuterium gas is cooled down by helium through heat exchangers 
and liquefied inside the cell. \


Starting from room temperature it takes around 9 hours to have a full liquid hydrogen cell -- cooldown takes approximately 5  hours to liquefy.
15 minutes are needed to empty the cell using an electric heater and about 2 hours to refill.




\section{Photon tagging system and beam operation}
\label{sec:Tagger}

There are two main methods of producing high energy $\gamma$-rays beams: Compton scattering of laser
light off an electron beam \cite{Arut:1963,Milburn:1963}, or electron bremsstrahlung from a radiator
\cite{Diambrini:1968,Timm:1969}. 
Both techniques can produce linearly polarised photon beams, either by using a polarised laser source or a diamond radiator to produce coherent bremsstrahlung radiation.
The two techniques are complementary: the high degree of polarisation and the minimum level of low energy background obtained via Compton backscattering are compensated by the higher intensity and higher energy obtained with the coherent bremsstrahlung. Both methods produce a continuous energy spectrum, therefore tagging is mandatory to access the initial state value
of the photon energy.
BGOOD uses the coherent bremsstrahlung technique with the 3.2 GeV unpolarised electrons from ELSA.


\subsection{Operation principle}
\label{subsubsec:OperationPrinciple}
\begin{figure}[htbp]
	\begin{center}
		\vspace{-0cm}
		\resizebox{0.5\textwidth}{!}{\includegraphics{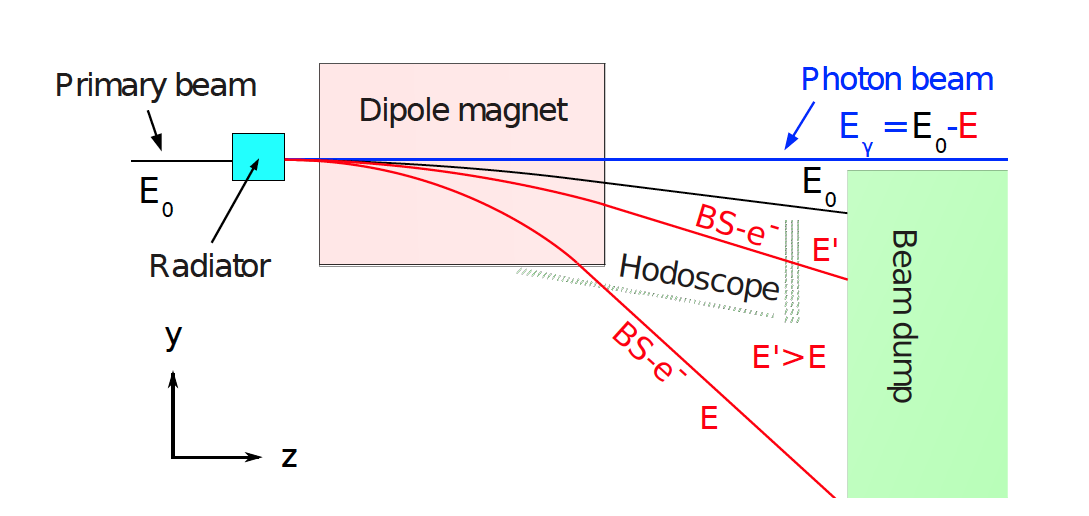}}\vspace{-0.1cm}
		\caption{Components and working principle of the photon tagger at the BGOOD experiment.}
		\label{fig:op_principle}
	\end{center}
\end{figure}

The Photon Tagger setup is used to assign the correct energy and production
time to the bremsstrahlung photons.  Timing coincidences with other detectors then relates the bremsstrahlung photons to corresponding hadronic reactions, where final state particles are identified in other detectors.
The working principle of the Photon Tagger is illustrated in fig.~\ref{fig:op_principle}.\

To determine the energy of the post-bremsstrahlung electron, $E_{e^-}$, the electron momentum is analysed by a magnetic spectrometer
consisting of a dipole magnet and a detector hodoscope. According to its momentum and to the integrated
magnetic field strength, $\int B \mathrm{d}l$, the electron trajectory curves in the magnetic field into
the hodoscope. From the detection position in the hodoscope, the energy  $E_{e^-}$, and therefore the energy of the correlated bremsstrahlung photon can be determined. The electrons
which are not involved in the bremsstrahlung process enter the beam dump.

The dipole magnet is connected to the exit beam pipe under vacuum.  
A vertical strip of kapton foil covers a space at the downstream side of the dipole magnet to allow electrons to pass out of the vacuum to either the hodoscope, or, if no bremsstrahlung process occurred, to the beam dump.

\subsection{Tagging hodoscopes}
\label{subsubsec:TaggingHodoscopes}
\begin{figure}[htbp]
\begin{center}
\vspace{-0cm}
\resizebox{0.5\textwidth}{!}{\includegraphics{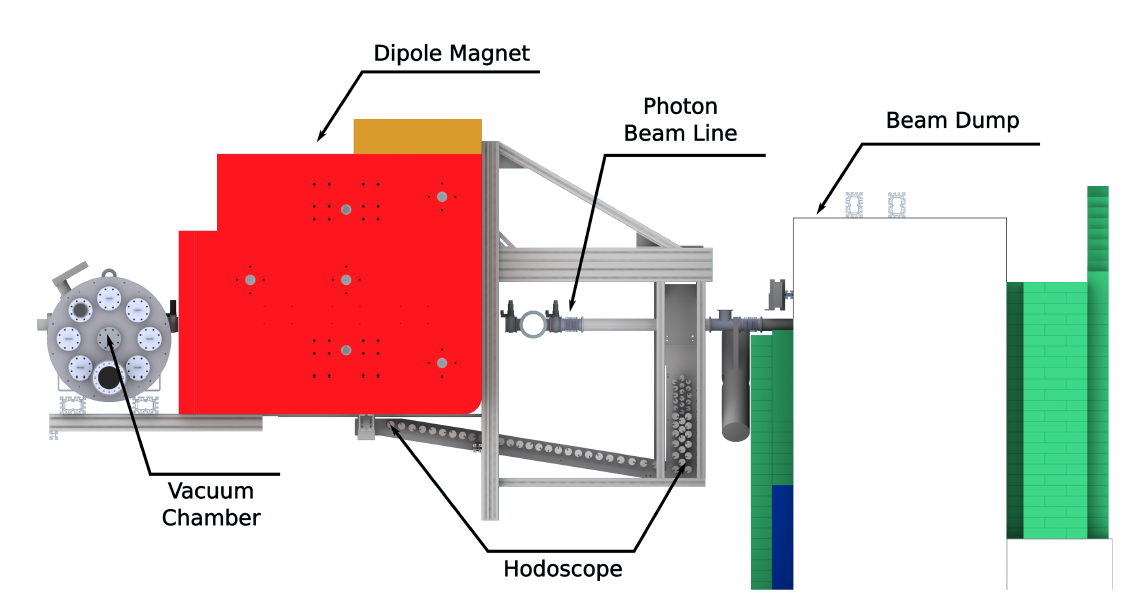}}\vspace{-0.1cm}\\
  \caption{Scheme of the photon tagger system. Main components are the vacuum chamber including
the bremsstrahlung radiators and beam monitoring tools, the dipole magnet and the detector hodoscope.}
 \label{fig:hodoscope}
\end{center}
\end{figure}

Most of the components of the photon tagger system of the BGOOD experiment are illustrated in fig.~\ref{fig:hodoscope}
which shows a technical scheme of the full setup.  The single components are described in this subsection.

\subsubsection{Radiators}
\label{subsubsec:Radiators}
Several bremsstrahlung radiators and tools for beam monitoring
purposes are mounted on a goniometer system. This is installed within a vacuum
chamber and placed in front of the dipole magnet. A list of the installed radiators
and beam monitoring tools is given in table~\ref{tab:radiators}. In total, four different amorphous radiators of different
thickness and materials and one crystalline radiator are installed. Amorphous radiators are used to measure
incoherent bremsstrahlung spectra to normalise the measured bremsstrahlung spectra of a diamond.
The normalised diamond spectra are used to compute the degree of linear polarisation.  The kapton radiator is used to understand the incoherent spectra originating from the kapton foil the diamond radiator is mounted on.

\begin{table}[htbp]
\centering\begin{tabular}{l l}
\hline
Material &  Thickness / $\mu$m	 \\ 
\hline
Copper       &50\\
Copper      &100\\
Kapton		&125\\
Copper+Kapton    		&67+65\\
Diamond+Kapton&560+65\\
Nickel-Steel wires & 400\\
Chromox & 1000\\

\hline
\end{tabular}
\caption{Radiators and beam monitoring tools used for the BGOOD experiment.}
\label{tab:radiators}
\end{table}

To determine the shape and the position of the beam profile, a fluorescent screen made of $\mathrm{Al_{2}O_{3}Cr}$, also
known as Chromox is used.  This can be placed with an angle of $45^{\circ}$ relative to the electron
beam direction, allowing the fluorescent spot to be observed with a camera perpendicular to the beam direction.

\begin{figure}[htbp]
\begin{center}
\vspace{-0cm}
\resizebox{0.5\textwidth}{!}{\includegraphics{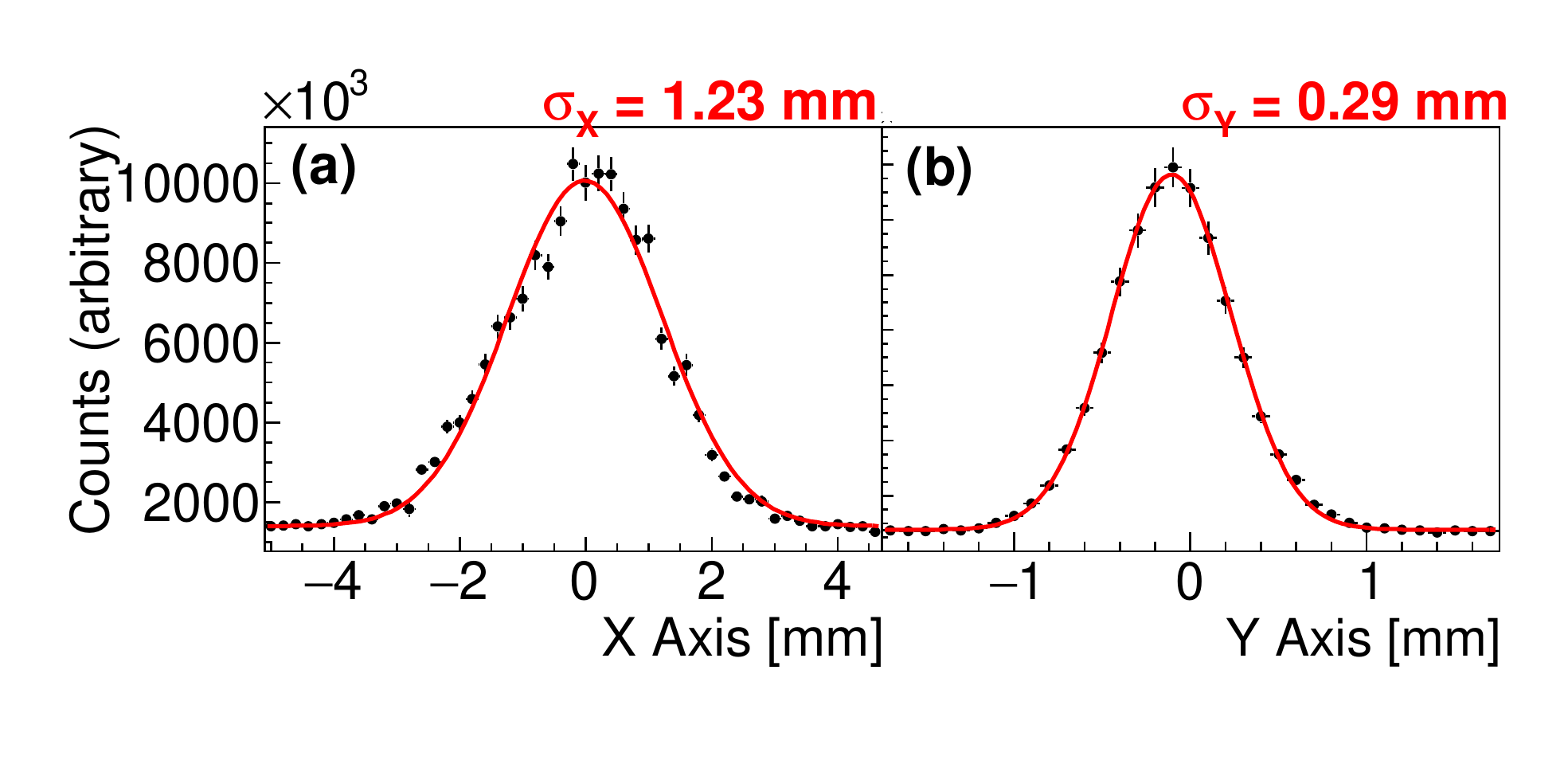}}\vspace{-0.1cm}
  \caption{Measured primary electron beam profile and position performed with two wire scans. Each distribution is fitted with a Gaussian function, the sigma labelled inset.}
 \label{fig:scan}
\end{center}
\end{figure}

For a more precise analysis of the beam profile, two nickel-steel wires are included in
the goniometer system. These wires can be moved orthogonally with respect to the reference electron
beam direction along the horizontal or vertical axis. While moving the wires stepwise through the beam,
the post-bremsstrahlung electrons are detected by the tagging hodoscope. Hence, an iterative scan of
the beam profile is possible (see fig.~\ref{fig:scan}) by plotting the total tagging rate in dependence of the position of the wire.

\subsubsection{Goniometer}
\label{subsubsec:Goniometer}

\begin{figure}[htbp]
\begin{center}
\vspace{-0cm}
\resizebox{0.5\textwidth}{!}{\includegraphics{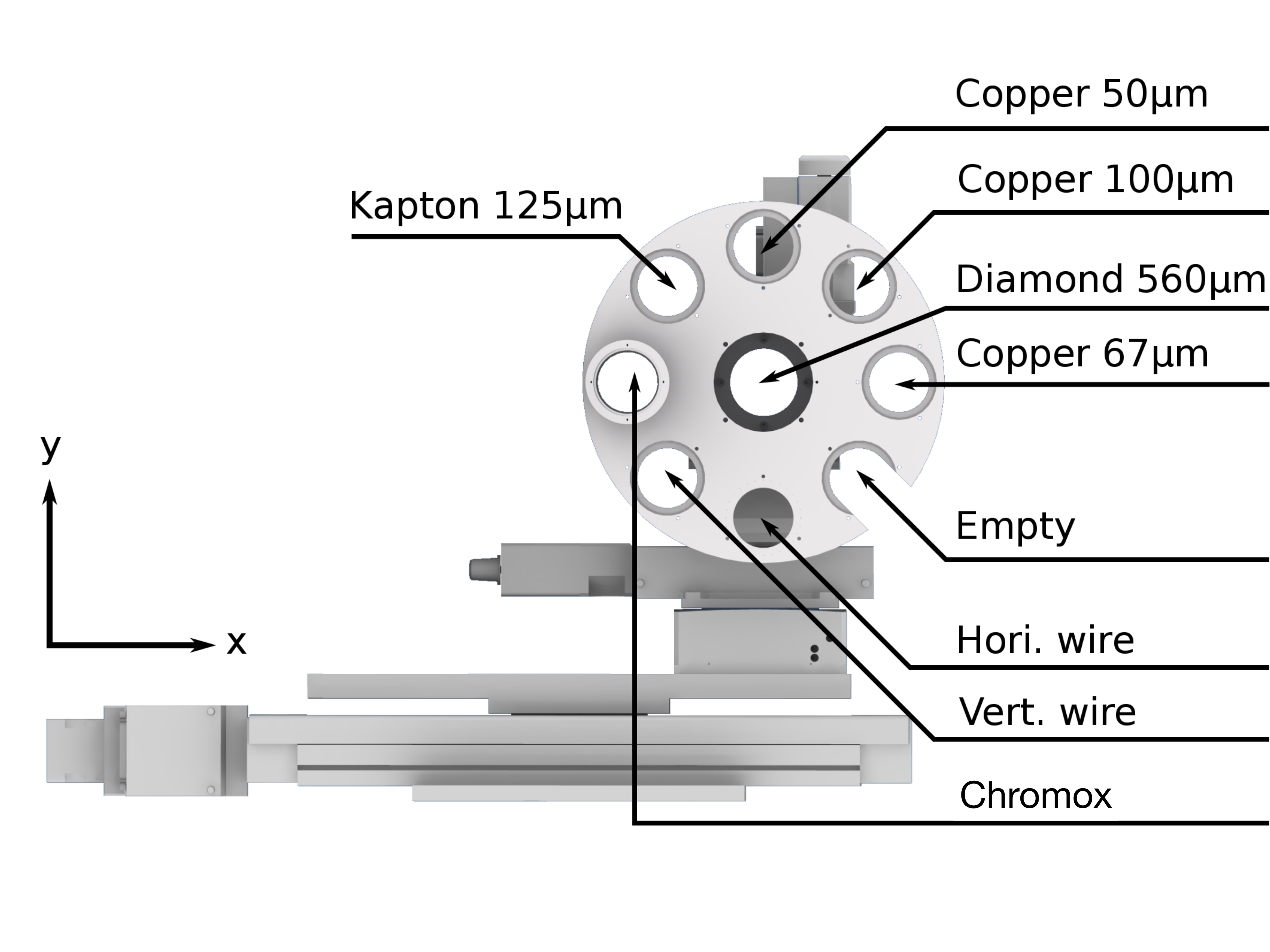}}\vspace{-0.1cm}
  \caption{Front view of the goniometer containing the bremsstrahlung radiators and alignment tools.}
 \label{fig:goniometer}
\end{center}
\end{figure}

The goniometer system is used to choose between different radiators and beam monitoring
tools. However, the main task of the goniometer system is the precise alignment of the diamond
radiator. The energetic range for the production of linearly polarised
photons and subsequently the degree of polarisation of the produced photon beam is sensitive to the
relative position of the diamond with respect to the direction of the primary electron beam. Therefore,
a goniometer system with a high accuracy in positioning and reproducibility is required. The commercial
goniometer system from Newport consists of five motorised positioning units\footnote{Two linear motors, models UTS150PP and UZS80PP, and three rotational motors, models URS100BPP, BGS80PP and RVS80PP.  The goniometer uses an XPS type controller.}. This setup
allows translations horizontally and vertically relative to the reference electron beam direction and rotations
along all three spatial axes. The current setup of the goniometer system is illustrated in fig.~\ref{fig:goniometer}.

\subsubsection{Tagging magnet and collimators}
\label{subsubsec:TagMagColl}

The dipole magnet is used to bend the trajectory of the post-bremsstrahlung electrons according to
their momentum onto the tagging hodoscope. Electrons which are not involved in the
bremsstrahlung process are directed into the beam dump.
 The dipole magnet is from
Brown-Bovery Switzerland (type MC) which can be operated with currents up to 1500\,A, corresponding
to a maximum field value of $B = 2.0$\,T. The deflection angle of the primary electron beam is 
$7.74^{\circ}$, which determines the geometry of the electron beam dump. 
For highly relativistic electrons, the deflection
depends only on the ratio $B/E_0$, where $E_0$ is the electron beam energy.  This deflection angle is reached for $B/E_0= 0.43$\,T/GeV.
Since primary electron beam energies up to 3.2\,GeV are used, magnetic field values up to $1.376$ T are
necessary.\
At a distance of 3.25\,m to the radiator, a system of two collimators
and one permanent magnet is installed. The collimators are made of lead and are both 20\,cm long.
The first collimator has an inner diameter of 7\,mm. Its purpose is to reduce the incoherent contribution in the
bremsstrahlung spectrum when using a diamond radiator for the production of a linearly polarised photon
beam, resulting in an enhancement in the degree of polarisation (described in sect.~\ref{subsec:Beam_pol}). Due to the collimation
of the photon beam, electron-positron pairs are produced. Hence, a permanent magnet is placed directly
after the first collimator to clean the photon beam from charged particles. The second collimator with
an inner diameter of 12\,mm is placed after the cleaning magnet to stop high energetic charged particles which
could not be removed from the vacuum beam guidance by the cleaning magnet. 

\subsubsection{Tagger hodoscope}
\label{subsubsec:ScintillatorLadder}

\begin{figure}[htbp]
	\begin{center}
		\vspace{-0cm}
		\includegraphics[trim={0cm 15cm 0 0cm},clip=true,width=\linewidth]{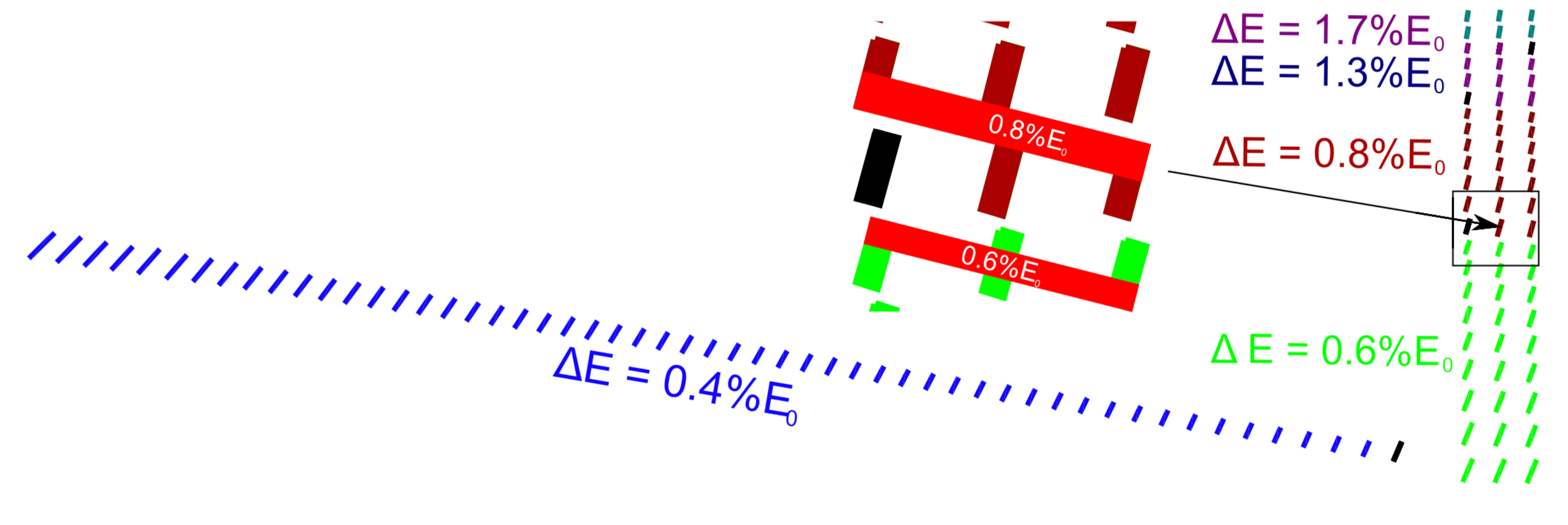}
		\caption{Schematic showing the plastic scintillators of the Tagger hodoscope.  The different colours have respective energy resolutions, $\Delta E$ labelled.  The zoomed in region demonstrates the spatial overlap principle used create coincidence channels.  Figure adapted from ref.~\cite{Bella}.}
		\label{fig:taggerhodoscope}
	\end{center}
\end{figure}

The hodoscope of the tagging system, shown schematically in fig.~\ref{fig:taggerhodoscope}, consists of 120 plastic scintillators covering an
energy range of 10\,\%$E_0$ to 90\,\%$E_0$. It is split into two parts: 54 scintillators are inserted into the horizontal
part and the remaining 66 scintillators are installed in the vertical part. Two adjacent scintillators spatially
overlap by 55\,\%. Such overlaps are considered as a coincidence channel providing the momentum of
the detected post-bremsstrahlung electron. Requiring the coincidence of two neighbouring scintillators
as a trigger condition reduces the background and improves the energy resolution. The energy width of a coincidence channel varies from
0.40\,\%$E_0$ in the horizontal part and 0.60\,\%$E_0$ to 1.70\,\%$E_0$ in the vertical part (see sect. 4.4.4). The scintillator bars of the hodoscope are read out via photomultipliers from one side. For geometrical reasons, two different kinds of photomultipliers are used, namely ET-Enterprise 9111 and Hamamatsu R7400U. A time resolution of 210 ps for the ET9111SB and of 180 ps for the Hamamatsu R7400U is achieved (including the electronics). 
A complete description of the construction, installation and commissioning of the tagging hodoscope can be found in ref.~\cite{Bella}.

\subsubsection{{ARGUS} scintillating fibre detector}
\label{subsubsec:ARGUS}

\begin{figure}[htbp]
\begin{center}
\vspace{-0cm}
\resizebox{0.3\textwidth}{!}{\includegraphics{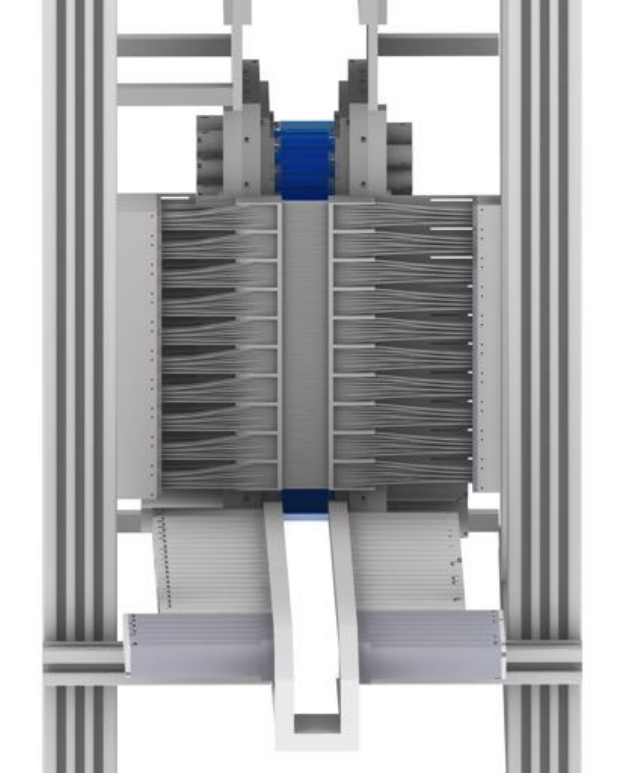}}\vspace{-0.1cm}
  \caption{CAD rendering of the tagging system and ARGUS, viewed in the beam direction.}
 \label{fig:argus}
\end{center}
\end{figure}

The measurements intended at BGOOD foresee the use of a photon beam linearly polarised in
an energy range between 1.2\,GeV and 1.8\,GeV. For these photon energies, the corresponding
post-bremsstrahlung electrons are deflected in the direction of the vertical part of the tagger, shown in fig.~\ref{fig:hodoscope}.
To monitor the stability of the position of the coherent bremsstrahlung peak 
during data-taking, a better position resolution in this vertical part is required. This is achieved
by a scintillating fibre detector called {\it ARGUS} which is placed in front of the vertical part of
the tagger with respect to the flight direction of the post-bremsstrahlung electrons (see fig.~\ref{fig:argus}).

The detector covers an energy range from 30\,\% to 66\,\% of the
incoming ELSA electron energy and consists of 480 double-clad BCF-10 round scintillating fibres, 2\,mm diameter, from Saint Gobain, which are arranged in three layers. The readout is performed with 30 Hamamatsu
H6568 multi-anode photomultipliers with 16 channels each. An energy resolution of 0.08\,\%
to 0.45\,\% of the incoming electron energy is achieved.  A spatial coincidence of the post-bremsstrahlung electron trajectory is required between ARGUS and the Photon Tagger Hodoscope.  Time information is determined by the  Photon Tagger Hodoscope, which has a better time resolution, compared to approximately 0.7\,ns for ARGUS.

%

\subsection{Flux monitors}
\label{subsubsec:FluxMonitors}

The Flux Monitor ({\it FluMo}) and the Gamma Intensity Monitor ({\it GIM}) are placed at the end of the BGOOD experiment, at approximately 720\,cm and 780\,cm downstream from the target respectively~\cite{Zimmermann}.
These detectors measure the photon flux, which is required to perform
normalisations, for example, for cross section measurements. The GIM is a fully absorbing lead glass detector, with a height, width and depth of $ 14 \times 14 \times 28$\,cm.  Incoming photons produce electromagnetic showers which are detected by measuring the produced Cherenkov light.  This provides short light pulses compared, for example, to scintillation light, which is crucial for high rate stability in determining the photon flux.  Additional rate stability tests were performed with pulsed LED light, and in the experimental area using comparisons to the extracted ELSA current and the measured Photon Tagger rate.  See ref.~\cite{Zimmermann} for details.
The GIM is subject to radiation damage and therefore cannot be kept continuously in the
photon beam. The relative measurement of the photon flux is thus obtained with the FluMo detector.

FluMo consists of five plastic scintillators, with a height, width and depth of $ 5.0 \times 5.0 \times 0.5$\,cm, in series in the beam to detect photons which converted into electron-positron pairs. This measures only a fraction of the  brems\-strahlung photons with respect to the GIM.  Photon flux measurements with FluMo require the knowledge of its detection efficiency, which is
determined with data taken by triggering on the GIM detector in dedicated low intensity runs. Coincidences between the tagger system
and the GIM are identified at a software level. For each coincidence, it is determined if the FluMo detected an electron-positron pair originating from a bremsstrahlung photon. The discriminator
thresholds have to be set in order to assure the detection of both minimum ionising particles.
The energy resolution is sufficient to identify and disregard single electron or positron hits, and to identify events where both the electron and positron from pair production impinge upon the detector.

Flumo and GIM use ET-Enterprise 9111513 and Hamamatsu R2083 photomultipliers respectively, and the signals are electronically processed with jDisc.


\subsection{Tagging system coincidence electronics}
\label{subsubsec:TaggerElectronics}


The front-end electronics consist of an active splitter
and amplifier stage, a dual-threshold discriminator and shaper stage, and an FPGA stage including the
trigger logic and a TDC. In addition, the FPGA allows time alignment of the signals coming from the
several channels of the tagger hodoscope and delivers the scaler information for single and coincidence
channels.



Each analogue signal of the photomultipliers is sent to a splitter and amplifier board (AFA: Analogue Fan-out Amplifier) which is
developed in-house \cite{Messi}. Two outputs with different amplifications (x1, x2) relative to the input signal
are available. The x2 amplified signal is sent to the discriminator. If required, the x1 output is given to
an ADC for HV adjustments of the photomultipliers, otherwise it is terminated with a 50 $\mathrm{\Omega}$ resistor.
The measured time jitter between input and output is less than 15 ps.\

The amplified signal is then sent to a dual-threshold discriminator and shaper board called B-FrED
(Big-Front End Discriminator)
which is also developed in-house. Each board has 16 input channels and 16 LVDS output channels.
The output signals have a fixed length of 3.2\,ns. Since originally TDC and trigger logics were separated,
the B-FrED output has a fan-out of two. The double pulse time resolution of the discriminator amounts to 7.5\,ns (which is comparable to the pulse decay time)
and the time jitter between input and output is less than 10\,ps. Using a micro-controller, the thresholds
of each channel can be set individually via Ethernet.  Depending upon position, the rates in individual channels are approximately 3 to 10\,MHz.  Further details are found in ref.~\cite{Bella}. 

One of the LVDS outputs of the B-FrED is sent to the FPGA board containing the TDC and trigger
logic.  For the tagging system, the board is equipped with an
IO LEMO mezzanine card, used for triggers and service purposes and two LVDS input mezzanine cards
to connect the output of the B-FrEDs. 

An additional logic unit checks for coincidences between two signals originating from adjacent tagger
channels. The time window in which a coincidence between two signals is defined can be set to 1.7,
2.5, 3.3 or 4.1\,ns.
To align the logical signals, a time delay in steps of 833\,ps can be set.
The output signal of this coincidence unit is given as the OR of all 119 coincidence
channels and is used as a trigger. In addition to the trigger and TDC information, life-time
gated scalers for single and coincidence channels are available.

\section{Trigger system}
\label{sec:TriggerDAQ}
The trigger logic for the BGOOD experiment is split into two parts. At first, local trigger signals for each participating sub-detector are created. The second step of the trigger logic is contained in the global trigger module. This module receives the local trigger signals of the sub-detectors. If the trigger condition is met, the trigger latch is set and the global busy is asserted. This prevents any further generation of trigger signals. When all sub-detectors are ready for the next trigger, the latch is reset and the system can accept the next trigger.  The design of the DAQ enables readout parallel to data taking.

\subsection{Local trigger}
\label{subsec:LocalTrigger}
A short overview of the sub-detectors capable of providing a local trigger and their trigger conditions is given in table~\ref{tab:loc_triggers}. These triggers indicate that one or more particles have been detected by the corresponding detector. The trigger conditions vary between the different sub-detectors according to detector type and geometry.

\begin{table}[htbp]
\centering\begin{tabular}{l l}
\hline
Sub-detector &  Trigger condition 	 \\ 
\hline
Tagger       & Coincidence of adjacent channels\\
GIM      &Detector hit \\
FluMo      &Detector hit \\
 Scintillator barrel &OR of all channels\\
SciFi    		&OR of all channels\\
BGO Rugby Ball &Energy sum (multiple thresholds) \\
SciRi  & OR of all channels \\
ToF & Coincidence between PMTs in one bar\\

\hline
\end{tabular}
\caption{Local triggers from sub-detectors.}
\label{tab:loc_triggers}
\end{table}

\subsubsection{Tagger}
\label{subsubsec:LT-Tagger}
The tagger trigger is implemented on Spartan\textsuperscript{\textregistered}6 FPGA modules. The firmware on these modules forms the coincidences between adjacent scintillators.  The trigger output is then determined by the OR of all coincidences. As the
coincidences as well as the OR are implemented in clocked logic, the output signal can only occur in fixed intervals at six steps relative to the
200\,MHz clock, providing a step width of $\simeq  833$\,ps. 


\subsubsection{GIM and FluMo}
\label{subsubsec:LT-GIMFluMo}

As the GIM detector has only one single channel the trigger is simply implemented using a discriminator. A discriminated signal of one of the FluMo scintillators is used for tagger time alignment data.

%
\subsubsection{BGO Rugby Ball}
\label{subsubsec:LT-BGO}
The trigger condition for the BGO Rugby Ball is given by a threshold on the
total energy deposited in the calorimeter. This is realised by building the analogue sum of all analogue signals of the detector and discriminating upon the amplitude. To create the analog sum the signal of each crown of 32 channels is
summed in the CAEN SY493 mixer modules. The 15 sums are then combined using two stages of LeCroy 428F Linear Fan-In Fan-Out modules. The analog output of the second Fan-In Fan-Out module is then passed to two discriminators (a jDisc and a constant fraction discriminator). The thresholds of the discriminators are set according to the required energy deposit in the BGO Rugby Ball, typically 120\,MeV and 75\,MeV.



%
\subsubsection{SciFi and SciRi}
\label{subsubsec:LT-SciFi}
A trigger on the SciFi detector is implemented as a separate OR of the horizontal and vertical fibres. To avoid unnecessary signal splitting a trigger output was added to the FPGA module containing the TDC firmware.  This output gives the OR of 96 channels. To create the OR of all 640 channels the signals are then ORed using NIM modules.
SciRi creates a trigger in the same way and is formed by the OR of the 96 channels.
%
\subsubsection{ToF Walls}
\label{subsubsec:LT-ToF}
The trigger condition on the ToF walls is given by a hit in any of the scintillator bars. As the bars are read out on both sides, the meantime between the two signals is used to reduce noise and improve the trigger timing.  Using the adjustable delay timing, time differences between channels can be compensated. For each of the ToF walls a separate output signal is generated and sent to the global trigger module.


%
\subsection{Global trigger}
\label{subsec:GlobalTrigger}
The global trigger module,  implemented as a VME FPGA board,  combines information provided by the individual local triggers. The board provides 32 LVDS input channels for the local triggers. Before the input signals are passed to the logic part of the trigger module, the signals
are sampled with a resolution of 833 ps (the same as the Tagger in sect.~\ref{subsubsec:LT-Tagger}). Using shift registers the signals can be delayed by up to
375 ns in steps of the sampling resolution.  This allows
the adjustment of the trigger timings by software and removes the need for additional delay cables in the trigger signal paths.  In addition to the configurable delay, the gate length for all inputs can be set independently of the input signal length.

The digitised time information is then analysed in the logic part of the trigger module. To allow for a flexible setup covering the interesting trigger combinations,
the logic is separated into two steps. In the first step, the so called primary logic, 16 individual logic blocks can be configured. These blocks build the logical AND or the logical OR on any subset of the inputs. In addition, inputs may be declared as VETO inputs to suppress the output of the combination. A
second stage of logic blocks takes the output of the primary logic blocks and processes them in the same way as the primary logic processes the input signals. These 6 secondary logic blocks allow for the creation of independent, complex trigger conditions. The
output of the secondary logic blocks is then prescaled and ORed to generate the final trigger attempt.

In addition to these data trigger attempts, the global trigger module features several other trigger possibilities not based on the detectors. An integrated clock and fixed life time trigger provide minimum bias triggers for studying the behaviour of the detectors as
well as the data acquisition system. These are included in the OR of the secondary logic blocks. A special scaler trigger configured to a rate of 20\,Hz is used to initiate the readout of the scaler modules and additionally for studying systematics.  The start and end of spill signals provided by the
accelerator are also used to generate trigger signals. These are used for automatically changing parameters during special runs. These scaler and spill triggers are identified in the signal distributed through the synchronisation system, allowing all sub-detectors to act
accordingly. As the special triggers must not be discarded, they are queued instead of ignored if they occur during the dead time and are sent as soon as the DAQ can handle the next trigger.

The Global Trigger module provides built-in scalers for all input channels as well as the outputs of the secondary trigger logic blocks. This allows the continuous monitoring of the rates of the incoming signals and of the trigger attempts.  The incoming signals and the logic outputs are aquired by a TDC.


By accepting all digital local trigger outputs, the global trigger module allows for an almost arbitrary selection of coincidences, logical ORs and vetoes, providing a very flexible mechanism for data taking, calibrations and preselection of final states.
Using a condition as a veto is possible, with optional prescaling factors.



In the following we provide an overview for the most important trigger conditions used during physics data taking. All conditions contain the Tagger detector, which is crucial for coincidence time selection of hadronic events, and as a start time for the ToF walls.  All the trigger rates were obtained using a beam current of $\sim$ 1200\,pA at 3.2\,GeV electron energy, and a 6\,cm long  hydrogen target.

\begin{itemize}
\item Tagger AND BGO Rugby Ball (high threshold) is the most open trigger condition only requiring the BGO Rugby Ball
and Tagger local triggers, achieving a trigger attempt rate of $\sim$ 0.75~kHz. This
condition uses the local trigger with the high threshold on the energy sum of the BGO Rugby Ball
calorimeter of approximately 120 MeV, just below the $\pi^0$ mass.

\item Tagger AND  SciRi AND BGO Rugby Ball (low threshold) is the trigger condition chosen to enhance the acquisition of hadronic events with low energy deposition in the central detectors and charged particles, for example protons, in the intermediate angular region covered by the SciRi detector.  The combination with the Tagger trigger and the BGO Rugby Ball trigger with low threshold reduces the rate of the combined trigger to about 400\,Hz which is feasible for the data acquisition without significantly increasing the dead time.

\item Tagger AND SciRi AND SciFi AND ToF is the only standard data-taking trigger condition not requiring the BGO Rugby Ball, thus also enhancing the detection of hadronic reactions for which little to no energy is deposited in the central calorimeter.  The requirement on correlated hits in the different angular regions covered by SciRi and the forward detectors reduces the overall trigger attempt rate to approximately 250\,Hz.

\item Tagger AND SciFi AND ToF AND BGO Rugby Ball (low threshold) selects hadronic
events with low energy deposition in the central detectors and forward
going charged particles reaching the ToF Walls.  The trigger attempt rate for this condition is less than 150\,Hz.

\end {itemize}

The global trigger module performs the OR of the above conditions providing the {\it Physics Trigger} condition. The total trigger attempt rate for the Physics Trigger condition is approximately 1.5\,kHz while the effective trigger rate is approximately 1\,kHz, the reduction mainly being caused by combinatorial overlap.  Typical detector, trigger rates and gate widths are summarised in table~\ref{table:triggers}.

\begin{table}[htbp]
	\centering\begin{tabular}{l l l}
	\hline
	Detector & Triggered rate & Gate width  \\
		\hline
		Tagger       & 25\,MHz & 1\,ns\\
		BGO     & 1400\,Hz & 40\,ns\\
		ToF Wall 1      & 40\,kHz  &50\,ns\\
		SciFi (horizontal fibres) & 160\,kHz & 30\,ns\\
		& \\
		 \multicolumn{3}{ c }{Physics Triggers (all include the Tagger)}   \\
		 & \\
		BGO (high)   		& 750\,Hz &\\
		SciRi \& BGO (low) & 400\,Hz & \\
		SciRi \& SciFi \& ToF & 250\,Hz & \\
		SciFi \& TOF \& BGO (low) & 150\,Hz &\\
		
		\hline
	\end{tabular}
	\caption{Examples of typical trigger rates and gate widths for detectors and Physics Triggers.  All Physics Triggers require a coincidence with the Photon Tagger.   BGO (low/high) corresponds to the low or high energy threshold in the BGO Rugby Ball.  Only gate widths for individual detector triggers are given.}
	\label{table:triggers}
\end{table}


\section{Data acquisition system}
\label{sec:DAQ}

The BGOOD DAQ was derived from the existing DAQ of the Crystal Barrel experiment \cite{Schmidt04} and 
although it was significantly modified and extended, it shares the same overall design. A complete
description can be found in ref.~\cite{DanielThesis}.
To adequately manage the data provided by the different sub-detectors, the system
features a distributed readout, as well as a low level global trigger.  The global trigger receives inputs from local triggers which process the signals
provided by a sub-detector. The output of the global trigger is then distributed to the
readout modules by a synchronisation system. The readout hardware is located in
several crates, each containing its own CPU. The crates, including the readout modules,
as well as the acquisition software running on the CPUs are referred to as LEVBs.

\begin{figure}[htbp]
\begin{center}
\vspace{-0cm}
\resizebox{0.5\textwidth}{!}{\includegraphics{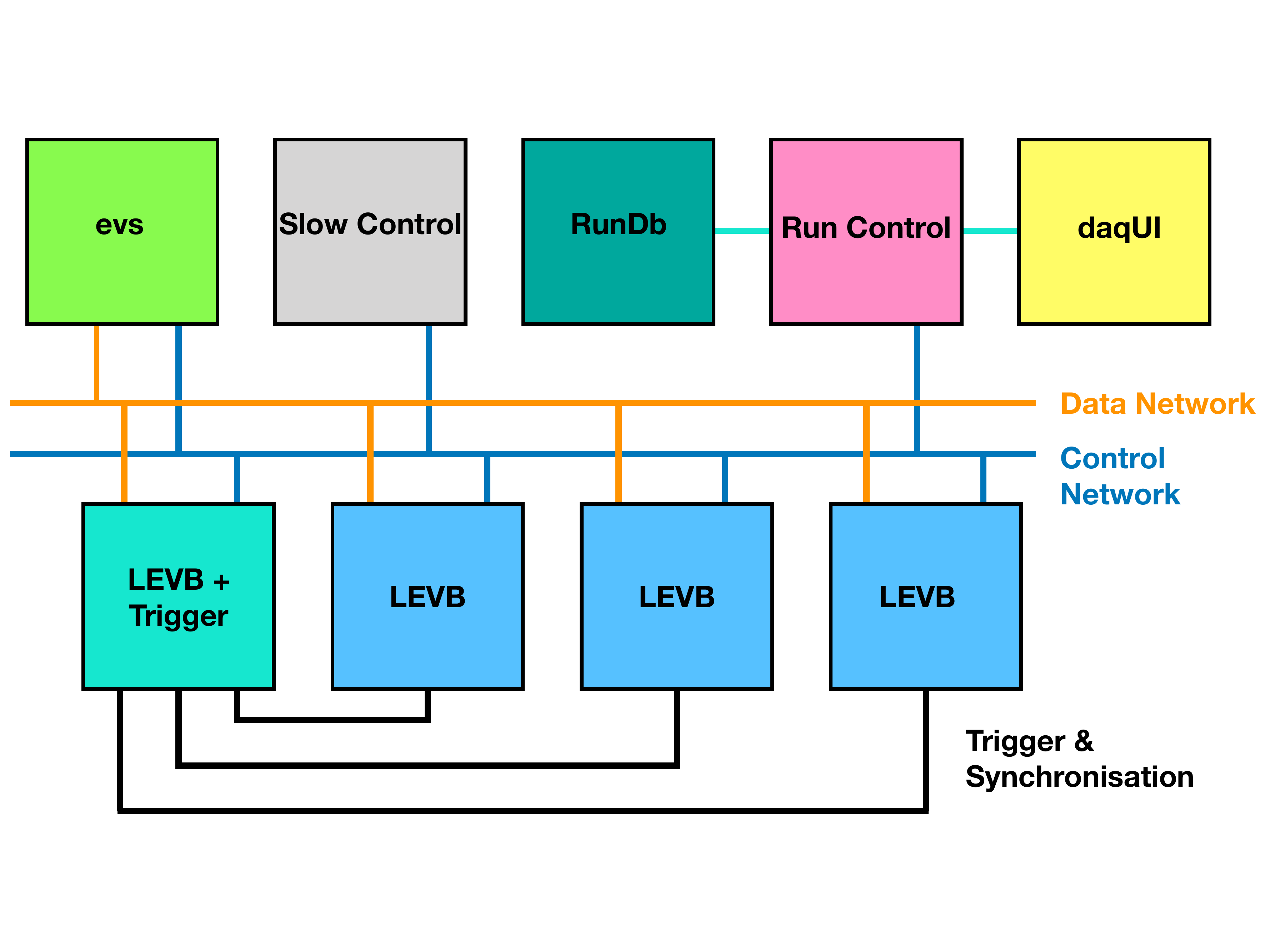}}\vspace{-0.1cm}
  \caption{Overview of the DAQ components and connections.}
 \label{fig:DAQ_overview}
\end{center}
\end{figure}

The structure of the DAQ system is illustrated in fig.~\ref{fig:DAQ_overview}.
The DAQ system is split into the event saver (evs), the runcontrol, the user interface
(daqUI), the database (runDb) and the readout machines per sub-detector (LEVBs). The daqUI
allows the user to control the DAQ while the runDb provides a log of all runs acquired.
The daqUI has no direct connection to the readout machines, but only connects to the
central runcontrol. The runcontrol and the remaining computers are connected using
two ethernet networks. The data network is used to send the detector data
from the LEVBs to the event saver. The control network carries the control commands
between the different components and provides an interface to the slowcontrol. The
trigger information is sent using dedicated synchronisation lines from the LEVB
containing the global trigger module to all other LEVBs.
Events are buffered per LEVB module and sent to the evs asynchronously. 

A major advantage of the DAQ is that trigger attempts are still accepted while the readout is still in progress.
This is achieved as most hardware modules have a dual-page layout and can keep data buffered for readout while continuing the measurement.
The dead time therefore only depends on the slowest hardware module and the time structure of the accepted triggers, and is not linearly correlated with the accepted trigger rate. 
Due to the random nature of the event topology, the trigger dead time varies per event and is of the order of 200 to 400\,$\mathrm{\mu}$s.

\section{Beam performance}
\label{sec:Beam_perf}

The BGOOD tagged photon beam routinely achieves rates of 25\,MHz both with amorphous and with diamond radiators. Two orthogonal polarisation planes are used alternatively and an automatic switch between the planes is performed approximately every 20 minutes between data taking runs during normal operation.

\subsection{Tagging efficiency}

\begin{figure}[htbp]
	\begin{center}
		\vspace{-0cm}
		\resizebox{0.45\textwidth}{!}{\includegraphics{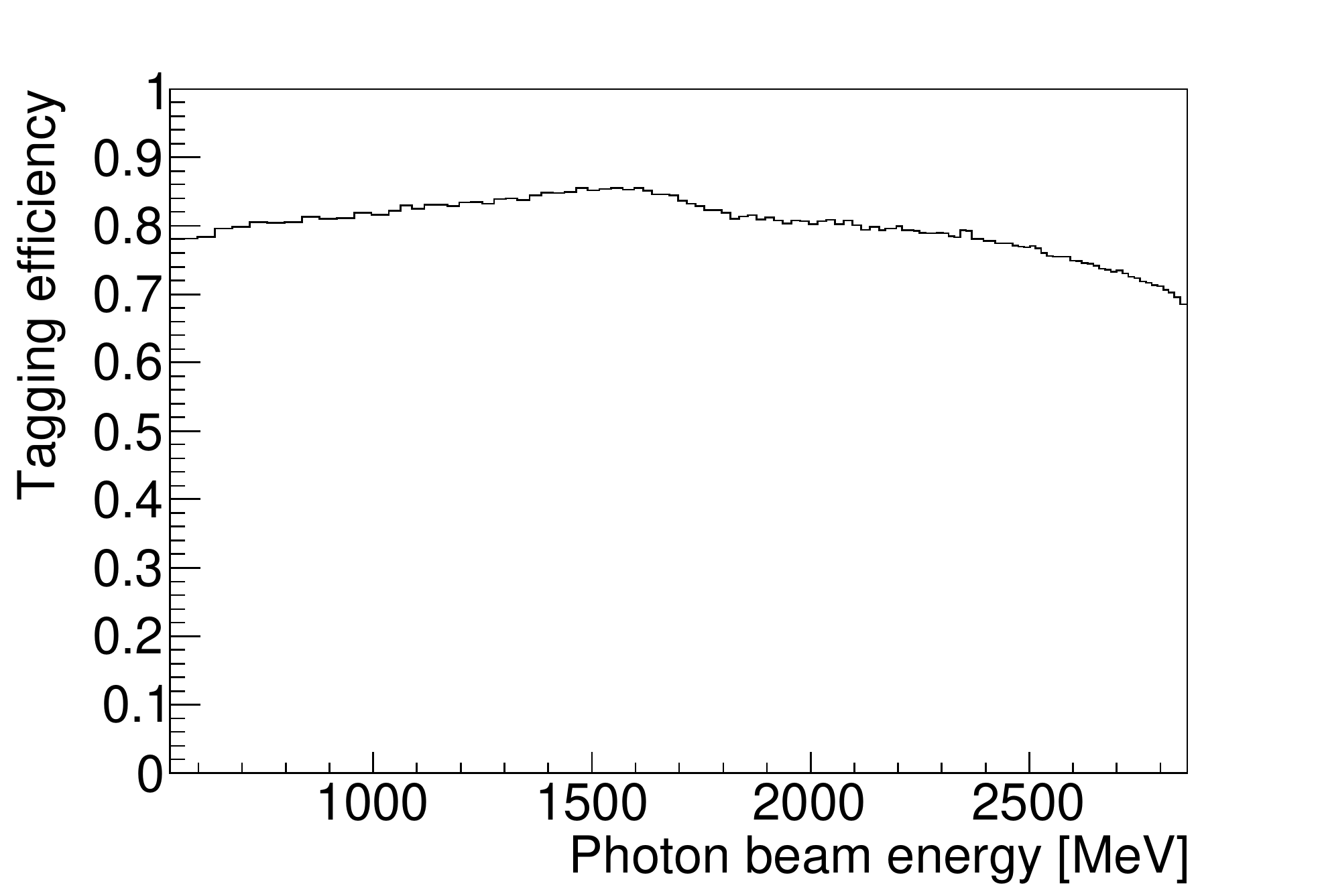}}\vspace{-0.1cm}
		\caption{Tagging efficiency spectrum.}
		\label{fig:taggingefficiency}
	\end{center}
\end{figure}

The tagging efficiency is a measure of the fractional amount of energy tagged photons which pass the collimator and impinge upon the target, compared to the total amount measured in the Photon Tagger.  This is continually monitored during data taking via the relative rates between FluMo (after calibration with GIM using special low rate runs) and the Photon Tagger.  Figure~\ref{fig:taggingefficiency} shows a typical tagging efficiency spectrum with a diamond radiator, which is at a level of approximately 80\,\%.  The small increase at 1600\,MeV is due to the increase of the degree of linear polarisation at this energy (see sect.~\ref{subsec:Beam_pol}), changing the angular distribution of the beam and increasing the number of photons passing the collimator.

\subsection{Crystal radiator alignment and beam polarisation}
\label{subsec:Beam_pol}

Coherent bremsstrahlung  occurs when there is constructive interference between the bremsstrahlung process on atoms in a periodical lattice structure such as a diamond.  The recoil momentum is then absorbed by the entire crystal.  It can be shown (see ref.~\cite{Timm:1969}) that the 
lower in energy the coherent bremsstrahlung peak is relative to the end-point, the larger the degree polarisation and enhancement is.  
The width of the coherent peak depends upon the incident electron beam divergence and multiple scattering in the radiator, and therefore the thickness of the radiator.  The rate however decreases with a thinner radiator.
With additional consideration of cost and availability, the 560\,$\mu$m thick diamond radiator used at BGOOD was chosen as an optimal compromise for these factors.

\begin{figure}[htbp]
\begin{center}
\vspace{-0cm}
\resizebox{0.4\textwidth}{!}{\includegraphics{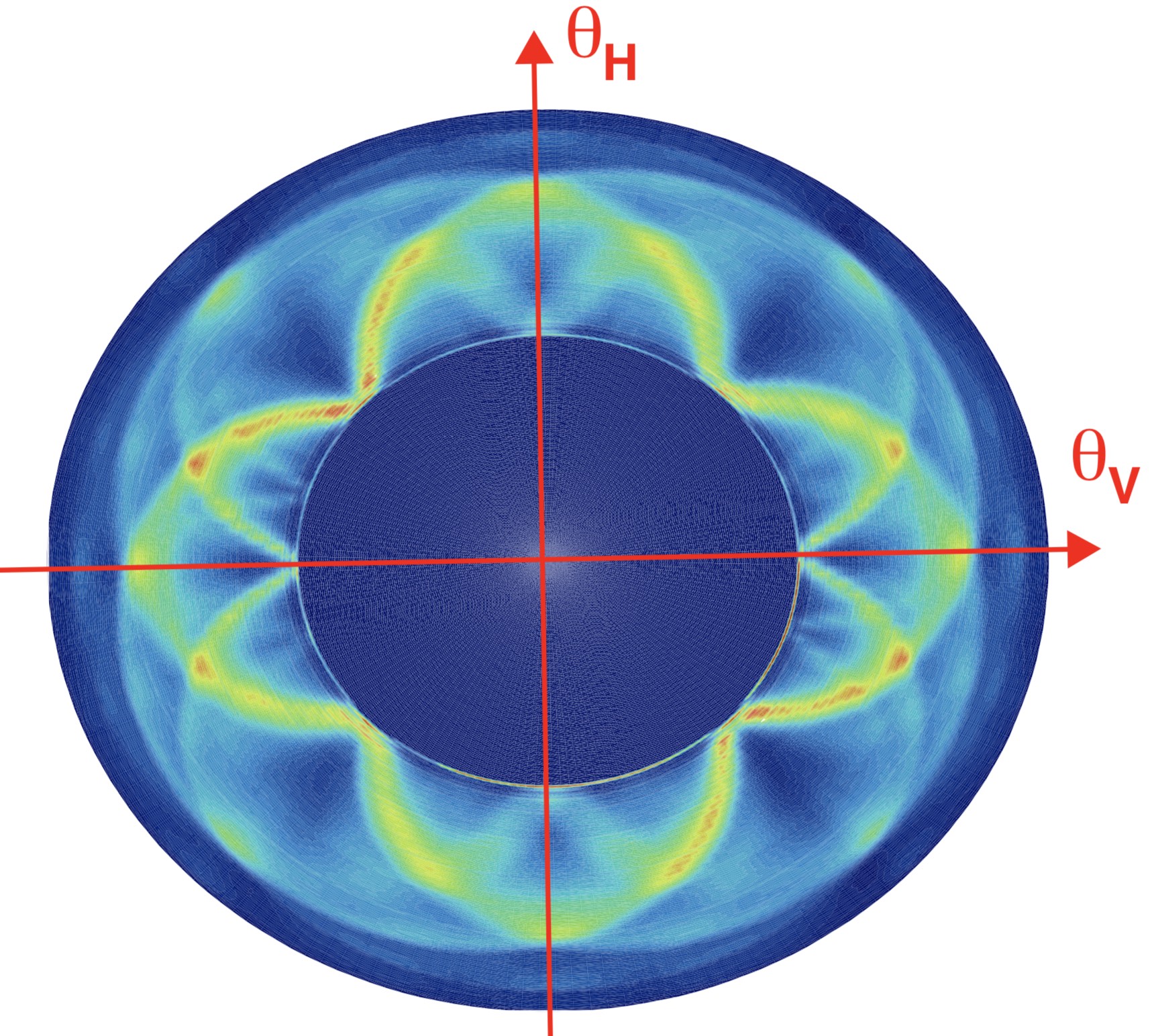}}\vspace{-0.1cm}
  \caption{Stonehenge plot determined using the Photon Tagger, of a crystal with planes well aligned with the beam momentum.  The axes $\theta_\mathrm{H}$ and $\theta_\mathrm{V}$ indicate rotations around the X and Y axes respectively (the two axes orthogonal to the beam direction).}
 \label{fig:Stonehenge}
\end{center}
\end{figure}

To obtain the highest possible degree of polarisation from coherent bremsstrahlung, the diamond crystal has to be aligned precisely with respect to primary electron beam momentum. This task is performed by manoeuvring the goniometer system according to the indications obtained through a {\it Stonehenge} plot~\cite{Livingston:2008hv}, described in detail for the BGOOD experiment in ref~\cite{Bella}. An example of such a plot for a crystal well aligned with the beam momentum is given in fig.~\ref{fig:Stonehenge}.

\begin{figure}[htbp]
	\begin{center}
		\vspace{-0cm}
		\resizebox{0.45\textwidth}{!}{\includegraphics{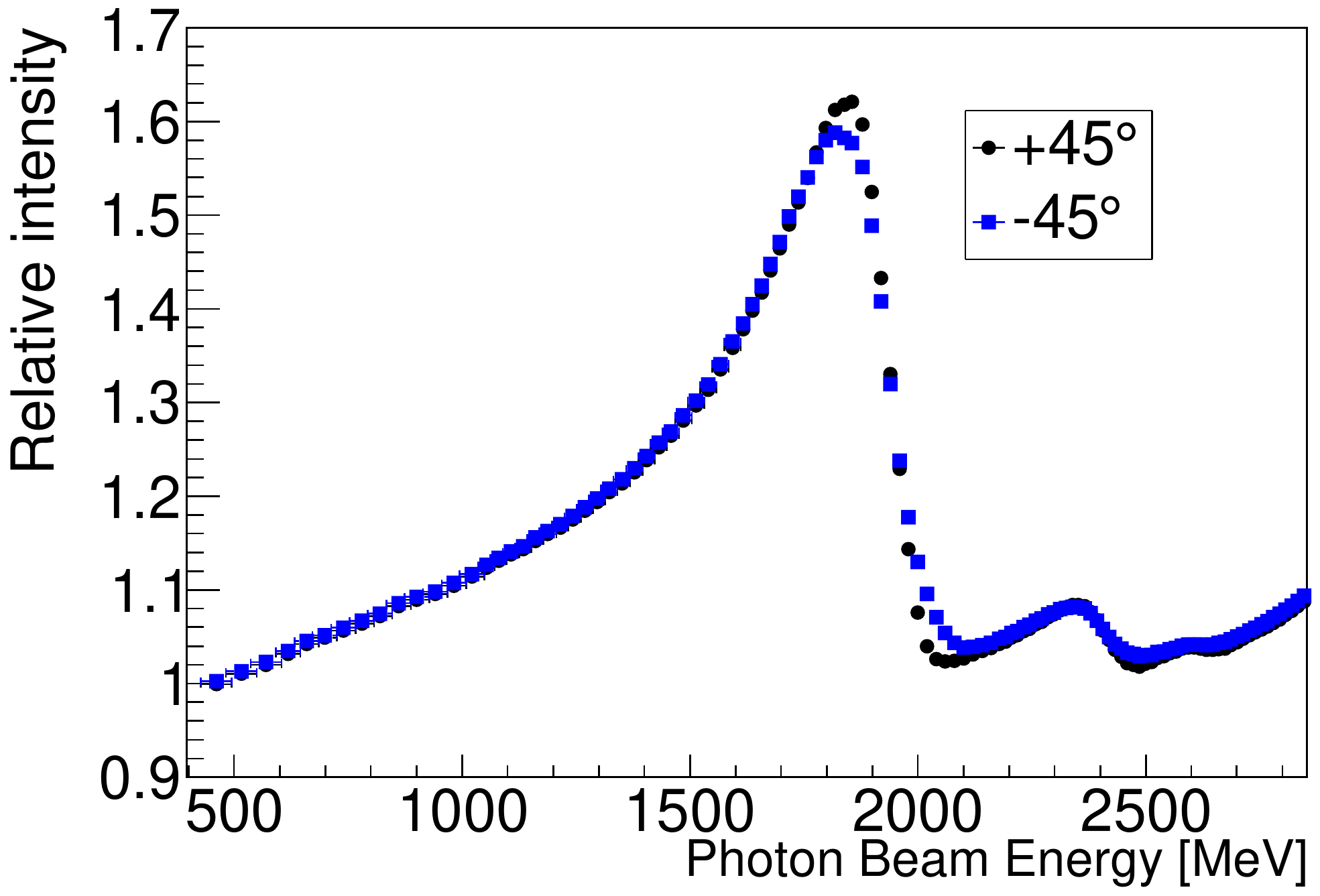}}\vspace{-0.1cm}
		\caption{Normalised diamond bremsstrahlung spectra for azimuthal orientations of $\pm 45^{\circ}$.}
		\label{fig:polppolm}
	\end{center}
\end{figure}

The divergence of the ELSA electron beam is larger in the horizontal direction than in the vertical one. Therefore, in order to assure that the coherent bremsstrahlung spectra have the same shape for both polarisation directions, the optimal azimuthal orientation of the planes is $\pm 45^{\circ}$. Figure~\ref{fig:polppolm} shows the comparison of the normalised diamond bremsstrahlung spectra for both polarisation planes. The spectra generally agree, with a small deviation between 1800\,MeV and 2100\,MeV.

During long data taking periods a displacement of the primary electron beam may occur, changing the position of the coherent edge in the bremsstrahlung spectrum. Hence, the spectrum is monitored to be able to react on such variations. For the monitoring, the diamond bremsstrahlung spectrum is observed with a frequency of 20\,Hz simultaneously to the data taking. Each bremsstrahlung spectrum is normalised by a reference copper bremsstrahlung spectrum which is recorded every eight hours.

The distribution of the normalised
spectrum in the region of the main peak is fitted by the empirical function in eq.~1.
\begin{equation}
f (x) = \mathrm{pol}(0) + (1- \mathrm{Erf}((x-p_0)/p_1))\cdot \mathrm{pol}(3)
\end{equation}\label{eq:cohedge}
$p_0$ is the position halfway down the falling edge of the coherent peak in the brems\-strahlung spectrum, $p_1$ the width of the
error function and pol(0) and pol(3) are polynomials of zeroth and third order.

\begin{figure}[htbp]
\includegraphics[trim={0cm 0 0 0cm},clip=true,width=0.48\textwidth]{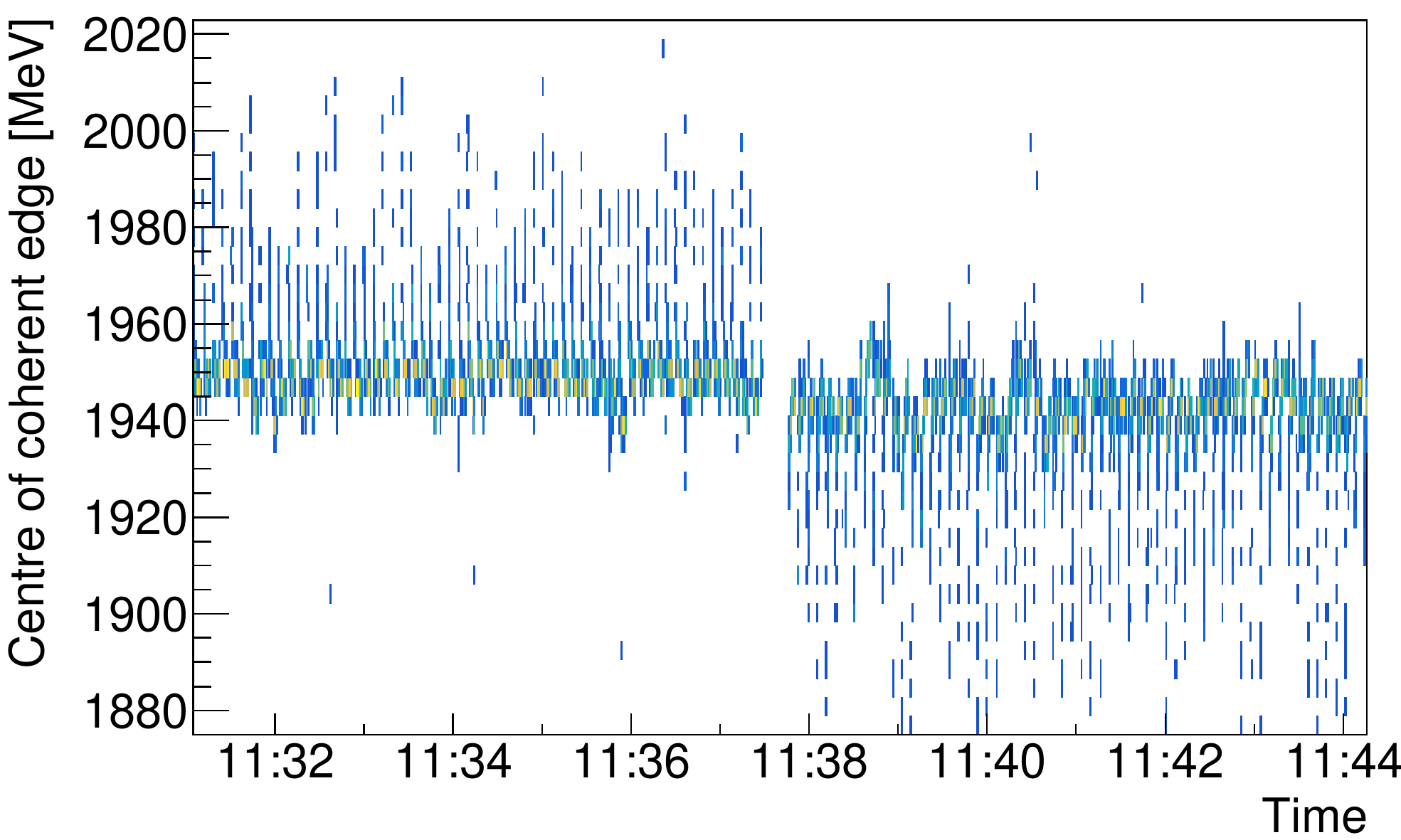}\\
	\includegraphics[width=0.48\textwidth]{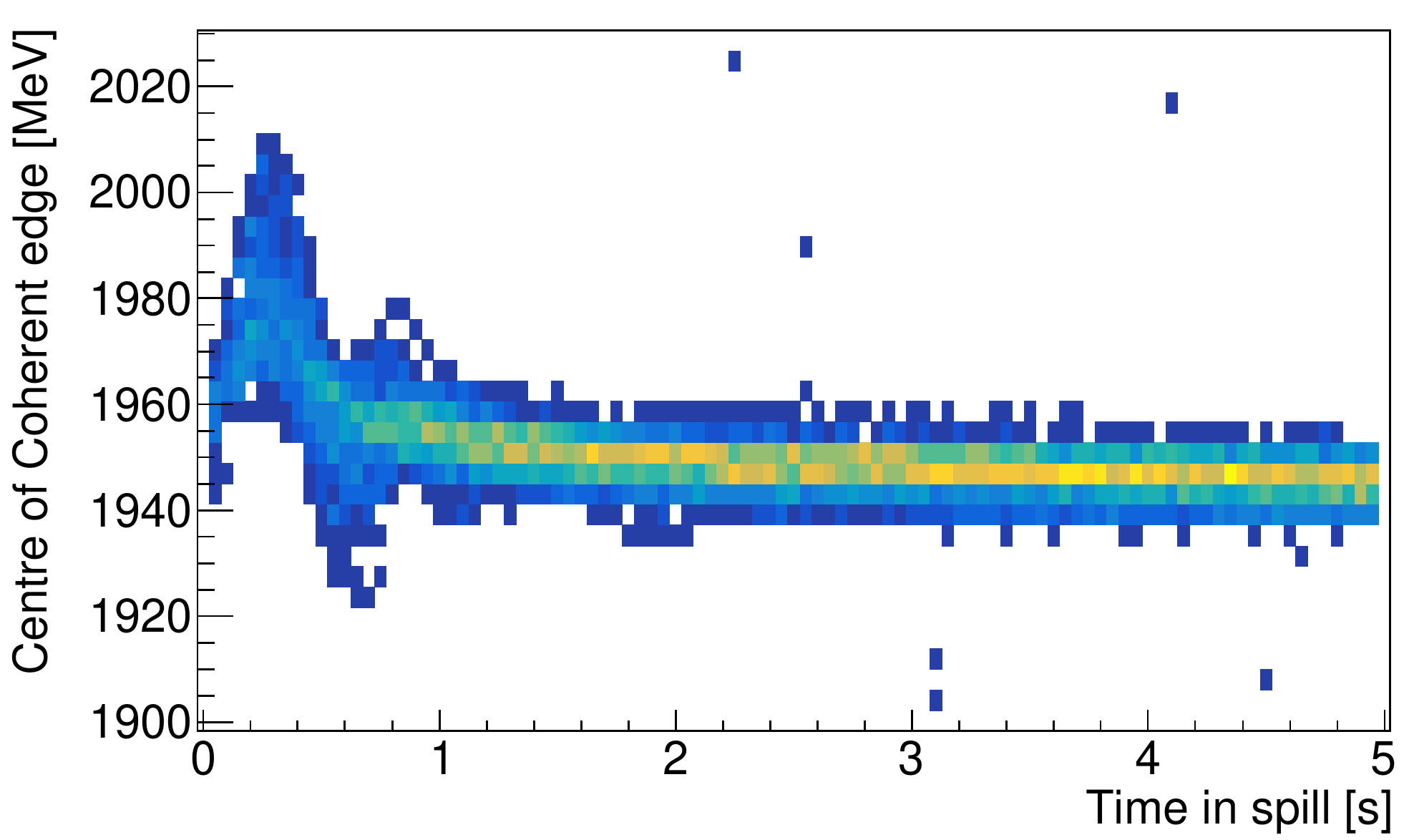}
	\caption{Online monitoring of the position of the coherent edge, using the fitted coherent edge position, $p_0$, from the empirical fit given in Eq.~1.  Top: during a data taking period. A change of polarisation state was made at 11:38.
		Bottom: During the extraction of the beam from ELSA for the $-45^{\circ}$ polarisation plane.}\label{fig:Coherent_edge}
\end{figure}

The stability of the coherent edge position for two successive polarisation directions is shown in fig.~\ref{fig:Coherent_edge} (top).
The coherent edge position is monitored throughout the spill extraction. There is an intrinsic instability during the first 0.5\,s of the spill, while in the remaining part of the spill, the position is found to be stable  to within approximately 0.5\,\% of the photon energy. The data collected in the first half second of the spill are therefore not used in the determination of beam asymmetries (see fig.~\ref{fig:Coherent_edge} (bottom)).

\begin{figure}[htbp]
\begin{center}
\resizebox{0.45\textwidth}{!}{\includegraphics{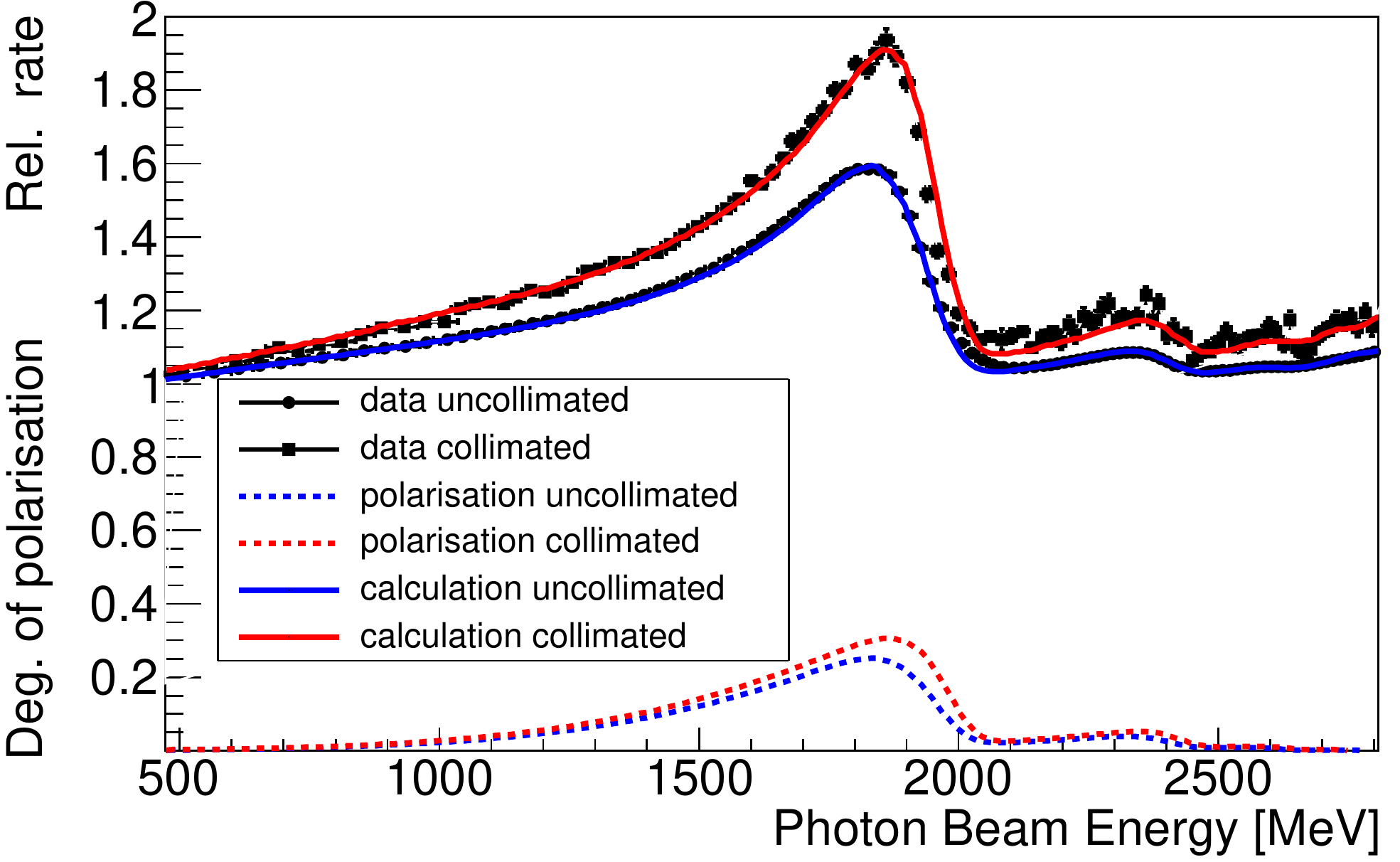}}
  \caption{Relative uncollimated and collimated diamond bremsstrahlung spectra compared to calculations performed
with COBRIS. The dashed lines show the corresponding expected degree of polarisation. By collimating the produced bremsstrahlung photon beam, the degree of polarisation can be significantly increased.}
 \label{fig:Coll_uncoll}
\end{center}
\end{figure}

The degree of polarisation is determined on a run by run basis. The diamond bremsstrahlung spectra are normalised with amorphous
bremsstrahlung spectra and compared to analytical calculations performed with a program that  accounts for experimental effects such as primary electron beam divergence, primary electron beam spot size and multiple scattering in the radiator (COBRIS~\cite{Bella}). Moreover, the coherent part of the spectrum is enhanced with tighter beam collimation. In fig.~\ref{fig:Coll_uncoll} the results for uncollimated and collimated spectra are shown and compared to the calculation.

\subsection{The ARGUS spectrum}
\label{subsec:Argus_sp}

\begin{figure}[htbp]
\begin{center}
\vspace{-0cm}
\resizebox{0.45\textwidth}{!}{\includegraphics{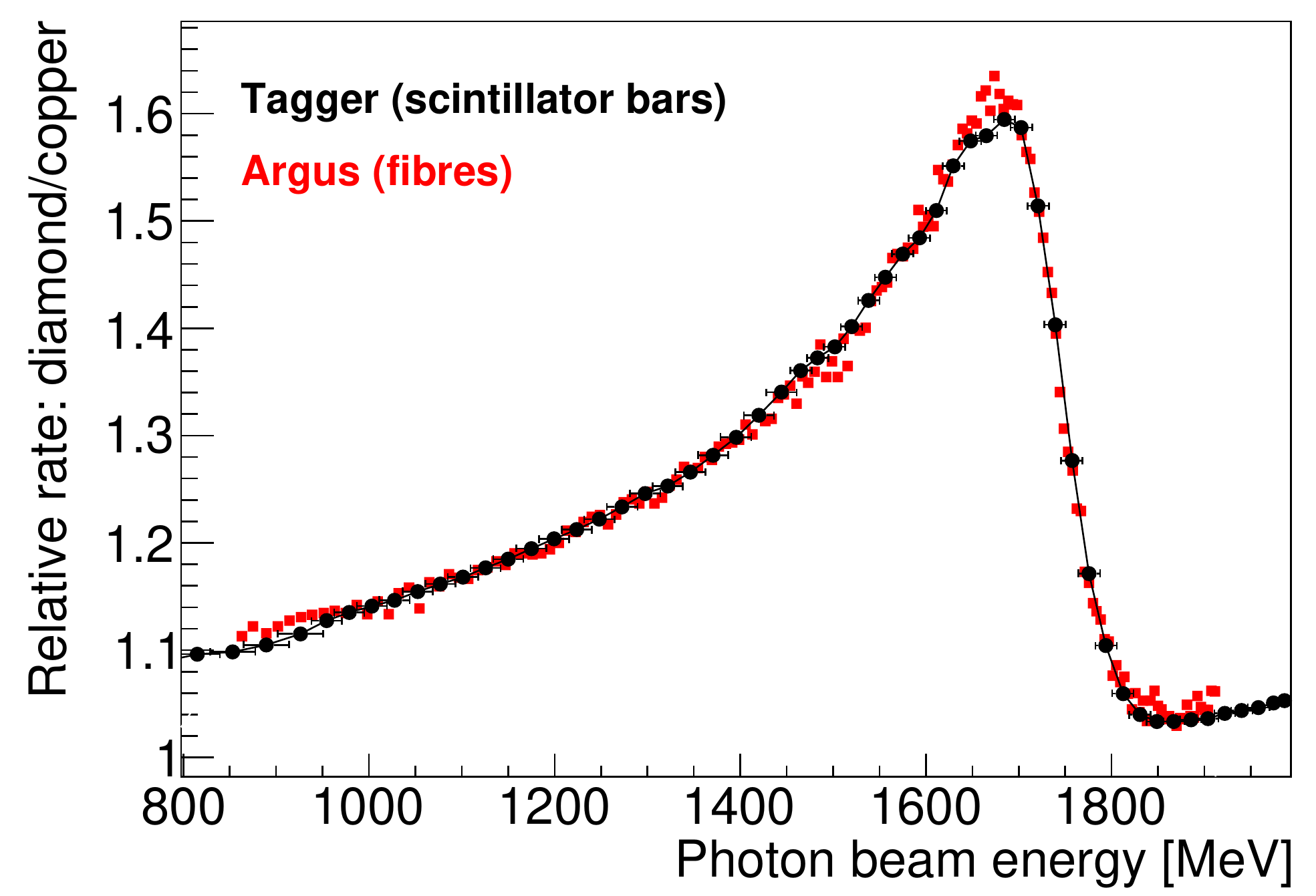}}
  \caption{Energy spectrum measured by the ARGUS detector and compared to the spectrum measured by the tagger.  A line connects the tagger spectrum data to guide the eye over the falling edge. The data used to produce this figure is different than the data shown in previous figures in sec.~\ref{subsec:Beam_pol}, where the peak of the coherent edge was set at a different energy.}
 \label{fig:argus_spectrum}
\end{center}
\end{figure}

The ARGUS detector, described in \ref{subsubsec:ARGUS}, with a $\sim$5\,MeV energy resolution, allows a finer energy binning of the photon spectrum around the polarisation peak.  This can be used either for a more precise determination of the degree of polarisation, or for a more precise determination of the initial state. The response of ARGUS is shown in fig.~\ref{fig:argus_spectrum} where the measured spectrum is compared with the one obtained by the Tagger alone.

\section{Detector performance}
\label{sec:Det_perf}

In this section the performance of the BGOOD detector is presented. The data were obtained during the commissioning stage of the experiment when a 6\,cm liquid hydrogen target was used and the ELSA energy was set between 2.4 and 3.2\,GeV.  The four physics triggers listed in table~\ref{table:triggers} were used in combination and modelled in all simulated data presented.


%
\subsection{Central detector}
\label{subsec:CentralDetector}

\subsubsection{Detection of $\pi^0
$, $\eta$ and $\eta^{\prime}$ mesons in the BGO Rugby Ball}
\label{subsubsec:pi0-eta-BGO}

\begin{figure}[htbp]
\begin{center}
\vspace{-0cm}
\resizebox{0.45\textwidth}{!}{\includegraphics{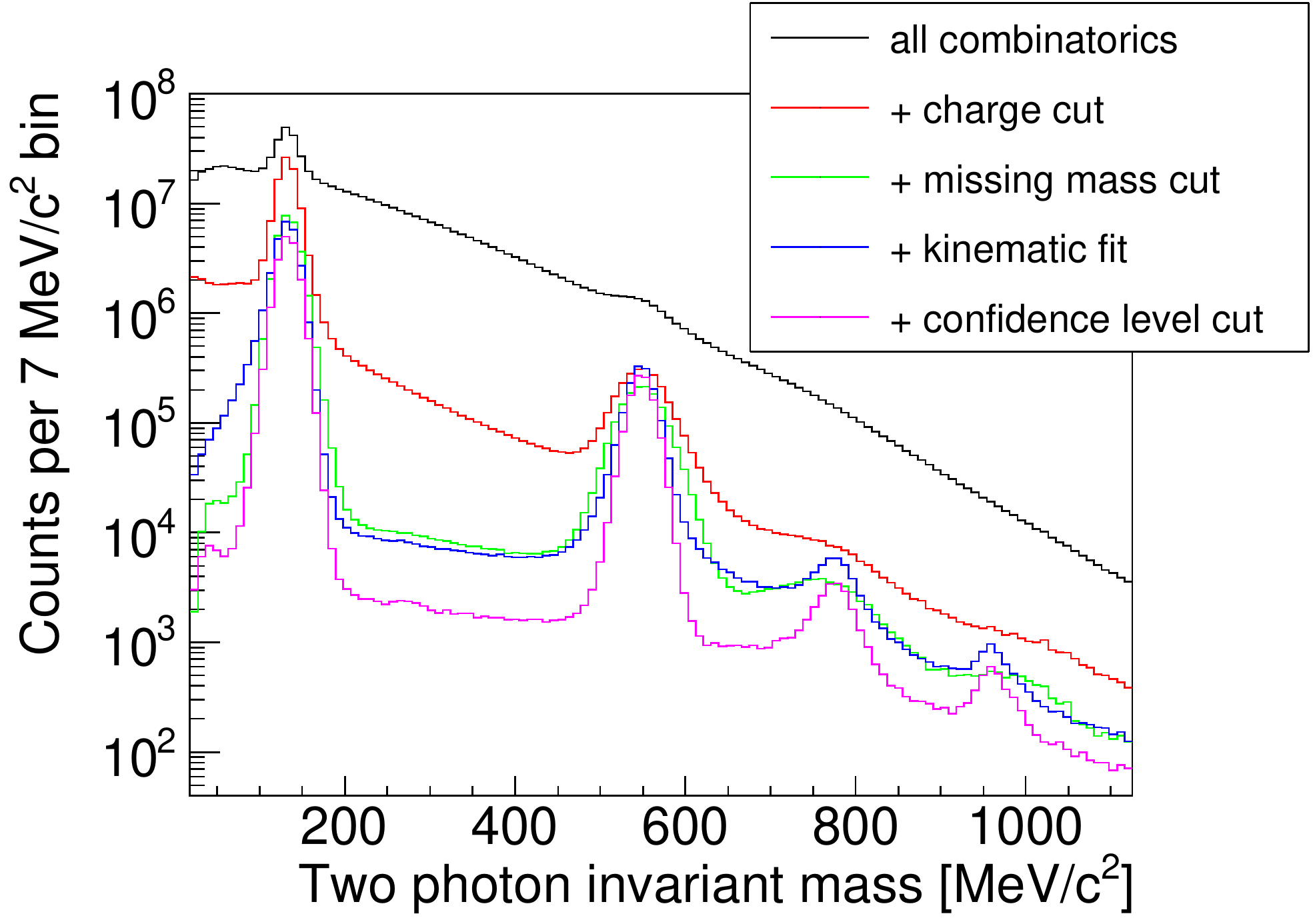}}\vspace{-0.1cm}
  \caption{Invariant mass of two photons measured by the BGO Rugby Ball  in coincidence with a recoil proton.  The different selection criteria listed in the legend accumulatively required (from top to bottom, see the text for details).}
 \label{fig:TwoGamma}
\end{center}
\end{figure}

Mesons decaying into two photons are reconstructed via their invariant mass, $m_{\gamma\gamma}$.   An example is shown in fig.~\ref{fig:TwoGamma}.  Different selection criteria were used to enhance the signal to background.  The black line includes all combinations of two clusters, where the missing momentum (the proton candidate) is in the same position as a detected charged particle.  The red line additionally requires that the clusters were identified as neutral particles via no coincidence with the Scintillator Barrel.  The green line includes these criteria, plus an additional selection of the missing momentum mass being consistent with that of a proton and only two neutral clusters in total.
The blue line applies a kinematic fit to this selection, requiring full three momentum conservation for the detected two photons and proton.  Finally, the magenta line applies a 0.9 confidence level cut upon the kinematic fit results.

The $\pi^0$ and the $\eta$ peaks are clearly visible with an estimated background of $\sim 2\cdot 10^{-4}$ and $\sim 5\cdot 10^{-3}$ respectively. At higher energy (around 800\,MeV/c$^2$), the signal corresponding at the $\pi^0\gamma$ decay of the $\omega$ meson where either the electromagnetic shower of two photons overlap, or a low energy photon is missed, is clearly visible as well. The $\eta^{\prime} \rightarrow 2\gamma$ decay is also visible despite the low branching ratio.  Gaussian function fits to the $\pi^0$ and $\eta$ invariant mass distributions yield sigma values of approximately 12 and 15\,MeV/c$^2$ respectively.  This resolution is a convolution of the position and energy resolution of the decay photon momenta.

\subsubsection{Charge particle identification in the BGO Rugby Ball}
\label{subsubsec:chargedbgo}

\begin{figure}[htbp]
	\begin{center}
		\vspace{-0cm}
		\resizebox{0.45\textwidth}{!}{\includegraphics{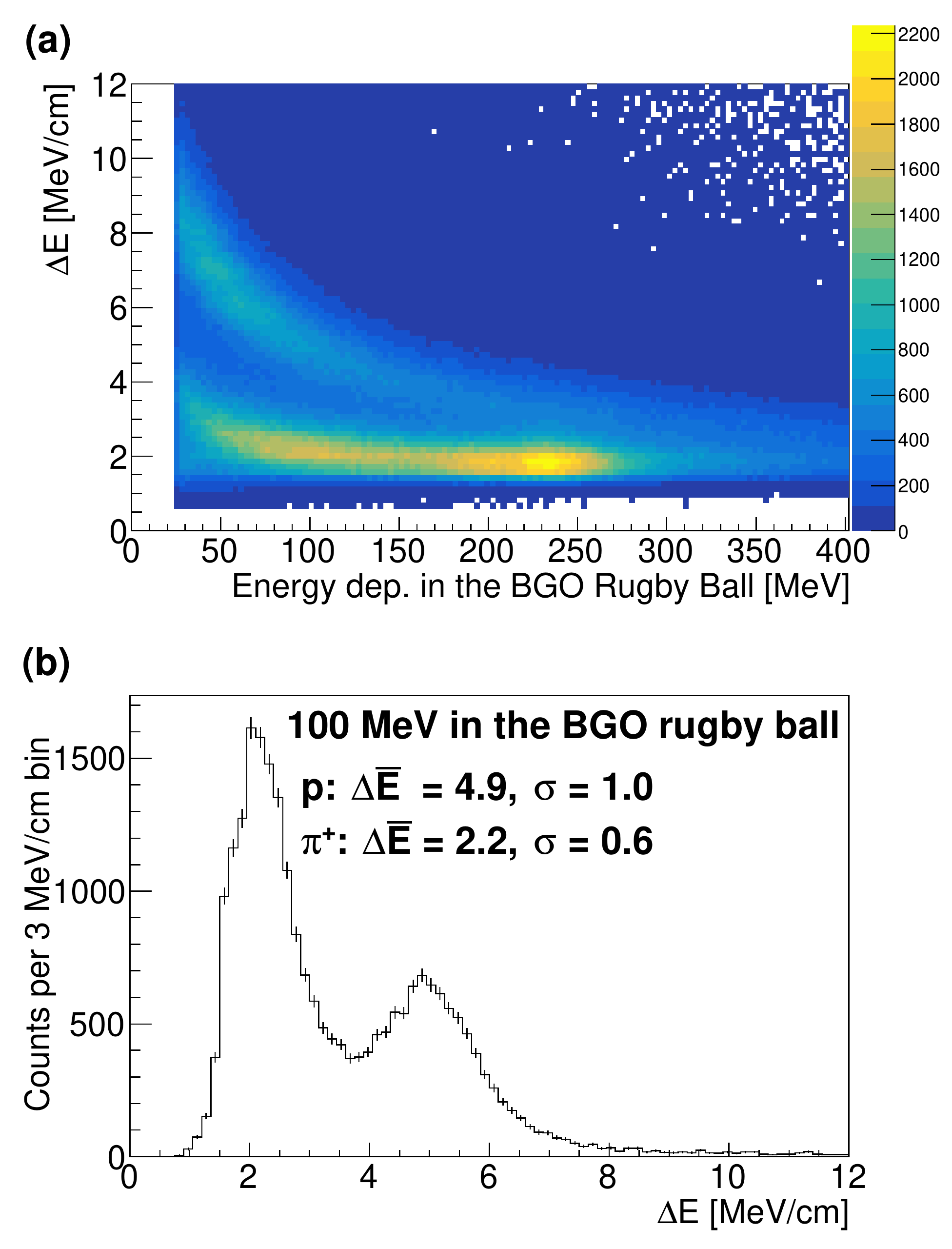}}\vspace{-0.1cm}
		\caption{(a) Charged particle identification by comparing the energy deposition in the Plastic Scintillator ($\Delta E$) to the energy deposition in the BGO Rugby Ball.  (b) $\Delta E$ for 100\,MeV energy deposition in the BGO Rugby Ball.  Approximate Gaussian mean and sigma values for protons and $\pi^+$ labelled inset in units of MeV/cm.}
		\label{fig:DeltaE_E}
	\end{center}
\end{figure}

Shown in fig.~\ref{fig:DeltaE_E}, comparison of the partial energy deposition in the Scintillator Barrel ($\Delta E$) compared to the energy deposited in the BGO Rugby Ball yields characteristic loci to distinguish between protons and $\pi^+$.  A small energy correction is applied to the measured energy in the BGO Rugby Ball to account for energy loss as charged particles traverse the target material and holding structure of the BGO crystals.  This is particle dependent and determined using simulated data.

%
%
\subsubsection{$K^+$ detection in the BGO Rugby Ball}
\label{subsubsec:Kplus-BGO}

A technique for the clean identification of $K^+$ in segmented calorimeters which was developed with the Crystal Ball calorimeter~\cite{Jude:2013jzs}
is employed with the BGO  Rugby Ball.  The method uses the fast time resolution per crystal to identify the time delayed weak decay of the $K^+$, 
vastly increasing the acceptance for strangeness photoproduction channels and vector mesons with hidden strangeness.
$K^+$ stop in the BGO crystals with momentum below 790\,MeV/c, and decay via two main decay modes: $K^+\rightarrow \mu^+\nu_\mu$ and $K^+\rightarrow \pi^+\pi^0$ (muonic and pionic decays respectively) with a lifetime of approximately 12\,ns.
From spatial and timing coincidences, a cluster of adjacent crystals with energy depositions can be split into two: a first (incident) sub-cluster from stopping the $K^+$ and a second sub-cluster from the subsequent decay.
  
\begin{figure}[htbp]
   \begin{center}
   		 \resizebox{0.45\textwidth}{!}{\includegraphics{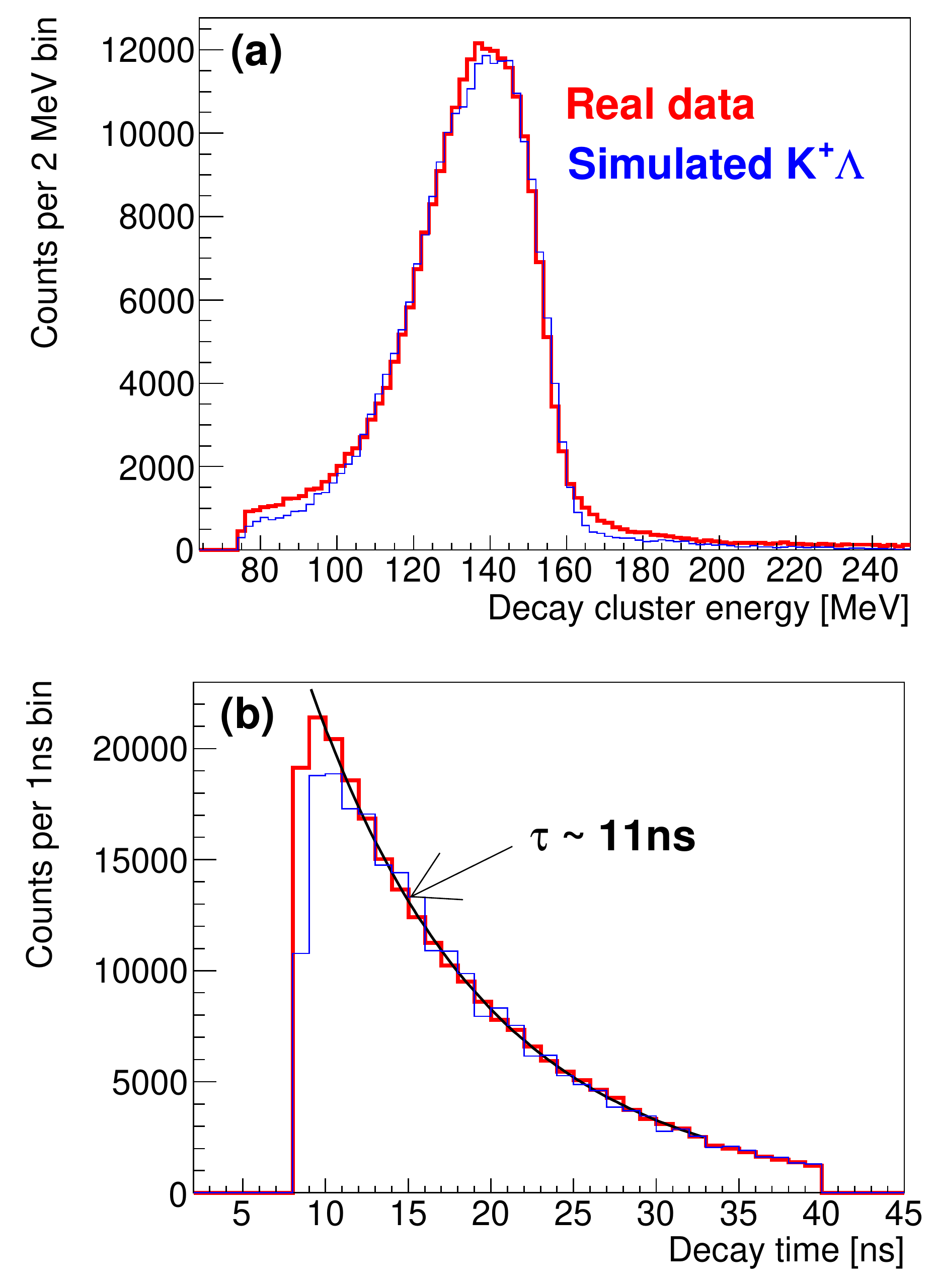}}
   
   \vspace{-0.1cm}
   \caption{Identification of $K^+$ in the BGO Rugby Ball via the time delayed, weak decay for real and simulated data (thick red and thin blue lines respectively).  (a) Energy deposition from $K^+$ decay, with a peak at 153\,MeV corresponding to the energy released in the decay $K^+\rightarrow \mu^+\nu_\mu$.  (b) Time between stopping $K^+$ and the subsequent decay.  A fitted exponential function gives a lifetime close to the expected $K^+$ lifetime.}
\label{fig:ESideTSide}
\end{center}
\end{figure}

For a $K^+$ decaying muonically at rest, an energy deposition from the $\mu^+$ of 153\,MeV is expected from the $\mu^+$ passing through adjacent crystals.  There is excellent agreement between this for both real and simulated data shown in fig.~\ref{fig:ESideTSide}(a).  Timing cuts and the relatively short triggered time range ensure no additional energy is included from the $\mu^+$ decay.
Figure~\ref{fig:ESideTSide}(b) shows the time difference between incident and decay sub-clusters.  An exponential fit yields a lifetime of approximately 11\,ns, with close agreement between real and simulated data.

The efficiency of this $K^+$ detection technique was determined as approximately 5 to 6\,\% using simulated data, which is consistent with the description in \cite{Jude:2013jzs}.

\begin{figure}[htbp]
	\begin{center}
		\resizebox{0.45\textwidth}{!}{\includegraphics{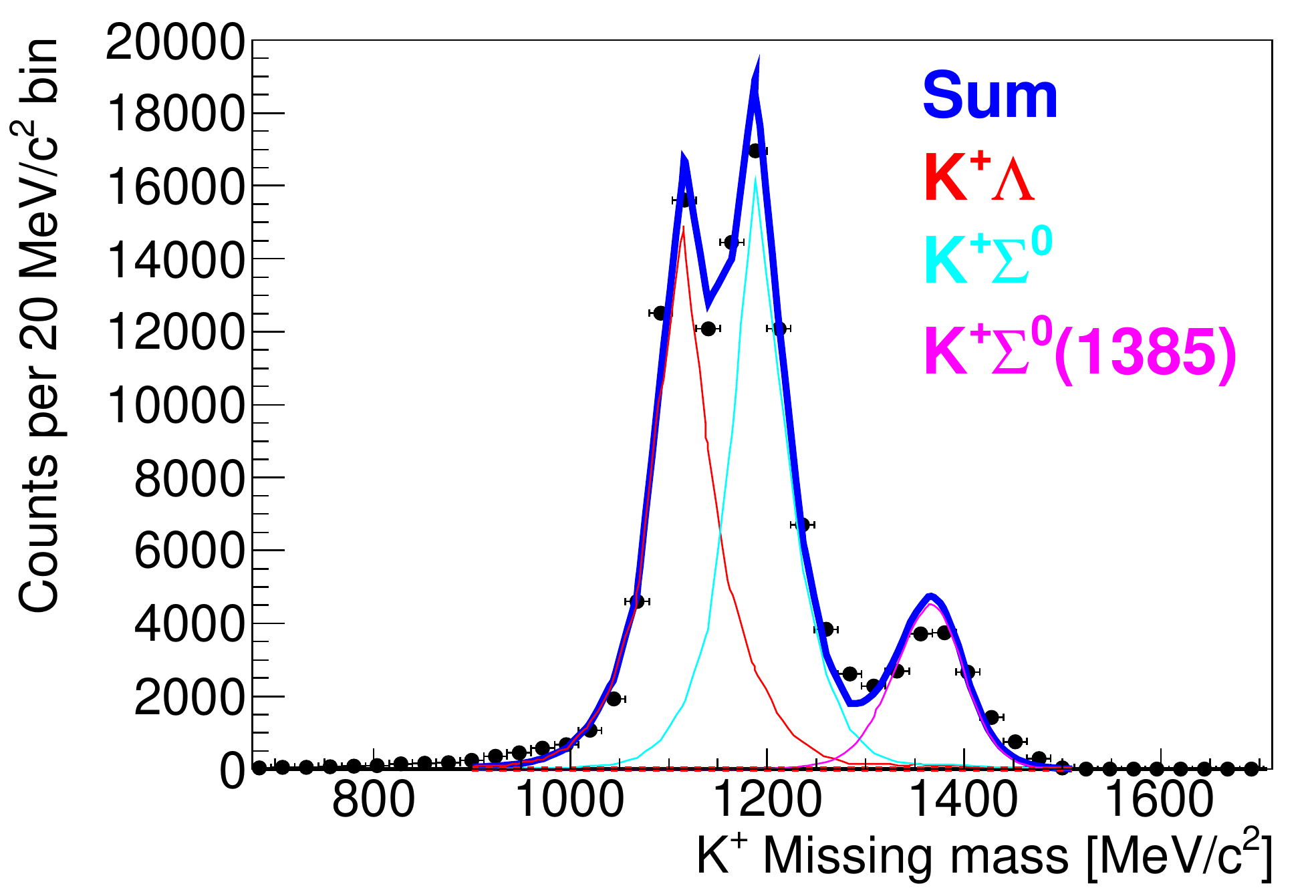}}
		
		\vspace{-0.1cm}
		\caption{Missing mass recoiling from $K^+$ identified via the time delayed weak decay in the BGO Rugby Ball for photon beam energies between 950 and 1800\,MeV.  An approximate fit using simulated data was made (thick blue line).  The contributing three channels are labelled inset in hyperon mass order.}
		\label{fig:centralmissingmass}
	\end{center}
\end{figure}

The $K^+$ four-momentum is determined from the incident sub-cluster energy and position, assuming the $K^+$ originated from the target centre.  The mass recoiling from $K^+$ candidates identified via this technique is shown in fig.~\ref{fig:centralmissingmass}, with good agreement between real and simulated data.

\subsubsection{Central tracking with the MWPC}
\label{subsubsec:TrackingMWPC}

The MWPC is used to track charged particles from the target fiducial volume to the BGO Rugby Ball.
This enables reaction vertex reconstruction, improving momentum resolution.  
The MWPC can also be used to identify particle decay vertices outside of the target volume, for example $K^0_S\rightarrow \pi^+\pi^-$, where $c\tau \approx 2.7$\,cm.

\begin{figure}[htbp]
	\begin{center}
		\resizebox{0.45\textwidth}{!}{\includegraphics{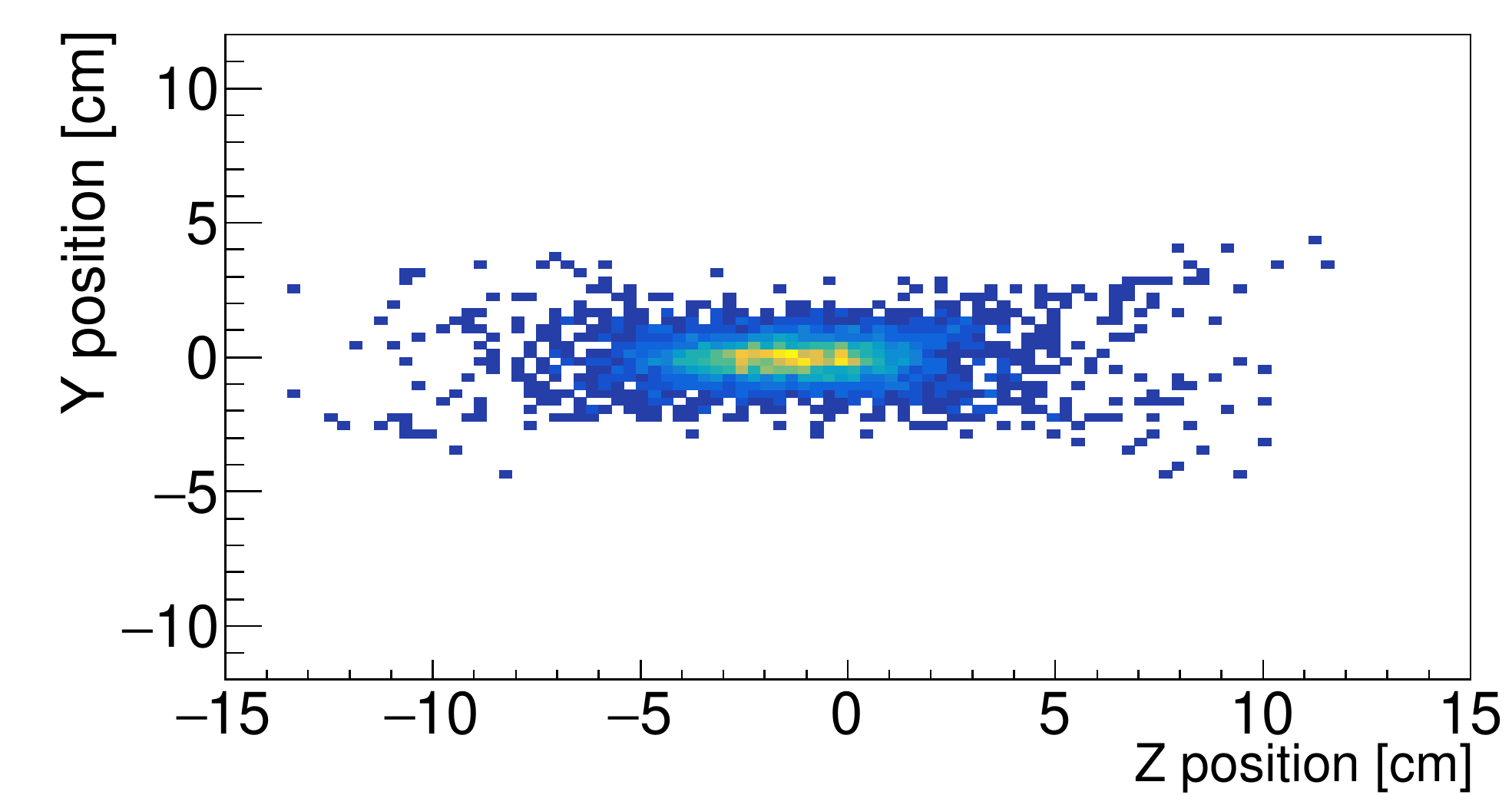}}
		\vspace{-0.1cm}
		\caption{Reconstruction of the target fiducial volume by track identification using the MWPC.
		The Z and Y axis correspond to the beam direction and the orthogonal vertical direction, respectively.}
		\label{fig:mwpcanalysis}
	\end{center}
\end{figure}

Figure~\ref{fig:mwpcanalysis} shows the fiducial target volume (liquid hydrogen, length 6\,cm) described by charged particle track identification using the MWPC.  To ensure a hadronic event occurred within the target, a $\pi^0$ was identified in the BGO Rugby Ball.  The centre of the target at approximately $-1.5$\,cm is consistent with the known target position for this data set. 



%

%
\subsection{The forward spectrometer}
\label{subsec:PerformanceForwardSpectrometer}
%
%
\subsubsection{Tracking and momentum reconstruction}
\label{subsubsec:Tracking-MomentumReconstruction}

Charged particle momentum in the forward spectrometer is determined by the deflection of the trajectory through the open dipole magnetic field.
This trajectory is determined by the position reconstruction at the MOMO and SciFi detectors before the magnetic field, and the Drift Chambers after.  
Particle identification is achieved by the combination of the momentum and $\beta$ of the particle, determined from the ToF walls.
  Described below is the default algorithm and selection criteria, which were optimised using both real and simulated data.   

Tracks of particles originating from the target are identified using all combinations of MOMO and SciFi clusters and retained if they originated from the target.  As described in sec.~\ref{subsubsec:MOMO-SciFi}, MOMO can be neglected, and tracks formed from SciFi and the target can be used to increase the track finding efficiency.

Tracks between clusters in the three ToF walls at the most downstream end of the forward spectrometer are created.  These are combinations of one, two or three clusters from different walls which are spatially in coincidence.

Combinations of Front Tracks and ToF tracks are made if the extrapolated Front Track extends to the ToF Track in the vertical plane, where there is no Lorentz force, and a straight trajectory can be assumed.  
 
\begin{figure}[htbp]
 \begin{center}
\vspace{-0cm}
       \resizebox{0.5\textwidth}{!}{\includegraphics{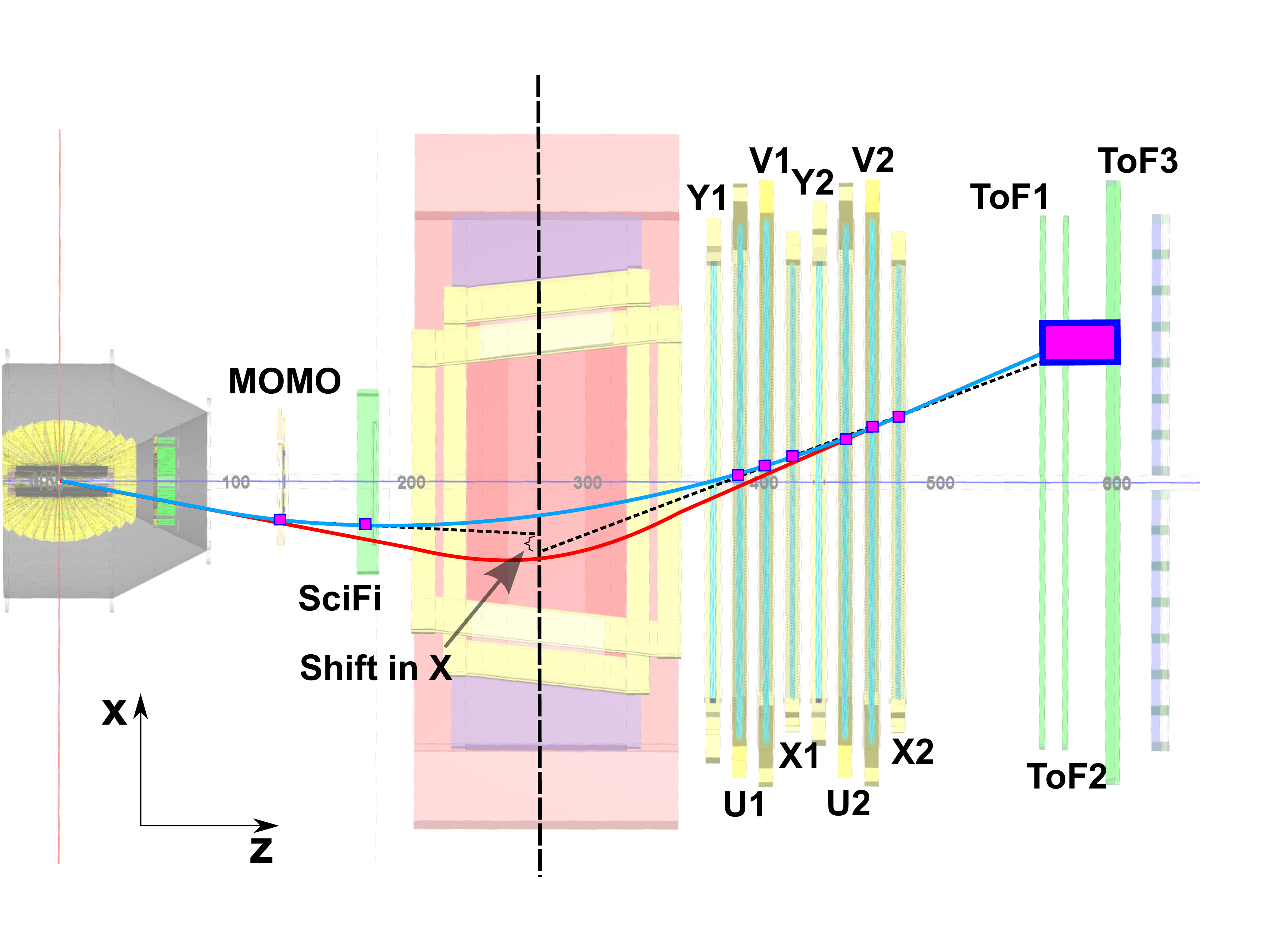}}\vspace{-0.1cm}\\
\caption{Charged particle trajectory in the forward spectrometer.  Purple boxes indicate cluster positions in the labelled detectors.
The red line is the first trajectory estimation assuming a constant ``box'' field within the open dipole magnet.
The blue line is the corrected trajectory when accounting for field non-uniformity and fringe fields (the difference is exaggerated in the figure so that it is visible).
The dotted black lines are extrapolations of the upstream and downstream parts of the trajectory outside of the magnetic field.  These lines do not quite meet at the centre of the magnet due to the fringe fields.  The gap between these lines is labelled ``Shift in X".  Colour image online.}
   \label{fig:fortrack1}
\end{center}
\end{figure}

Figure~\ref{fig:fortrack1} depicts the full track finding procedure in the X-Z plane, where the deflection of the particle trajectory in the magnetic field occurs.
The direction and extent of this deflection is used to determine the particle charge and a first estimation of the particle momentum, assuming a uniform magnetic field within the magnet, and zero field outside (referred here as a ``box'' field).  This is shown by the red line in fig.~\ref{fig:fortrack1}.

The track finding procedure must account for the non-uniform magnetic field, with fringe fields existing beyond the magnet yoke, and so a correction is made to the trajectory as a function of momentum.
This was calculated using simulated data and an accurate three dimensional description of the magnetic field (sect.~\ref{subsubsec:Magnet}).  An exaggerated example of a corrected trajectory is shown by the blue line in fig.~\ref{fig:fortrack1}.  Drift chamber clusters are subsequently included in the track if this trajectory passes close to them.

 A linear fit to the track direction is made in the Z-Y plane (where there is no deflection due to the magnetic field) to MOMO, SciFi, all Y-orientated drift chambers and ToF clusters, using both the position and error information.  
The fit result is then used to determine the X-coordinate of the clusters of the U- and V-orientated drift chambers included in the track.
A second linear fit to the track direction is applied in the Z-X plane downstream from the magnetic field to all drift chamber and ToF clusters. The two fits in these orthogonal planes are combined to describe the particle trajectory after deflection in the magnetic field.

Due to the comparatively low spatial resolution of the ToF clusters compared to the drift chambers, occasionally single ToF tracks are used in multiple combined tracks.  
In these occasions, the track with the most drift chambers included is retained.  
A final selection criterion requires there to be at least three drift chamber clusters included in every track.

An accurate determination of the magnitude of the particle momentum and trajectory is now made for all selected tracks.  The initial direction of the particle trajectory is fixed from the  front track direction (MOMO and SciFi clusters).
Starting from the reaction vertex, a charged particle is assumed to have the initial momentum estimation.  This particle is ``stepped'' through the forward spectrometer in small intervals.
During each interval, the Lorentz force experienced by the particle and the expected energy lost due to the material traversed is determined, and the particle momentum is altered accordingly.
The interval lengths are calculated dynamically, for example,  reducing in size through the magnetic field, and increasing in size in regions of no field.
This ensures accurate momentum reconstruction without excessive computational time.
This process continues until the particle reaches the ToF walls, 
where a minimisation technique compares the calculated trajectory to the error weighted cluster positions in the track.  
This is repeated with iterative changes in the initial particle momentum of 0.25\,\% in order to improve the agreement between the calculated and fitted trajectories. 
As the energy loss is different for different particle species, the process is repeated for charged kaons, charged pions, and electrons/positrons using the proton momentum as a starting value.

$\beta$ is determined for each ToF wall which was included in the track, accounting for the length of the trajectory.  The mean average $\beta$, weighted by the time resolution of each ToF wall is then used.

\subsubsection{Forward Spectrometer performance}
\label{subsubsec:ForwardParticleID}

\begin{figure*}[htp]
		 \resizebox{\textwidth}{!}{\includegraphics{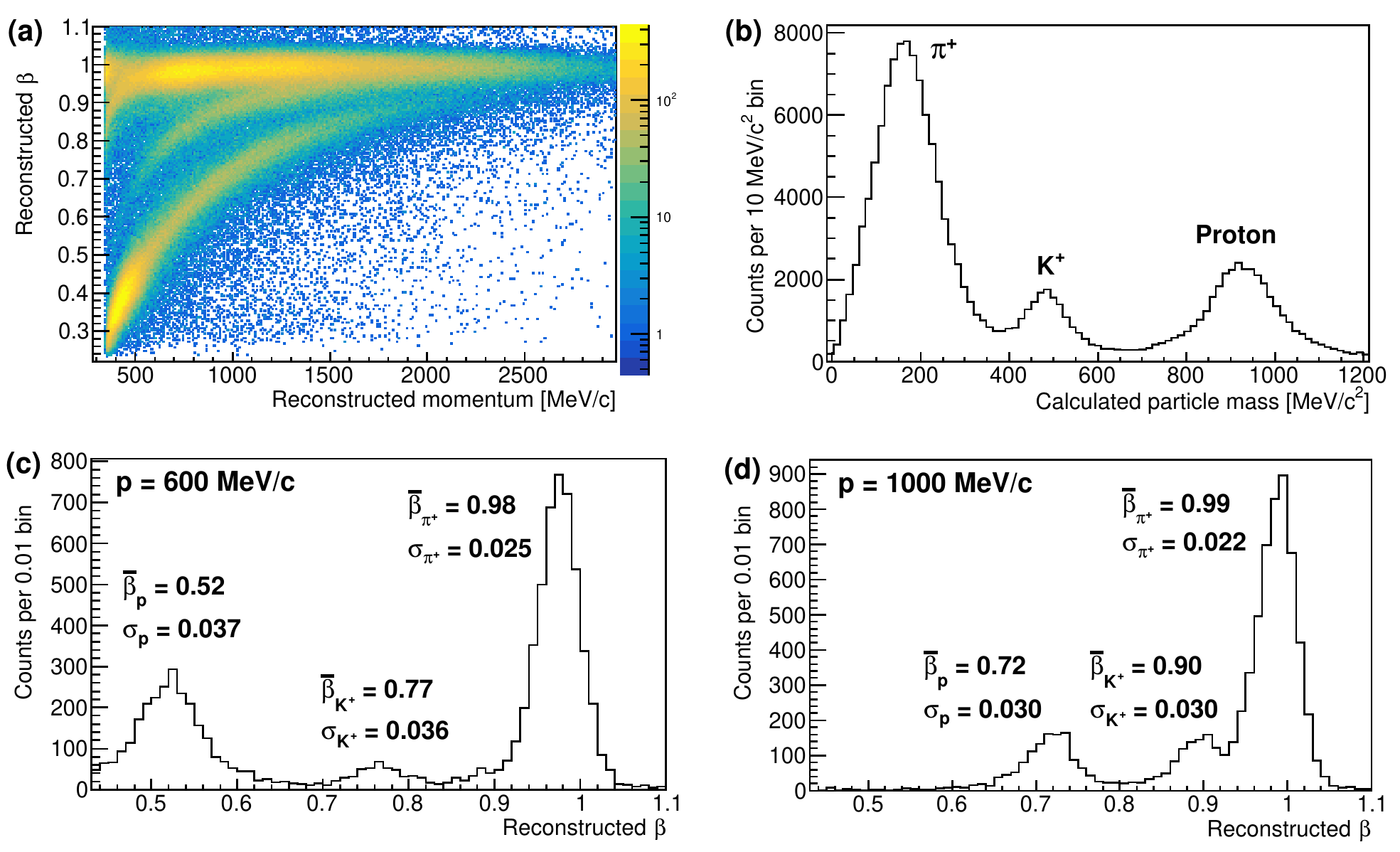}\vspace{-0.1cm}}
		\caption{Charged particle identification in the forward spectrometer (only positively charged particles shown).  To accentuate the contribution from $K^+$ and to remove background from non-hadronic reactions, a $\pi^0$ was required to be identified in the BGO Rugby Ball, a photon beam energy over 900\,MeV and a total energy deposition in the BGO Rugby Ball smaller than 250\,MeV.
			(a)  Particle $\beta$ versus momentum.  Characteristic loci corresponding to $\pi^+$, $K^+$ and protons are evident.
			(b) Calculated particle masses for particles with momentum between 600 and 1000\,GeV/c.
		(c, d) Particle $\beta$ for momentum 600\,MeV/c and 1000\,MeV/c.  Approximate Gaussian mean ($\bar\beta$) and sigma ($\sigma$) values for different particle types are labelled inset.}
		\label{fig:forwardparticleidentification}
\end{figure*}

Figure~\ref{fig:forwardparticleidentification}(a) shows particle $\beta$ plotted against the average momenta over the trajectory, accounting for energy loss for positively charged particles.  Characteristic loci of $\pi^+$, $K^+$ and protons are evident.  At $\beta$ close to one, there is background under the $\pi^+$ contribution from positrons originating from the beam.  The calculated particle masses of the particles are shown in fig.~\ref{fig:forwardparticleidentification}(b).
Figures~\ref{fig:forwardparticleidentification}(c) and (d) demonstrates the good separation of particle types for low and high momentum for hadron physics studies typical at BGOOD.  At a momentum of 600\,MeV/c, the proton$- K^{+}$ and $K^{+}-\pi^+$ signal separations are both approximately $7\sigma$.
At a momentum of 1000\,MeV/c, the separations are approximately $6\sigma$ and $3\sigma$ respectively.

%

\begin{figure}[htbp]
\vspace{-0cm}
   \resizebox{0.45\textwidth}{!}{\includegraphics{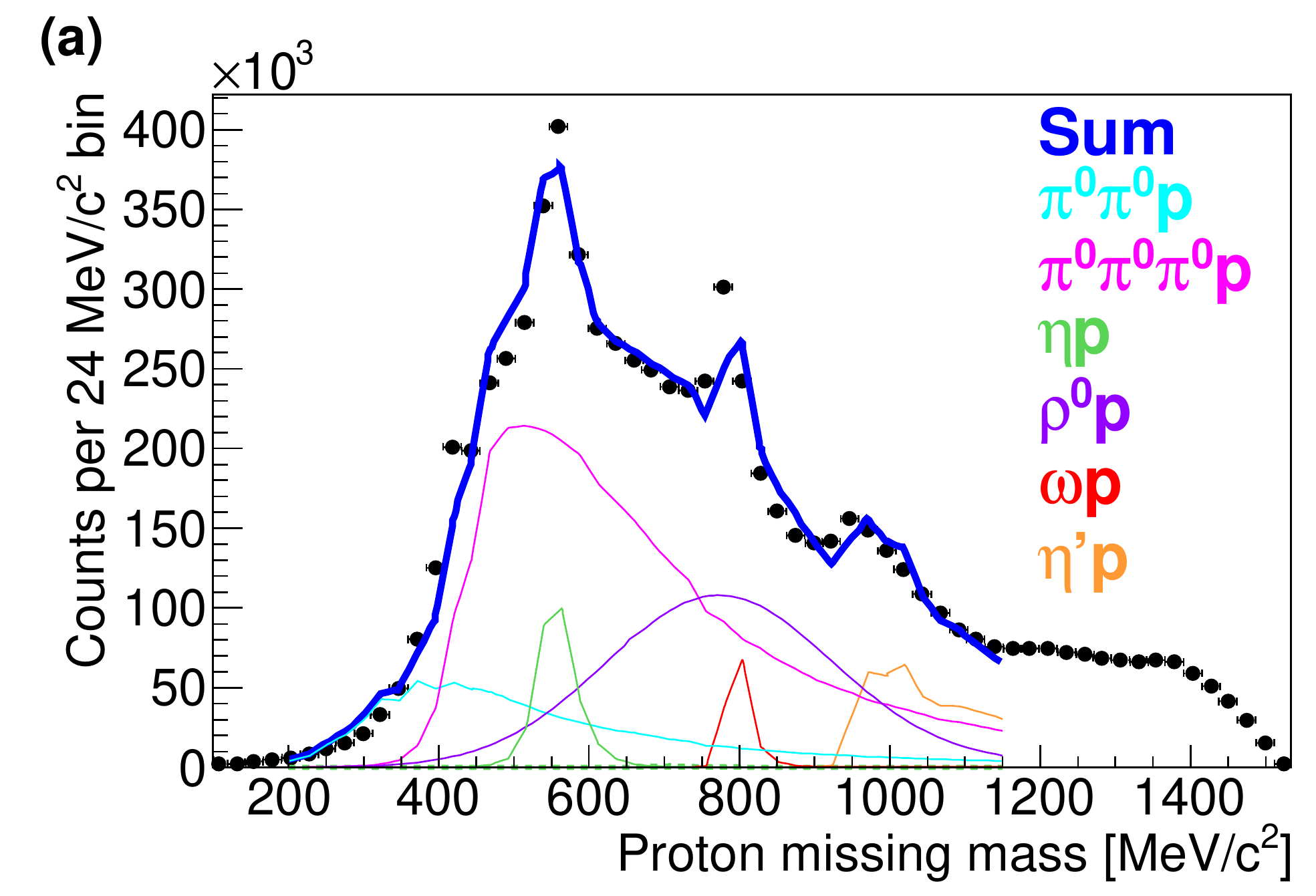}}\\
      \resizebox{0.45\textwidth}{!}{\includegraphics{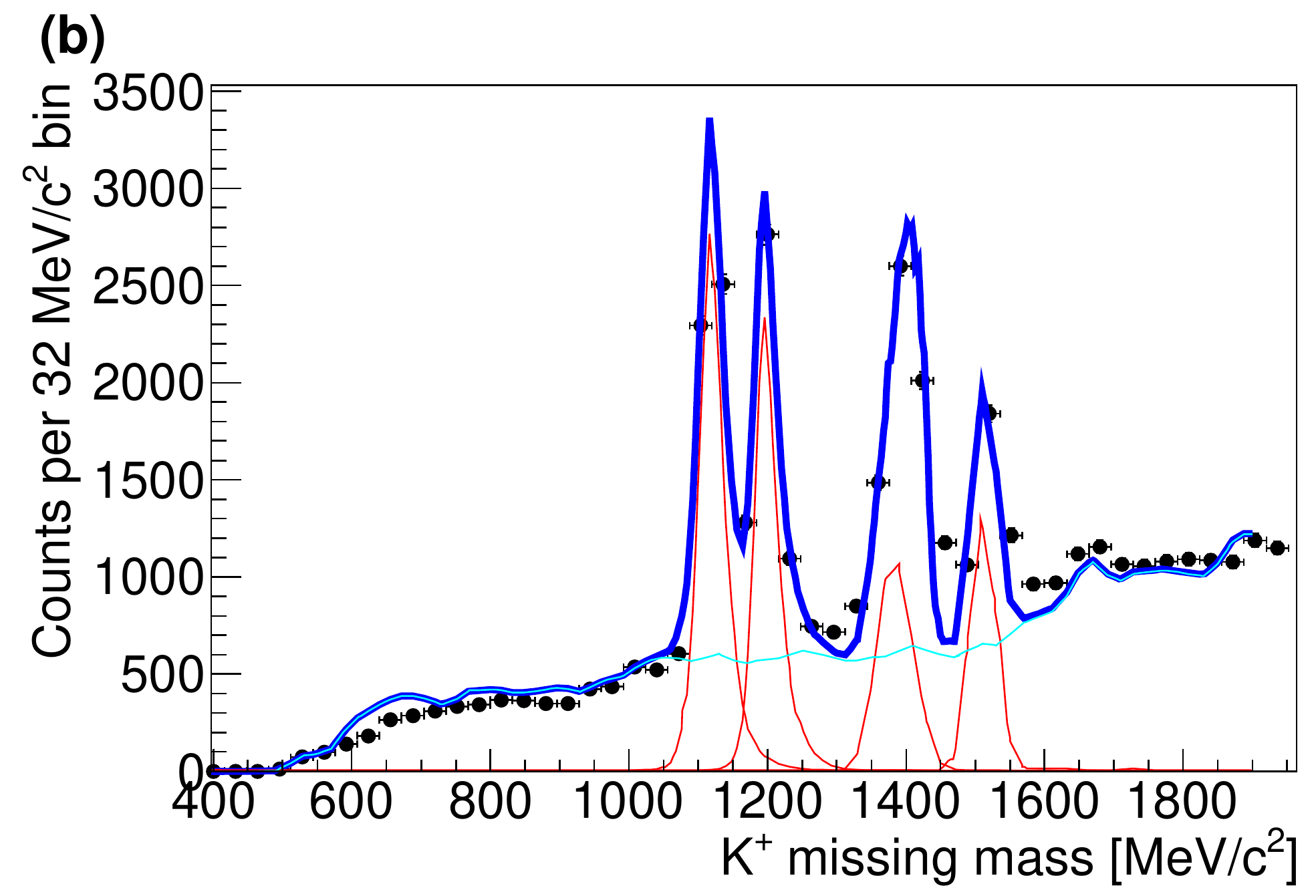}}

   \caption{Top: Missing mass spectrum from protons detected in the Forward Spectrometer.  An approximate fit using simulated data highlights the most prominent single and multiple meson final states.  The thick blue line is the total fit and the contributing channels are labelled in ascending thresholds, with corresponding spectra starting at ascending missing mass. 
Bottom: Missing mass spectrum $K^+$ detected in the Forward Spectrometer.  The spectrum was fitted with simulated photoproduction channels (in mass order): $K^+\Lambda$, $K^{+}\Sigma^0$, $K^+\Sigma^0(1385)/\Lambda(1405)$ (mass degenerate) and $K^+\Lambda(1520)$ (red lines).  The background is described by the cyan line and the summed total fit by the thick blue line.}
   \label{fig:forwpmissmass}
\end{figure}

The missing mass spectrum obtained by selecting protons in the forward spectrometer is shown
in fig.~\ref{fig:forwpmissmass}(a). The peaks corresponding to the different mesons photoproduced are clearly visible.
Figure~\ref{fig:forwpmissmass}(b) shows the $K^+$ missing mass signal for $K^+$ with momenta below 1\,GeV/c.  Peaks corresponding to $\Lambda$, $\Sigma^0$, $\Lambda$(1405)/$\Sigma^0$(1385) (almost mass degenerate), and $\Lambda$(1520) are immediately visible.

  Simulated data yielded a polar angle resolution (in
  $\sigma$) of 0.3$^\circ$, and using the maximum magnetic field strength
  in the open dipole, the momentum has a constant resolution (in
  $\sigma$) of 3\,\%.  Using both simulated and real commissioning data with different open dipole field strengths, it was found that the momentum resolution is approximately linear with the magnetic field strength.  The $\beta$ resolution is approximately 2.4\% close to $\beta = 1$.  This arises from a combination of the Photon Tagger and TOF wall time resolutions.

The efficiencies, as a function of particle position and $\beta$, of all detectors in the forward spectrometer were measured and implemented in the simulation.
This was achieved by identifying the final states of either $p\eta$ or $p\pi^0$, where the decay photons from the meson are identified in the BGO Rugby Ball, and the proton traversed the forward spectrometer. 
This provided events clean of background with a forward proton candidate.
Comparisons were made using simulated data to ensure the measured efficiency was not affected by geometry or particles stopping in upstream detector components.

\begin{figure}[htbp]
     \begin{center} 
   \resizebox{0.45\textwidth}{!}{\includegraphics{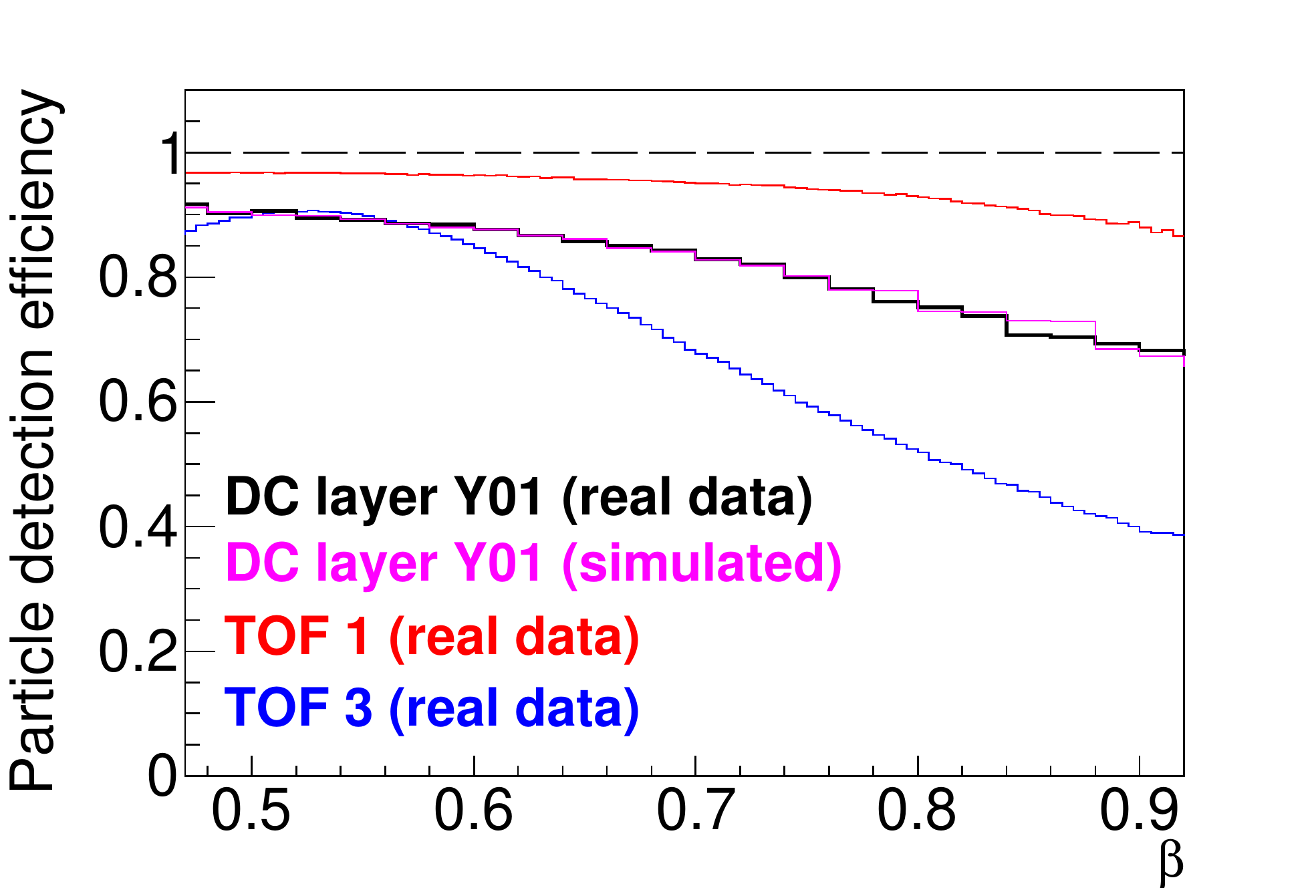}}\vspace{-0.1cm}
   \caption{Example efficiency calculations for the forward spectrometer as a function of particle $\beta$.  Drift chamber layer Y01 (one of 16) shown for real and simulated data (black and magenta lines respectively).  ToF wall 1 and 3 shown in red and blue respectively.}
   \label{fig:DCEffExample}
\end{center}
\end{figure}

The drift chamber efficiency was determined for each layer of each drift chamber.  
An example of the efficiency determination as a function of particle $\beta$ is shown for one layer in fig.~\ref{fig:DCEffExample}.  

ToF walls 1 and 2 exhibited efficiencies over 85\% for all $\beta$, whilst ToF wall 3 has a lower efficiency (shown in fig.~\ref{fig:DCEffExample}).
The efficiency of track identification is significantly higher, as only one of the three walls is required.

SciFi and MOMO efficiencies were determined as 97.5\% and 80\% respectively.

%
%
%

\section{Examples of results}
\label{sec:Examples_results}

This section presents some ``bench mark" results from data collected in a 21 day run during August 2017 with a 6\,cm liquid hydrogen target and a 3.2\,GeV electron beam energy. 

\subsection{Bench mark cross sections}
\label{BMCS}

\begin{figure}[htbp]
    \begin{center}
\vspace{-0cm}
     \resizebox{0.48\textwidth}{!}{\includegraphics{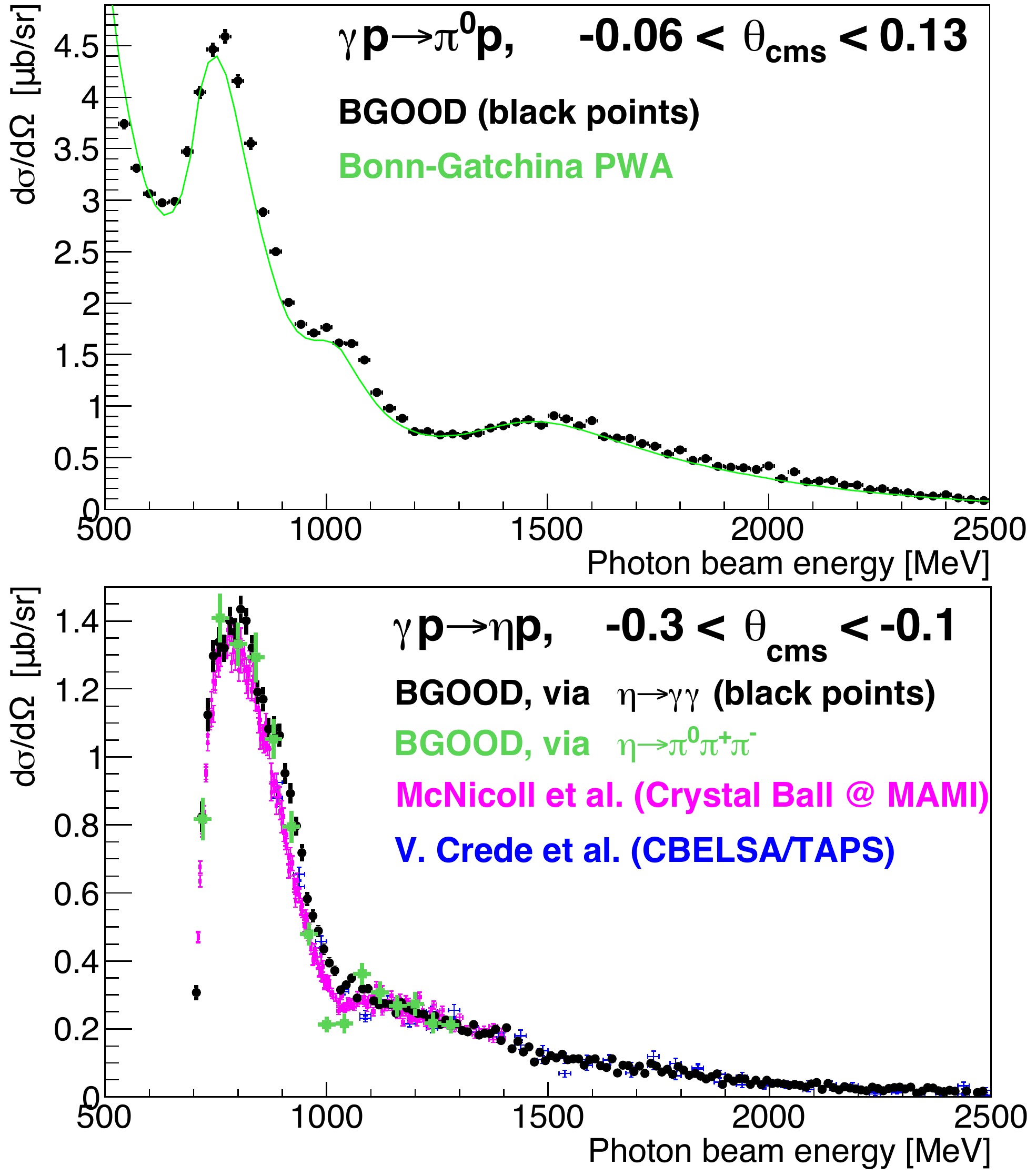}}\vspace{-0.1cm}
   \caption{Differential cross section for $\gamma p \rightarrow \pi^0 p$ (centre of mass $\pi^0$ polar angle interval, $\theta_\mathrm{cms}$ inset).  
The green line is a solution from the Bonn-Gatchina PWA~\cite{bnga}.  Bottom panel: differential cross section for $\gamma p \rightarrow \eta p$ ($\eta$ centre of mass polar angle interval, $\theta_\mathrm{cms}$ inset.).
The reaction has been reconstructed via both $\eta\rightarrow\gamma\gamma$ and $\eta\rightarrow\pi^0\pi^+\pi^-$ decay modes and compared to previous Crystal Ball data~\cite{McNicoll:2010qk} (magenta data points) and CBELSA/TAPS data~\cite{Crede:2009zzb} (blue data points).}
   \label{fig:csexamples}
\end{center}
\end{figure}

Figure~\ref{fig:csexamples}  shows examples of cross sections determined using the central detector region of BGOOD.  The differential cross section for $\gamma p \rightarrow \pi^0 p$ and $\eta p$, where the meson two photon decay is identified in the BGO are both measured with good agreement to existing data.  

The unique setup of BGOOD allows clean identification of the mixed charge decay, $\eta\rightarrow \pi^0\pi^-\pi^+$,
where charged particles are identified in the BGO Rugby Ball and SciRi.  SciRi only measures the direction of charged particles, which in combination with particle tracks measured using other detectors, is used to suppress background and is an input to a kinematic fit.  The kinematic fit (described in ref.~\cite{georgthesis}) was used to enhance the $\eta$ invariant mass signal, and the resulting cross section gives good agreement to previous datasets.

\begin{figure}[htbp]
	\begin{center}
		\vspace{-0cm}
	\resizebox{0.45\textwidth}{!}{\includegraphics{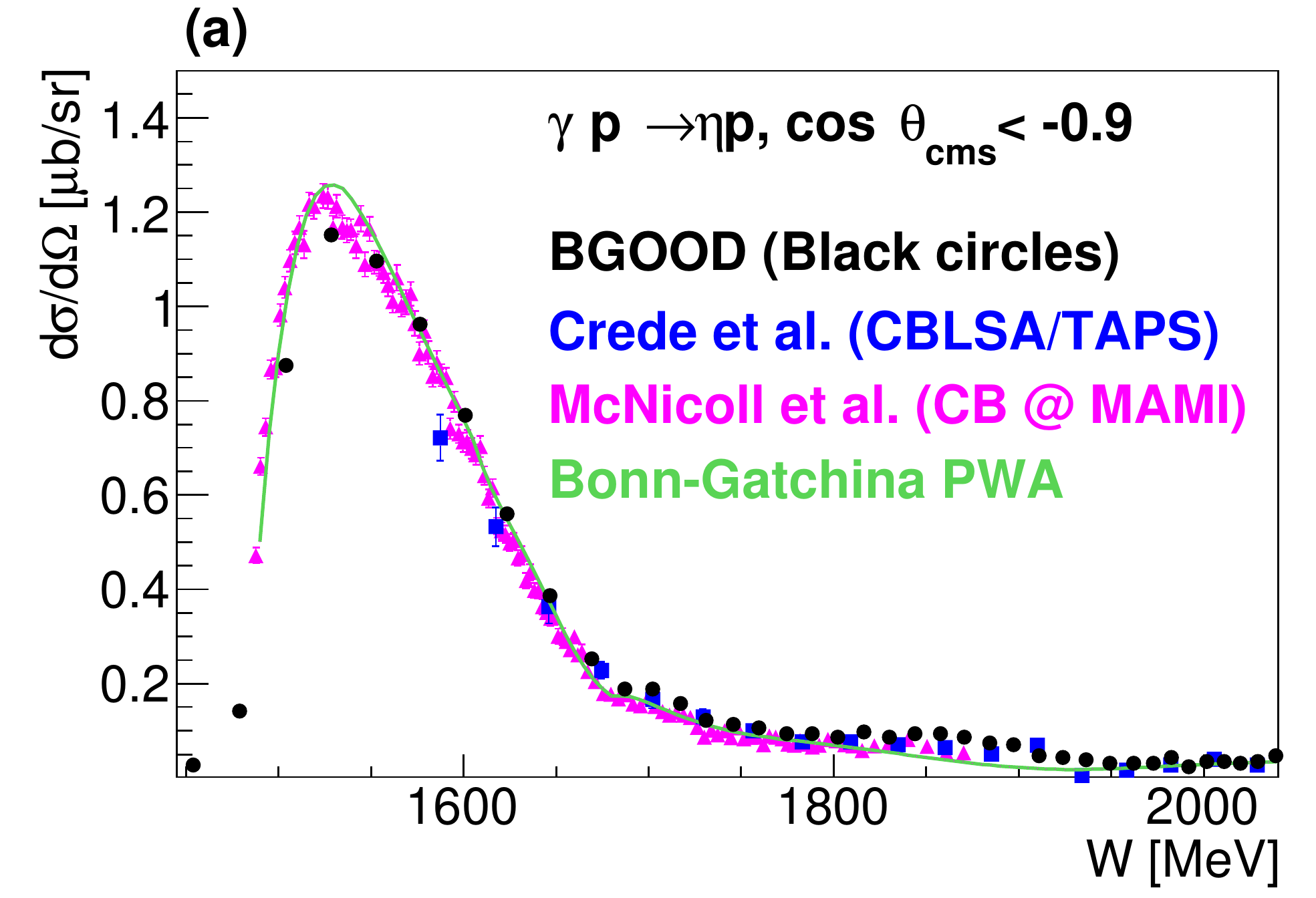}}\vspace{-0.1cm}
		\resizebox{0.45\textwidth}{!}{\includegraphics{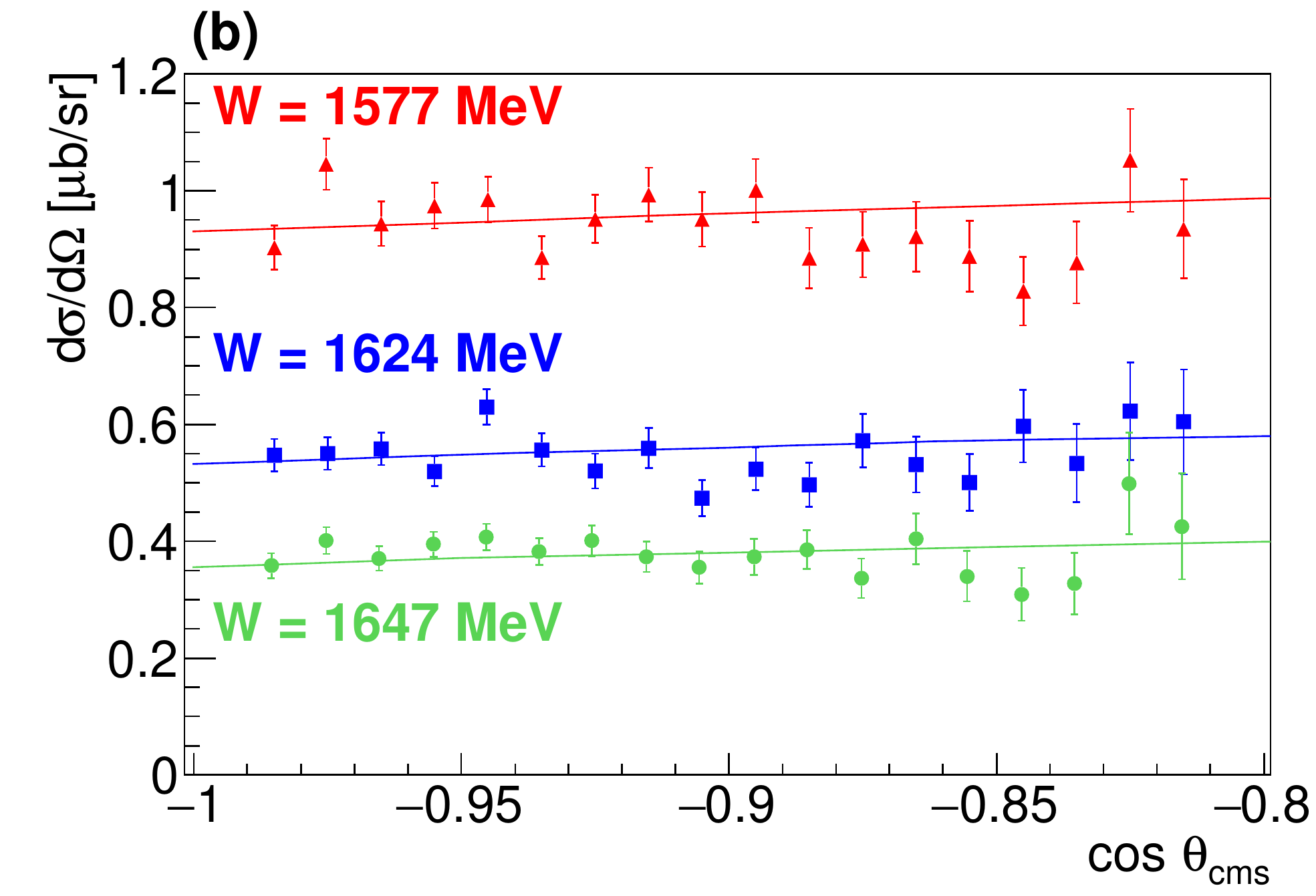}}\vspace{-0.1cm}
		\caption{$\gamma p \rightarrow \eta p$ differential cross section for $\eta$ polar angle $\cos\theta_\mathrm{cms} < -0.9$, when the proton is identified in the forward spectrometer.  (a) Plotted against the centre of mass energy, $W$.  Black points are BGOOD (statistical errors are smaller than the data points), previous data from McNicoll \textit{et al.}~\cite{McNicoll:2010qk} and Crede \textit{et al.}~\cite{Crede:2009zzb} in blue and red respectively.  The Bonn-Gatchina partial wave analysis solution~\cite{bnga} is shown in green.  (b) Angular distributions for three centre of mass energies, $W = 1577, 1624$ and 1647\,MeV shown in red triangles, blue squares and green circles respectively.  The coloured lines are the corresponding Bonn-Gatchina partial wave analysis solutions.}
		\label{fig:forwardetacs}
	\end{center}
\end{figure}

Figure~\ref{fig:forwardetacs}(a) shows the differential cross section for $\gamma p \rightarrow \eta p$ when the proton is identified in the forward spectrometer.  To provide a clean signal, the decay $\eta \rightarrow \gamma\gamma$ was additionally identified in the BGO Rugby Ball.
The forward track reconstruction and understanding of detector efficiencies yields good agreement with previous datasets. 
Figure~\ref{fig:forwardetacs}(b) shows the angular distribution covered by the forward spectrometer, for example centre of mass energies over the S$_{11}$(1535) resonance.  The expected, almost flat distribution is reproduced, with good agreement to the Bonn-Gatchina partial wave solution~\cite{bnga}, demonstrating the well understood acceptance over the forward spectrometer range.

\subsection{Bench mark beam asymmetries}
\label{BMBA}

\begin{figure}[htbp]
    \begin{center}
\vspace{-0cm}
     	\resizebox{0.5\textwidth}{!}{\includegraphics{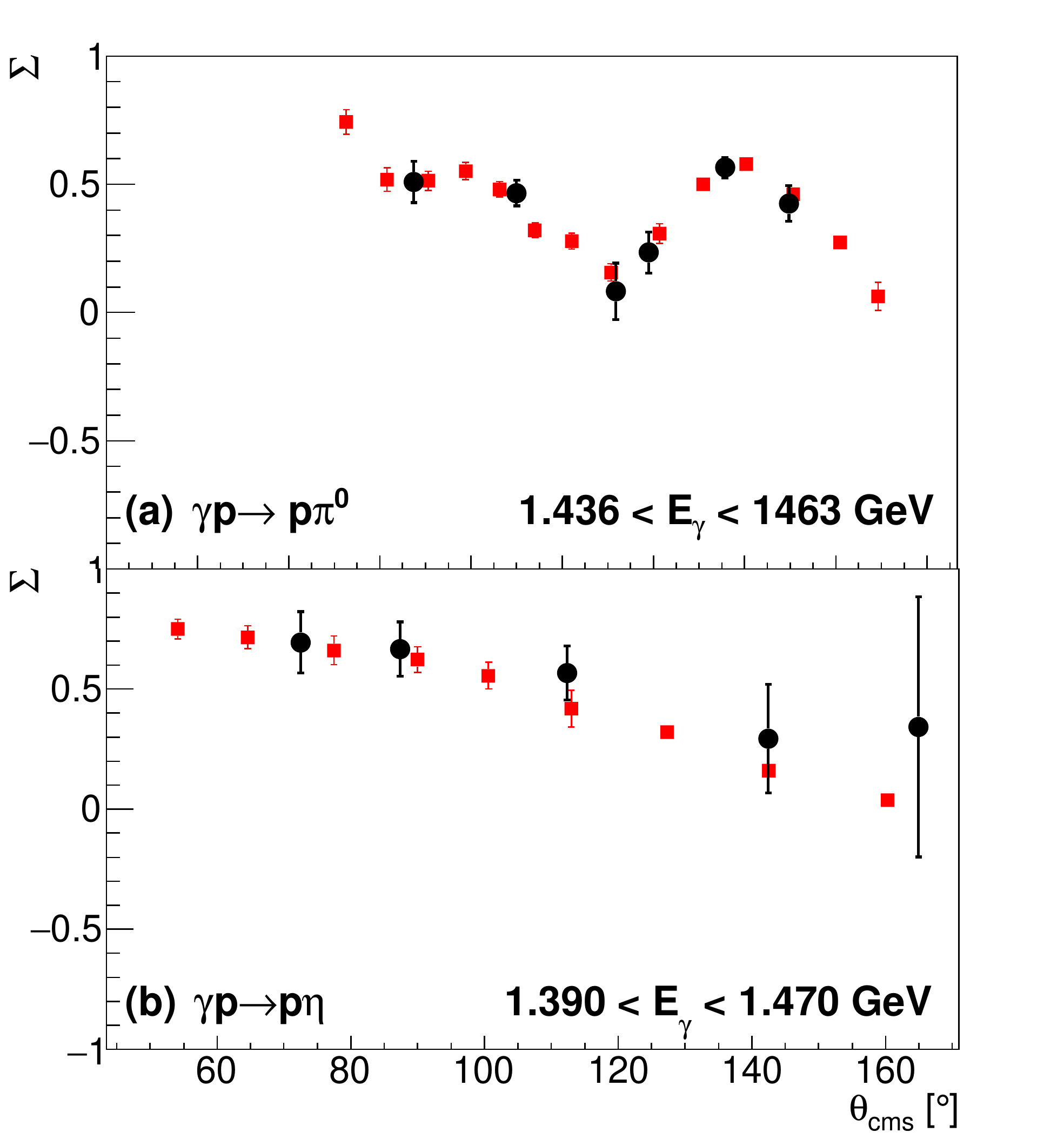}}\vspace{-0.1cm}
   \caption{Example of ``bench mark'' beam asymmetry results for photon beam energy, $E_\gamma$, intervals labelled inset: (a) $\pi^0$ photoproduction: BGOOD results (black circles) compared to GrAAL results (red squares)\cite{Bartalini:2005wx}. (b) $\eta$ photoproduction: BGOOD (black dots) compared to GrAAL (red squares)\cite{Bartalini:2007fg}.}
   \label{fig:baexamples}
\end{center}
\end{figure}

Shown in fig.~\ref{fig:baexamples}, the beam asymmetries, $\Sigma$, of the reactions \newline $\gamma + p \rightarrow \pi^0 (\eta) + p \rightarrow 2\gamma + p$  were measured at a photon beam energy of approximately 1.4\,GeV. The recoil proton was detected either in the Forward Spectrometer, in  SciRi, or in the BGO Rugby Ball, and three different analyses were performed according to the angular region of the detected recoil proton. The three analyses were found to be consistent in their angular overlap regions, both for the $\pi^0$ and for the $\eta$ channels.

The agreement between BGOOD and previous measurements is satisfactory within the error bars due to the limited statistics of the present analysis. This comparison verifies the technique of extraction of the degree of polarisation using the COBRIS software package, and confirms the good quality of the collected data.

\section{Summary}
\label{sec:Summary}

The BGOOD beam and experimental set-up have been described. The beam extracted from ELSA routinely produces a flux of $N_{\gamma}\simeq 2.5\cdot10^7$\,s$^{-1}$ tagged photons between 320 and 2880\,MeV. The photon beam can be polarised via the coherent bremsstrahlung process by using a diamond radiator.

The performances of the detector are as expected and the apparatus is ideal to perform measurements of final states with both charged and neutral particles. The OD spectrometer extends to small angles down to 1.5$^\circ$, providing access to  small momentum transfer regions.

The ``bench mark" measurements presented demonstrate the reliability of this versatile set-up.

\subsection*{Acknowledgements}
It is a pleasure to thank the ELSA staff for a reliable and
stable operation of the accelerator and the technical staff
of the contributing institutions for essential help in the
realisation and maintenance of the apparatus.

We thank DESY for the loan of the OD magnet, the IPN Orsay
for the ToF walls, the Crystal Barrel collaboration for the APDs for SciRi, GSI for their hospitality and crucial help in measuring the OD field maps and the CMS group of INFN-Bari (Italy) for
the MWPC FEC2 chips.

This work is supported by SFB/TR-16, DFG project numbers 388979758 and 405882627, the RSF grant number 19-42-04132, the Third Scientific Committee of the INFN and the European Union’s Horizon 2020 
research and innovation programme under grant agreement number 824093.

%

\end{document}